\documentclass{aa}

\usepackage{natbib}
\usepackage{amsfonts}
\usepackage{graphicx}
\usepackage{amssymb}
\usepackage{float}
\usepackage{amsmath}
\usepackage[hyphens]{url}
\usepackage{hyperref}
\usepackage{fontenc}

\begin{document}
\title{ICM cooling, AGN feedback, and BCG properties of galaxy groups}
\subtitle{Five properties where groups differ from clusters}
\author{V.~Bharadwaj\inst{1}, T.~H.~Reiprich\inst{1}, G.~Schellenberger\inst{1}, H.~J.~Eckmiller\inst{1}, R.~Mittal\inst{2}, H.~Israel\inst{3,1}}\institute{
  Argelander-Institut f\"ur Astronomie, Auf dem H\"ugel 71, 53121
  Bonn, Germany \and Center of Imaging Science, Rochester Institute of
    Technology, 54 Lomb Memorial Drive, Rochester, NY, USA 14623 \and Department of Physics, Durham University, South Road, Durham DH1 3LE, UK}
\date{Submitted}
\abstract{}
{ We aim to investigate cool-core and non-cool-core properties of
  galaxy groups through X-ray data and compare them to the AGN radio
  output to understand the network of intracluster medium (ICM) cooling and feedback by supermassive black holes. We also aim to
  investigate the brightest cluster galaxies (BCGs) to see how they are affected by cooling and heating processes, and compare the properties of groups to those of clusters.}  {Using
  Chandra data for a sample of 26 galaxy groups, we constrained the
  central cooling times (CCTs) of the ICM and classified the groups as
  strong cool-core (SCC), weak cool-core (WCC), and non-cool-core (NCC)
  based on their CCTs. The total radio luminosity of the BCG was
  obtained using radio catalogue data and/or literature, which in turn
  was compared to the cooling time of the ICM to understand the link
  between gas cooling and radio output. We determined K-band luminosities of the BCG with 2MASS data,
  and used a scaling relation to constrain the masses of the supermassive black holes, which
  were then compared to the radio output. We also tested for correlations between the BCG luminosity and the overall X-ray luminosity and mass of the group. The results obtained for the
  group sample were also compared to previous results for clusters.}  {The
  observed cool-core/non-cool-core fractions for groups are comparable
  to those of clusters. However, notable differences are seen: 1)~for
  clusters, all SCCs have a central temperature drop, but for groups
  this is not the case as some have centrally rising temperature
  profiles despite very short cooling times; 2)~while for the cluster
  sample, all SCC clusters have a central radio source as opposed to
  only 45\% of the NCCs, for the group sample, \textit{all} NCC groups have a
  central radio source as opposed to 77\% of the SCC groups; 3)~for clusters,
  there are indications of an anticorrelation trend between radio
  luminosity and CCT. However, for groups this trend is absent; 4)~the Indication of a
  trend of radio luminosity with black hole mass observed in SCC
  clusters is absent for groups; and 5)~similarly, the strong correlation
  observed between the BCG luminosity and the cluster X-ray luminosity/cluster mass weakens significantly for
  groups.}  {We conclude that there are important differences between
  clusters and groups within the ICM cooling/AGN feedback paradigm and speculate that more gas is fueling star formation in groups, than in clusters where much of the gas is thought to feed the central AGN.}

\titlerunning{ICM Cooling, AGN Feedback and BCG properties of Galaxy Groups}
\authorrunning{V.~Bharadwaj et al.}
\maketitle 

\section{Introduction}

The discovery that the cooling time of the intracluster medium (ICM)
in the centres of many clusters, the so-called cool-core clusters, is
much shorter than the Hubble time (e.g.~\citealt{ 1973ApJ...184L.105L,
  1977ApJ...215..723C, 1977MNRAS.180..479F}) led to the development of
the cooling flow model. In this model, as the gas cools
hydrostatically, it is compressed by the hot, overlying gas,
generating a cooling flow (see \citealt{1994ARA&A..32..277F} for a
review). After early indications from the advanced satellite for cosmology and astrophysics, i.e. ASCA (e.g.~\citealt{2001PASJ...53..401M}), the high-spectral resolution data from the reflection grating spectrometer (RGS) instrument on
XMM-Newton have shown that the actual mass deposition rates fall short
of the predictions by an order of magnitude. These data showed that not
enough cool gas was present in the cool-core clusters
(e.g.~\citealt{2001A&A...365L.104P,2002ApJ...579..600X,2001A&A...365L..87T,2001ASPC..234..351K,refId,2003ApJ...590..207P,2006PhR...427....1P,2008MNRAS.385.1186S}). %
Optical and UV data revealed the same level of discrepancy between the
expected and observed star formation rates
(e.g.~\citealt{1989AJ.....98.2018M,1995MNRAS.276..947A,2003ApJ...594L..13E, 2006ApJ...652..216R,  2008ApJ...687..899R}).

Several different criteria have been used in the literature to
distinguish between cool-core~(CC) and non-cool-core~(NCC) clusters,
making it difficult to compare results. Some of the parameters used
are central entropy (e.g.~\citealt{ 2008ApJ...681L...5V}, \citealt{2009ApJS..182...12C}, \citealt{
  2011arXiv1106.4563R}), central temperature drop (e.g.~\citealt{
  2006MNRAS.372.1496S}, \citealt{ 2008ApJ...675.1125B}), classical
mass deposition rate (e.g.~\citealt{2007A&A...466..805C}), and central
cooling time (e.g.~\citealt{ 2010A&A...513A..37H}). \cite{
  2010A&A...513A..37H} analysed 16 parameters using the HIFLUGCS
sample \citep{ Reiprich-Boehringer:02} and concluded that for low-$z$
clusters, the central cooling time (CCT) was the best parameter to use to
distinguish between CC and NCC clusters.

The classical cooling flow model does not consider any heating
mechanism in addition to radiative cooling of the intracluster gas. In
recent years, several models incorporating heating mechanisms have
been explored, such as heating by supernovae, thermal conduction, and
active galactic nuclei~(AGN). Self-regulated AGN feedback has gained
favor in recent years (e.g.~\citealt{
  2002MNRAS.332..729C,2004ApJ...615..681R,2005ApJ...634..955V}). Excellent
correlations between X-ray deficient regions in the ICM, i.e.~cavities
and radio lobes, have given credence to this hypothesis (e.g.~\citealt{
  1993MNRAS.264L..25B, 2000ApJ...534L.135M, 2001ApJ...558L..15B,
  2004ApJ...616..178C, 2004ApJ...607..800B,  2008ApJ...686..859B}). 
\cite{ 2009A&A...501..835M} showed that the likelihood of a cluster
hosting a central radio source (CRS) increases as the CCT
decreases. The \textit{exact} details of the heating mechanism through
AGN are still unclear. Possibilities include heat transfer through
weak shocks or sound waves
(e.g. \citealt{2003MNRAS.344L..43F,2002ApJ...567L.115J,2006ApJ...638..659M}), cosmic ray interaction with the ICM \citep{ 2008MNRAS.384..251G},
and the $PdV$ work done by the expanding radio jets on the ICM (e.g.~\citealt{2004ApJ...615..675R}, see also \citealt{2007ARA&A..45..117M} for a review). 

Trying to understand the correlation between gas cooling in the intracluster medium (ICM)\footnote{We refrain from the abbreviation IGM in order to avoid confusion with the intergalactic medium.} and feedback processes on
the galaxy group scale is fundamental in our understanding of the
exact differences between clusters and groups. Galaxy groups, being
systems not as massive as clusters, have long been considered
scaled-down versions of clusters. The definition of a group
and cluster is extremely loose and a rule of thumb definition is to
designate systems comprising less than 50 galaxies as a group and
above 50 as a cluster, but in recent years there has been some suggestion
that groups cannot be simply treated as scaled-down versions of
clusters. For example, in clusters the ICM usually dominates the
baryonic budget, whereas in groups the combined mass of the member
galaxies may exceed the baryonic mass in the ICM
(e.g. \citealt{2009ApJ...703..982G}). Furthermore, the main cooling mechanism in groups (line emission) differs from that in clusters (thermal bremsstrahlung). In principle, feedback from an AGN would have a
greater impact on the group because of the smaller gravitational
potential. 

\cite{ 2011ApJ...726...86R} show that shock heating
perpetuated by an AGN explosion alone is enough to balance radiative
losses in the galaxy group NGC~5813. 
Groups such as HCG~62 \citep{2010ApJ...714..758G} show radio lobes corresponding to X-ray
cavities, 
which could indicate that the ICM cooling/AGN feedback on the group
regime is similar to that in galaxy clusters. However, as pointed out
by \cite{ 2009ApJ...704.1586S}, this is not so trivial and the situation is
complicated even more by the presence of the so-called coronae class of
kpc-sized objects representing emission from individual galaxies,
likely harbouring a supermassive black hole (SMBH).
\cite{2011MNRAS.415.1549G} argue through simulations that AGN feedback
on the group regime is persistent and delicate unlike in clusters. It is
worth stating here that the multitude of cool-core definitions used to
define clusters and comparisons between the cool-core properties has
not been thoroughly tested on the group regime with an objectively
selected group sample. Thus, it becomes highly imperative to see how groups are different to clusters
within the cool-core/feedback paradigm and to what extent; motivated by the basic fact that groups of galaxies are more numerous
than clusters (as evident from the shape of the cluster mass function). Most of the $10^{5}$ clusters to be detected
by the upcoming eROSITA X-ray mission will be groups of galaxies or
low-mass clusters \citep{2012MNRAS.422...44P}. The aim of upcoming cluster missions, eROSITA in particular, is to
perform precision cosmology using clusters as cosmological probes. The processes of ICM
cooling and AGN feedback could easily cause scaling
relations to diverge from their norm (e.g.~\citealt{
  2011A&A...532A.133M}), which in turn could undermine their utility
in cosmological applications.

The brightest cluster galaxies (BCGs) have a special role in the
aforementioned paradigm. These are highly luminous galaxies, usually
of giant elliptical and cD type, and are generally located quite close
to the X-ray peak ($< 50 h^{-1}$kpc in 88\% of the
cases, \citealt{ 2010A&A...513A..37H}). Elliptical galaxy properties, like the
optical bulge luminosity and velocity dispersion, have well defined
scaling relations, allowing one to indirectly estimate the mass of the
SMBH (e.g. \citealt{1995ARA&A..33..581K},
\citealt{2000ApJ...539L...9F}), which may be compared to the AGN
activity (e.g.~\citealt{ 2004ApJ...612..797F, 2009A&A...501..835M}). Brightest cluster galaxies also
help in studying the role of cooling gas in fueling star formation where the cooling gas at least partially forms new stars
(e.g.~\citealt{ 2010ApJ...719.1844H}). \cite{ 2009A&A...501..835M} also allude to the possibility of
different growth histories for SCC BCGs than for non-SCC
BCGs. An investigation of BCGs on the group regime and comparison to
cluster BCGs, factoring in the CC/NCC paradigm has not yet been
carried out.

In this paper, we attempt to address several questions related to gas
cooling properties, AGN feedback and BCG properties on the galaxy
group scale. It is organised as follows: Section~\ref{sample} 
deals with the sample selection and data analysis. Section~\ref{Results} contains
the results. A discussion of our results is carried out in Sect.~\ref{Discussion}. A short summary is presented in Sect.~\ref{Summary}. Throughout this work, we
assume a $ \Lambda$CDM cosmology with $ \Omega_{\mathrm{m}} = 0.3$, $
\Omega_{\Lambda} = 0.7$, and $ h = 0.71$, where $H_{0} = 100 h$
km/s/Mpc. Correlations between different physical parameters were
quantified with the help of linear regression analysis using the BCES
bisector code by \cite{ 1996ApJ...470..706A}, and the degree of the
correlations was estimated using the Spearman rank correlation
coefficient. All errors are quoted at the $1 \sigma$ level unless
stated otherwise.

\section{Sample selection and data analysis}
\label{sample}

\subsection{Sample selection}

The group sample used in this work is the same as used by \cite{
  Eckmiller}. The main aim of that study was to test scaling
relations on the galaxy group scale. The groups were selected from the
following three catalogues:
\begin{itemize}
\item NORAS: Northern ROSAT All Sky galaxy cluster survey by \cite{
    Boehringer-Voges-Huchra:00}.
\item REFLEX:~ROSAT-ESO Flux Limited X-ray galaxy cluster survey by
  \cite{ Boehringer-Schuecker-Guzzo:04}.
\item HIFLUGCS: Highest X-ray Flux Galaxy Cluster Sample by \cite{
    Reiprich-Boehringer:02}.
\end{itemize}

An upper-cut on the luminosity of $ 2.55 \cdot 10^{43} h^{-2}_{70}$
erg~s$^{-1}$ in the ROSAT band was applied to select only groups, and a lower redshift cut of $z > $ 0.01 was applied to
exclude objects too close to be observed out to a large enough
projected radius in the sky. This yielded a statistically complete
sample of 112 groups. However, not all of them have high-quality X-ray
data, and so only those groups with Chandra observations were
selected, giving a sample of 27 galaxy groups. One group, namely
IC~4296 was excluded as the observations were not suitable for ICM analysis, which resulted in a final sample of 26 groups. More details about the sample are provided in
\cite{Eckmiller}.

\subsection{Data reduction}

For the data reduction we used the CIAO software
package\footnote{\url{http://cxc.harvard.edu/ciao}} in version 4.4
with CALDB 4.5.0. Following the suggestions on the Chandra Science
Threads, we reprocessed the raw data (removing afterglows, creating the
bad-pixel table, applying the latest calibration) by using the
\verb+chandra_repro+ task. Soft-proton flares were filtered by
cleaning the lightcurve with the \verb+lc_clean+ algorithm (following
the steps in the Markevitch cookbook
\footnote{\url{http://cxc.harvard.edu/contrib/maxim/acisbg/COOKBOOK}}). All
lightcurves were also visually inspected afterwards for any residual flaring. Point sources
were detected by the \verb+wavdetect+ wavelet algorithm (the images were visually inspected to ensure that the detected point sources were reasonable) and excluded
from the spectral and surface-brightness analysis. The regions used
for the spectral analysis were selected by a count threshold of at
least 2500 source counts and fit by an absorbed (\textit{wabs}) APEC model, centred on the emission peak (EP), determined using the tool \textit{fimgstat}. Table 1 gives the co-ordinates of the EP in RA/DEC. The \cite{1989GeCoA..53..197A} abundance table was used throughout.

\subsubsection{Background subtraction}

For the background subtraction we followed the steps illustrated by e.g. \cite{2009ApJ...699.1178Z} and \cite{ 2009ApJ...693.1142S}
with some minor modifications.

The particle background, i.e. the highly energetic particles that interact
with the detector, was estimated from stowed events files distributed
within the CALDB. The astrophysical X-ray background was estimated
from a simultaneous spectral fit to the Chandra data and data from the
ROSAT all-sky survey (RASS) provided by Snowden's webtool\footnote{\url{http://heasarc.gsfc.nasa.gov/cgi-bin/Tools/xraybg}}, using the
X-ray spectral fitting package \textit{Xspec}. The components used to
fit the background were an absorbed power law with a spectral index of
1.41 (unresolved AGN), an absorbed APEC model (Galactic halo
emission) and an unabsorbed APEC model (Local Hot Bubble emission; see
\citealt{1998ApJ...493..715S}). The RASS data were taken from an
annulus far away from the group centre, where no group emission would
be present. On average we found temperatures of $0.25\,{\rm keV}$ for
the Galactic halo component and $0.11\,{\rm keV}$ for the Local Hot
Bubble emission.

The background for the surface-brightness analysis was estimated from the fluxes of the background models plus that from
the particle background estimated from the stowed events files
(weighted for each region according to the ACIS chips used, and
exposure corrected with the exposure maps created for the galaxy
groups in an energy range of 0.5-2.0 keV). These values were then subtracted from the surface brightness
profiles (SBPs), which were also obtained from an exposure corrected image
in an energy range of 0.5-2.0 keV.

\subsection{Surface brightness profiles and density profiles}
Centered on the EP, the SBP was fitted with either a
single or a double $\beta$ model given by
\begin{equation}
\Sigma = \Sigma_{0} \left [ 1+\left ( \frac{x}{x_{\mathrm{c}}} \right )^{2} \right ]^{-3\beta+1/2} \\
\end{equation}
or
\begin{equation}
\Sigma = \mathrm{\Sigma}_{01}\left[1+\left(\frac{x}{x_\mathrm{{c_{1}}}}\right)^{2}\right]^{-3\mathrm{\beta}_{1}+1/2} +  \mathrm{\Sigma}_{02}\left[1+\left(\frac{x}{x_\mathrm{{c_{2}}}}\right)^{2}\right]^{-3\mathrm{\beta}_{2}+1/2}~,
\end{equation}
respectively \citep{1976A&A....49..137C}.  
Here, $x_{\mathrm{c_{i}}}$ is the core radius. The density profile for both the models is given by:
\begin{eqnarray}
n &=& n_{0}\left [ 1+\left ( \frac{r}{r_{\mathrm{c}}} \right )^{2} \right ]^{\frac{-3\beta}{2}} \\
n &=& \left(
n_{01}^2\left[1+\left(\frac{r}{r_\mathrm{{c_{1}}}}\right)^2\right]^{-3\beta_1} +
n_{02}^2 \left[1+\left(\frac{r}{r_\mathrm{{c_{2}}}}\right)^2 \right]^{-3\beta_2}
\right)^{1/2}
\end{eqnarray}
where $n_{0} = \sqrt{n^{2}_{01}+n^{2}_{02}}$ is the central electron
density and $r_{\mathrm{c}}$ is the physical core radius. The central
electron density $n_{0}$ can be directly determined using the
formulae \citep{ 2010A&A...513A..37H}:
\begin{equation}
 n_{0} = \left ( \frac{10^{14}4\pi D_{\mathrm{A}}D_{\mathrm{L}}\zeta N}{\mathrm{EI}} \right )^{\frac{1}{2}}
\end{equation}

\begin{equation}
\label{n0} 
n_{0} = \left [\frac{10^{14}4\pi (\Sigma_{12}\mathrm{LI_{2}}+\mathrm{LI_{1}} )D_{\mathrm{A}}D_{\mathrm{L}}\zeta N}{\Sigma_{12}\mathrm{LI}_{2}\mathrm{EI}_{1}+\mathrm{LI}_{1}\mathrm{EI}_{2} }  \right ]^{\frac{1}{2}} .
\end{equation}
Here, $N$ is the normalisation of the APEC model in the innermost
annulus; $\zeta$ is the ratio of electrons to protons ($\sim 1.2$);
$\mathrm{\Sigma}_{12}$ is the ratio of the central surface brightness
of model-1 to model-2; $D_{\mathrm{A}}$ and $D_{\mathrm{L}}$ are the
angular diameter distance and the luminosity distance,
respectively; $\mathrm{EI}_{i}$ is the emission integral for model-i
and is defined as
\begin{equation}
 \mathrm{EI} = 2\pi  \int_{-\infty }^{\infty } \!  \int_{0}^{R} x\left(1+\frac{x^{2}+l^{2}}{x^{2}_{c}} \right)^{-3\beta} \mathrm{d}x\mathrm{d}l~,                                                                                                                                                                                                                                                                                                       
\end{equation}
where $R$ is the radius of the innermost annulus and $\mathrm{LI}_{i}$ is
the line emission measure for model-$i$ and is defined as
\begin{equation}
 \mathrm{LI}_{i} = \int_{-\infty}^{\infty} \left(1+\frac{l^{2}}{x^{2}_{c_{i}}} \right)^{-3\beta_{i}} \mathrm{d}l~.
\end{equation}
More details can be found in \cite{ 2010A&A...513A..37H}.

\subsection{Cooling times and central entropies}

The major focus of our work is connected to the CCT. This is
calculated using the formula:
\begin{equation*}
t_{\mathrm{cool}} = \frac{3}{2} \frac{(n_{\mathrm{e}}+n_{\mathrm{i}})kT}{n_{\mathrm{e}}n_{\mathrm{H}}\Lambda(T,Z)}
\end{equation*}

\begin{equation}
\label{CCTeq} 
\mathrm{CCT} = t_\mathrm{{cool}}(0) = \frac{3}{2}\zeta \frac{(n_\mathrm{{e0}}+n_\mathrm{{i0}})kT_{0}}{{n}^{2}_\mathrm{{e0}}\Lambda(T_{0},Z_{0})}  
\end{equation}
where ${n}_{i0}$ and ${n}_{e0}$ are the central ion and electron
densities, respectively, and $T_{0}$ is the central temperature. We note
that a bias due to different physical resolutions could be introduced 
arising because of different distances of the galaxy groups. Hence, we
took any parameter (except the central temperature) calculated at
$r=0$ to be the value at $r=0.004r_{500}$. The central temperature
$T_{0}$ is simply the temperature in the innermost bin in the
temperature profile. As in \cite{2010A&A...513A..37H}, $r_{500}$ was calculated from a scaling relation by \cite{1996ApJ...469..494E} and is given by \begin{equation}
                                                                                                                              r_{500} = 2\times \left (\frac{kT_{\mathrm{vir}}}{10~\mathrm{keV}}  \right )^{\frac{1}{2}}~,
                                                                                                                             \end{equation}
where the virial temperature was taken from \cite{Eckmiller} to calculate the $r_{500}$.                                                                                                                     

To ensure that the determination of the CCT is not strongly biased because of selection of annuli on the basis of a counts threshold, we performed tests for a few cases where the temperature and surface brightness annuli were increased by a factor of $\sim 3-4$. We did not identify any strong bias that could drastically affect our results.

The central entropy $K_{0}$, another important CC parameter, is
calculated as:
\begin{equation}
K_{0} = k T_{0} n^{-2/3}_\mathrm{e0}~.
\end{equation}

\subsection{Radio data and analysis}

All the radio data required for this work was either compiled from
existing radio catalogs or literature (references in Table \ref{radiodata}). We obtained data at several
frequencies between 10~MHz and 15~GHz.
The major catalogs used for this study were the NVSS
(1.4~GHz)\footnote{NRAO VLA Sky
  Survey-\url{http://www.cv.nrao.edu/nvss/}}, SUMSS
(843~MHz)\footnote{Sydney University Sky
  Survey-\url{http://www.physics.usyd.edu.au/sifa/Main/SUMSS}}, and VLSS
(74~MHz)\footnote{VLA Low frequency Sky
  Survey-\url{http://lwa.nrl.navy.mil/VLSS/}} catalogs.
  
Since this study involves radio sources associated with BCGs at the
center of the dark matter halo, it is imperative to set a criterion
for whether or not a group has a central radio source. Based on the
work of \cite{2007MNRAS.379..100E}, \cite{ 2009A&A...501..835M} suggest that a central radio source must
be located within 50~$h^{-1}_{71}$~kpc of the X-ray peak in order for
it to be categorised as a central radio source (CRS). We adopted the same
criterion in this work and discovered that most CRSs lie close to the EP (within a few kpc). Appendix~\ref{radicont} shows the location of the CRS with radio
contours overlaid on the optical images with the X-ray emission peak
also marked for most of the groups. For CRSs with extended emission, we considered the radio
emission from the lobes as well, since our goal is to obtain a correlation between the CCT
and the total radio emission from the central AGN.

Radio emission by AGN is characterised by synchrotron radiation
expressed as a power law relation given by $S_{\nu}
\varpropto \nu^{-\alpha} $, where $S_{\nu}$ is the flux
density at frequency $\nu$ and $\alpha$ is the spectral index. Much of
the synchrotron emission comes from the lower frequencies ($<
1.4~$GHz), making it highly important to obtain data on these
frequency scales. Moreover, a full radio spectral energy distribution
is advantageous since that allows spectral breaks and turn-overs to be
discovered. Spectral breaks indicate spectral aging and turn-overs
indicate self-absorption. Self-absorption is characterized by a
negative spectral index, particularly at lower frequencies. The
integrated radio luminosity between a pair of frequencies
$\mathrm{\nu}_{i}$ and $\mathrm{\nu}_{i+1}$ is given by
\begin{equation}
 L_{i+1} = 4\pi D^{2}_{L} \frac{S_{0}\mathrm{\nu}_{0}^{\mathrm{\alpha}_{i+1,i}}}{1-\mathrm{\alpha}_{i+1,i}}\left ( \nu^{1-\alpha_{i+1,i} }_{i+1}-\nu^{1-\alpha_{i+1,i} }_{i}   \right )~,
\end{equation}
where $S_{0}$ is the flux density at either frequency
$\mathrm{\nu}_{i+1,i}$ or $\mathrm{\nu}_{i}$,
$\mathrm{\alpha}_{i+1,i}$ is the spectral index between the two
frequencies, and $D_{\mathrm{L}}$ is the luminosity distance. To
calculate the total radio luminosity between 10~MHz and 15~GHz, the
spectral index at the lowest observed frequency was extrapolated to
10~MHz, and the spectral index at the highest observed frequency was
extrapolated to 15~GHz. The integrated radio luminosity was then
calculated as $L_{\mathrm{tot}} = \Sigma L_{\mathrm{i+1}}$. In the case of
unavailability of multi-frequency data, we assumed a spectral index of
1 throughout the energy range (e.g.~\citealt{
  2009A&A...501..835M}). This had to be done for 11 CRSs in the
sample. Table~\ref{radiodata} summarises the radio data.

\subsection{BCG data and analysis}

For the BCG analysis, we followed the same methodology as explained in
\cite{ 2009A&A...501..835M} and describe it here briefly.

The BCG near-infrared (NIR) $K$-band magnitudes
($km_\mathrm{ext}$) are obtained from the 2MASS Extended Source
Catalog \citep{2000AJ....119.2498J,2006AJ....131.1163S}, i.e. the
XSC. Redshifts were obtained from the NASA/IPAC Extragalactic Database
(NED). The magnitudes were corrected for Galactic extinction using
dust maps by \cite{1998ApJ...500..525S}. As these are extremely low
redshift galaxies, no $k$-correction was applied. The magnitudes were
then converted to luminosities under the Vega system, assuming an
absolute $K$-band solar magnitude equal to 3.32 mag
\citep{1997AJ....113.1138C}.

Studies like \cite{2003ApJ...589L..21M} and \cite{2007ApJ...663L..85B}
have established well-defined scaling relations between galaxies' NIR
bulge luminosity and the SMBH mass, consistent with results obtained
from velocity dispersions (e.g.~\citealt{ 2002ApJ...574..740T}). 

We use the scaling relation from \cite{2003ApJ...589L..21M} to obtain the SMBH mass,
\begin{equation}
 \mathrm{log_{10}}\left ( \frac{M_{\mathrm{BH}}}{M_{\odot}} \right )=a+b\left [ \mathrm{log}_{10}\left ( \frac{L_{\mathrm{BCG}}}{L_{\odot}} \right )-10.9 \right ]~,
\end{equation}
where $ a = 8.21\pm 0.07$ and $b = 1.13\pm 0.12 $. The derived SMBH
mass was compared to the integrated radio luminosity. The BCG
luminosities were compared to the global cluster properties, like the total X-ray
luminosity $L_{\mathrm{X}}$ and mass $M_{500}$.

\section{Results}\label{Results}

\subsection{Cool-core and non-cool-core fraction}\label{fractions}

\begin{figure}
 \includegraphics[trim = 20mm 10mm 22mm 10mm, clip, scale=0.40]{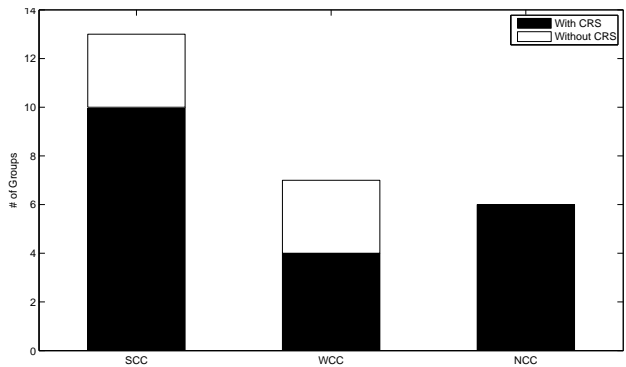}
\caption{Fraction of SCC, WCC, and NCC groups. Shaded regions are
  groups with central radio sources (see Sect.~\ref{CRSfract})}
\label{fig:fraction}
\end{figure}

\begin{figure}
\includegraphics[trim = 35mm 5mm 22mm 12mm, clip, scale=0.30]{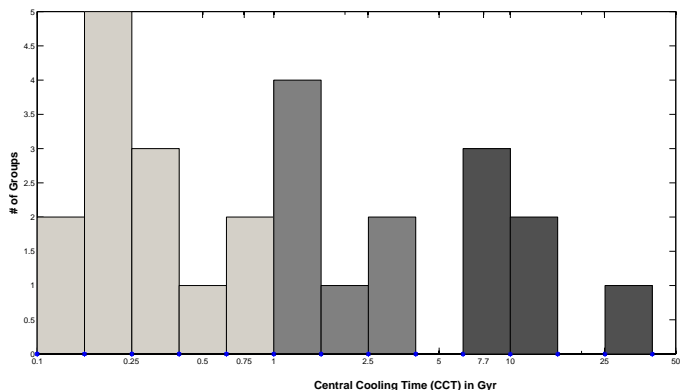}
\caption{Histogram of central cooling time. The increasing shade of gray indicates increasing cooling time. Light grey represents SCC groups, medium grey is for WCC groups, and dark grey is for NCC groups.}
\label{fig:CCThist}
\end{figure}

\cite{ 2010A&A...513A..37H} analysed 16 parameters using Kaye's
Mixture Mode (KMM) algorithm as described by \cite{1994AJ....108.2348A}, where
the CCT showed a strong trimodal distribution. Thus, the authors divided the clusters into three
categories: strong cool-core (SCC) clusters with CCTs below 1~Gyr, weak
cool-core (WCC) clusters with CCTs between 1~Gyr and 7.7~Gyr, and
non-cool-core (NCC) clusters with CCT above 7.7 Gyr. Using the same
classification system, we present our sample classified as SCC, WCC,
and NCC groups in Table~\ref{CCT}. This table also shows the central
electron density and the central entropy (Sect.~\ref{Entropy}). The
observed SCC fraction is 50\%, the WCC fraction is 27\%, and the NCC fraction
is 23\% (Fig.~\ref{fig:fraction}). A histogram showing the
distribution of the CCTs is shown in Fig.~\ref{fig:CCThist}.

\begin{table*}
\begin{center}
\caption{The columns are (1) group name, (2) co-ordinates (J2000) of the EP, (3) central electron
    density, (4) central temperature,
    (5) central cooling time, (6) central entropy, (7) cool-core type, and (8) presence or absence of a central radio source.}
\renewcommand{\arraystretch}{1.2}
\begin{tabular}{| c  c c  c  c c  c c|}
\hline \hline
Group name& EP (RA/DEC)&$n_{0}$ ($10^{-2}~\mathrm{cm}^{-3}$) &$T_{0}$ keV&CCT (in Gyr)&$ K_{0}$ ($\mathrm{keV} \mathrm{cm}^{2}$) &CC type& CRS?\\ \hline 
A0160& $01:12:59.67~+15:29:29.11$ &$0.380^{+0.068}_{-0.068}$&$2.38^{+0.218}_{-0.218}$&$8.04^{+1.26}_{-1.83}$ &$97.7^{+10.1}_{-13.7} $ & NCC & YES\\ 
A1177& $11:09:44.38~+21:45:32.81$ &$0.600^{+0.130}_{-0.130}$&$1.63^{+0.08}_{-0.08}$&$3.77^{+0.67}_{-1.04}$ &$49.4^{+6.05}_{-8.73} $ & WCC & NO\\ 
ESO55& $04:54:52.31~-18:06:54.29$ &$0.760^{+0.14}_{-0.14}$&$1.88^{0.13}_{-0.13}$ &$2.70^{+0.40}_{-0.57}$ &$48.6^{+5.18}_{-7.07}$ & WCC & NO\\ 
HCG62&$12:53:06.00~-09:12:11.57$ &$3.80^{+1.23}_{-1.23}$ &$0.813^{+0.008}_{-0.008}$ &$0.214^{+0.0523}_{-0.1023}$ &$7.19^{+1.23}_{-2.14}$& SCC & YES\\ 
HCG97&$23:47:23.03~-02:18:00.49$ &$1.44^{+0.19}_{-0.19}$ &$0.992^{+0.015}_{-0.015}$ &$0.977^{+0.114}_{-0.149}$ &$16.8^{+1.33}_{-1.66}$& SCC & NO \\ 
IC1262&$17:33:03.07~+43:45:34.88$ &$4.12^{+0.13}_{-0.13}$&$1.63^{+0.03}_{-0.03}$&$0.504^{+0.015}_{-0.016}$ &$13.7^{+0.280}_{-0.295}$& SCC & YES\\ 
IC1633&$01:09:56.07~-45:55:52.28$ &$0.38^{+0.0883}_{-0.0883}$&$2.82^{+0.13}_{-0.13}$&$8.20^{+1.54}_{-2.48}$ &$115^{+15.1}_{-22.3}$& NCC & YES\\ 
MKW4 &$12:04:27.14~+01:53:45.18$&$3.91^{+0.34}_{-0.34}$&$1.57^{+0.02}_{-0.02}$ &$0.258^{+0.0207}_{-0.0246} $ &$13.6^{+0.737}_{-0.852}$ & SCC & YES\\ 
MKW8&$14:40:42.99~+03:27:56.98$ &$0.34^{+0.0453}_{-0.0453}$&$3.69^{+0.18}_{-0.18}$&$10.1^{+1.19}_{-1.56}$ &$163^{+13.0}_{-16.3}$ & NCC & YES\\ 
NGC326&$00:58:22.82~+26:51:51.26$ &$0.34^{+0.079}_{-0.079} $&$1.87^{+0.09}_{-0.09}$ &$7.77^{+1.46}_{-2.35}$ &$82.7^{+10.8}_{-15.9} $& NCC & YES\\ 
NGC507&$01:23:39.93~+33:15:21.98$ &$1.09^{+0.13}_{-0.13}$&$1.22^{+0.02}_{-0.02}$&$1.22^{+0.13}_{-0.16}$ &$24.8^{+1.79}_{-2.19}$& WCC & YES\\ 
NGC533&$01:25:31.45~+01:45:32.69$ &$4.67^{+0.68}_{-0.68}$&$0.889^{+0.011}_{-0.011}$ &$0.191^{+0.0244}_ {-0.0327}$ &$6.85^{+0.59}_{-0.758}$& SCC & YES\\ 
NGC777&$02:00:14.94~+31:25:46.28$ &$4.99^{+1.17}_{-1.17}$&$1.21^{+0.05}_{-0.05}$ &$0.167^{+0.0318}_{-0.0513}$ &$8.92^{+1.17}_{-1.74}$& SCC & YES\\ 
NGC1132&$02:52:51.81~-01:16:28.85$ &$1.21^{+0.11}_{-0.11}$&$1.17^{+0.04}_{-0.04}$&$1.08^{+0.11}_{-0.09}$ &$22.2^{+1.25}_{-1.46}$& WCC & YES\\ 
NGC1550&$04:19:38.37~+02:24:38.92$ &$5.53^{+0.35}_{-0.35}$&$1.21^{+0.007}_{-0.007}$ &$0.231^{+0.014}_{-0.016}$ &$8.34^{+0.334}_{-0.371}$& SCC & YES\\ 
NGC4325&$12:23:06.52~+10:37:15.52$ &$3.39^{+0.25}_{-0.25}$&$0.899^{+0.009}_{-0.009}$&$0.244^{+0.017}_{-0.019}$ &$8.58^{+0.398}_{-0.449}$& SCC & NO\\ 
NGC4936&$13:04:17.08~-30:31:35.37$ &$0.620^{+0.14}_{-0.14}$&$0.949^{+0.039}_{-0.039}$&$1.54^{+0.26}_{-0.45}$ &$28.1^{+3.57}_{-5.23}$& WCC & YES\\ 
NGC5129&$13:24:10.08~+13:58:37.06$ &$3.30^{+0.74}_{-0.74}$&$0.894^{+0.026}_{-0.026}$&$0.298^{+0.054}_{-0.086}$ &$8.69^{+1.10}_{-1.60}$& SCC & YES\\ 
NGC5419&$14:03:38.77~-33:58:41.81$ &$0.210^{+0.086}_{-0.086}$&$2.09^{+0.097}_{-0.097}$ &$13.1^{+9.16}_{-3.81}$ &$127^{+26.1}_{-53.6}$& NCC & YES\\ 
NGC6269&$16:57:58.01~+27:51:15.07$ &$2.10^{+0.37}_{-0.37}$&$1.53^{+0.07}_{-0.07}$&$0.914^{+0.137}_{-0.197}$ &$20.1^{+2.06}_{-2.77}$& SCC & YES\\ 
NGC6338&$17:15:22.99~+57:24:39.06$ &$5.43^{+0.36}_{-0.36}$&$1.27^{+0.02}_{-0.02}$&$0.252^{+0.0157}_{-0.0179}$ &$8.86^{+0.371}_{-0.414}$& SCC & YES\\ 
NGC6482&$17:51:48.81~+23:04:18.19$ &$7.35^{+1.43}_{-1.43}$&$0.940^{+0.021}_{-0.021}$ &$0.134^{+0.0218}_{-0.0323}$ &$5.36^{+0.599}_{-0.831}$& SCC & NO\\ 
RXCJ1022&$10:22:09.98~+38:31:22.32$&$0.93^{0.17}_{-0.17}$&$1.99^{+0.07}_{-0.07}$ &$2.45^{+0.38}_{-0.55}$ &$45.0^{+4.76}_{-6.48}$& WCC & NO\\ 
RXCJ2214&$22:14:45.95~+13:50:23.76$&$1.09^{+0.23}_{-0.23}$&$1.12^{+0.05}_{-0.05}$ &$1.47^{+0.39}_{-0.25}$ &$22.8^{+2.73}_{-3.89}$& WCC & YES\\ 
S0463&$04:29:07.54~-53:49:39.44$ &$0.12^{+0.0270}_{-0.0270}$&$3.50^{+0.48}_{-0.48}$ &$32.6^{+5.99}_{-9.47}$ &$309^{+39.2}_{-57.4}$& NCC & YES\\ 
SS2B153&$10:50:26.12~-12:50:41.32$ &$7.5^{+1.29}_{-1.29}$&$0.997^{+0.012}_{-0.012}$ &$0.103^{+0.0151}_{-0.0214}$ &$5.61^{+0.563}_{-0.752}$& SCC & YES\\
\hline \hline
\end{tabular}
\label{CCT}
\end{center}
\end{table*}

\subsection{Temperature profiles}

The temperature profiles for all the groups, centered on the emission
peak, are shown in Appendix \ref{Tempprofiles}. We clearly see that
there is no universal inner temperature profile and the magnitude of the central
temperature drop, if present, varies considerably. \cite{
  2010A&A...513A..37H} clearly showed that \textit{all} SCC clusters
had a central temperature drop, indicating the presence of a cool
core. However, for our group sample, we find cases where despite
extremely short cooling times, a central temperature \textit{rise} in
the innermost region is seen. This stark contrast to the properties of
clusters is discussed in detail in Sect.~\ref{Tprofdiscuss}.

\subsection{Central entropy \texorpdfstring{$K_{0}$}{K0}}\label{Entropy}

\begin{figure}
\includegraphics[trim = 30mm 0mm 22mm 8mm, clip, scale=0.33]{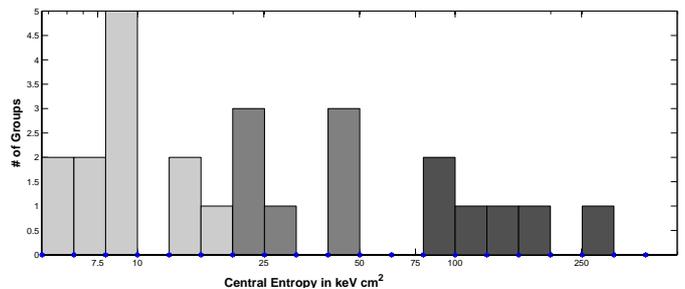}
\caption{Histogram of the central entropy. Light grey represents SCC groups, medium grey represents WCC groups, dark grey represents NCC groups.}
\label{fig:Khist}
\end{figure}

The central entropy is another parameter often used to classify CC/NCC
clusters (e.g.~\citealt{ 2008ApJ...687..899R}) and displays a tight
correlation with the CCT (\citealt{ 2010A&A...513A..37H}). This is not
surprising since the cooling time, $t_{\mathrm{cool}}$, is related to
the gas entropy, $K$, through the relation $t_{\mathrm{cool}} \propto
K^{3/2}/T$ for pure Bremsstrahlung. We also show a histogram of the distribution of the central
entropy in Fig.~\ref{fig:Khist}. The entropy values are given in
Table~\ref{CCT}.

The plot of the CCT and $K_{0}$ values is shown below in
Fig.~\ref{fig:entropy}. As expected, we see an excellent correlation
between the two quantities with a Spearman rank correlation coefficient of 0.99,
0.99, and 0.96 for all, SCC, and non-SCC groups respectively. A CCT of 1 Gyr corresponds to $\approx~20~\mathrm{keV~cm}^{2}$.

\begin{figure*}
\centering
\includegraphics[scale=0.50,angle=-90]{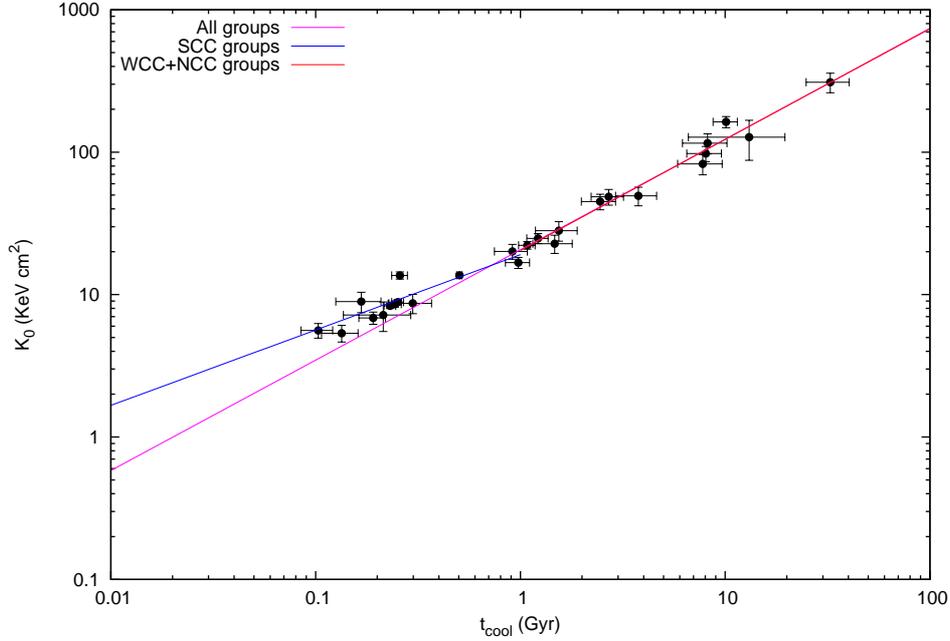} 
\caption{$K_{0}$ vs. CCT. Magenta is the best fit for all groups, blue
  for only SCC groups and red for WCC+NCC groups.}
\label{fig:entropy}
\end{figure*}

\subsection{Radio properties}

\subsubsection{CRS fractions-CC/NCC dichotomy}\label{CRSfract}

Table~\ref{CCT} also lists whether or not there is a central radio
source present in a group. We see that while \textit{all} NCC groups
have a CRS, the fractions of SCCs and WCCs containing a CRS are 77\%
and 57\%, respectively. The overall CRS fraction for CC groups is
70\%. Figure~\ref{fig:fraction} shows the CRS fractions in the group
sample.

\subsubsection{Total radio luminosity vs.~central cooling time}

\cite{ 2009A&A...501..835M} show that there is an anti-correlation
trend between the CCT and the total radio luminosity for CC clusters
(correlation coefficient of $-0.63$), which breaks down for cooling times shorter than 1~Gyr.

\begin{figure*}
\includegraphics[scale=0.35,angle=-90]{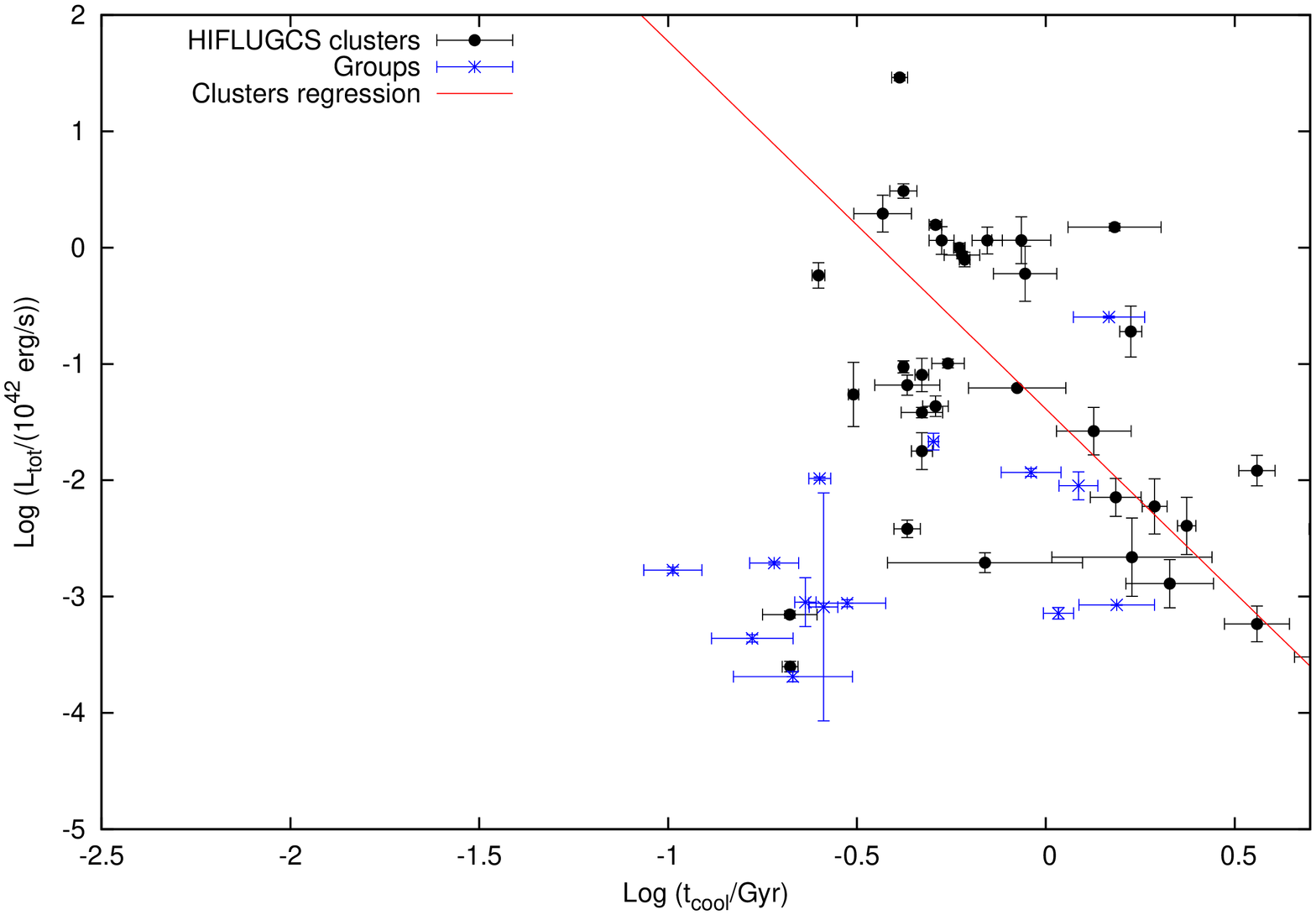}
\includegraphics[scale=0.35,angle=-90]{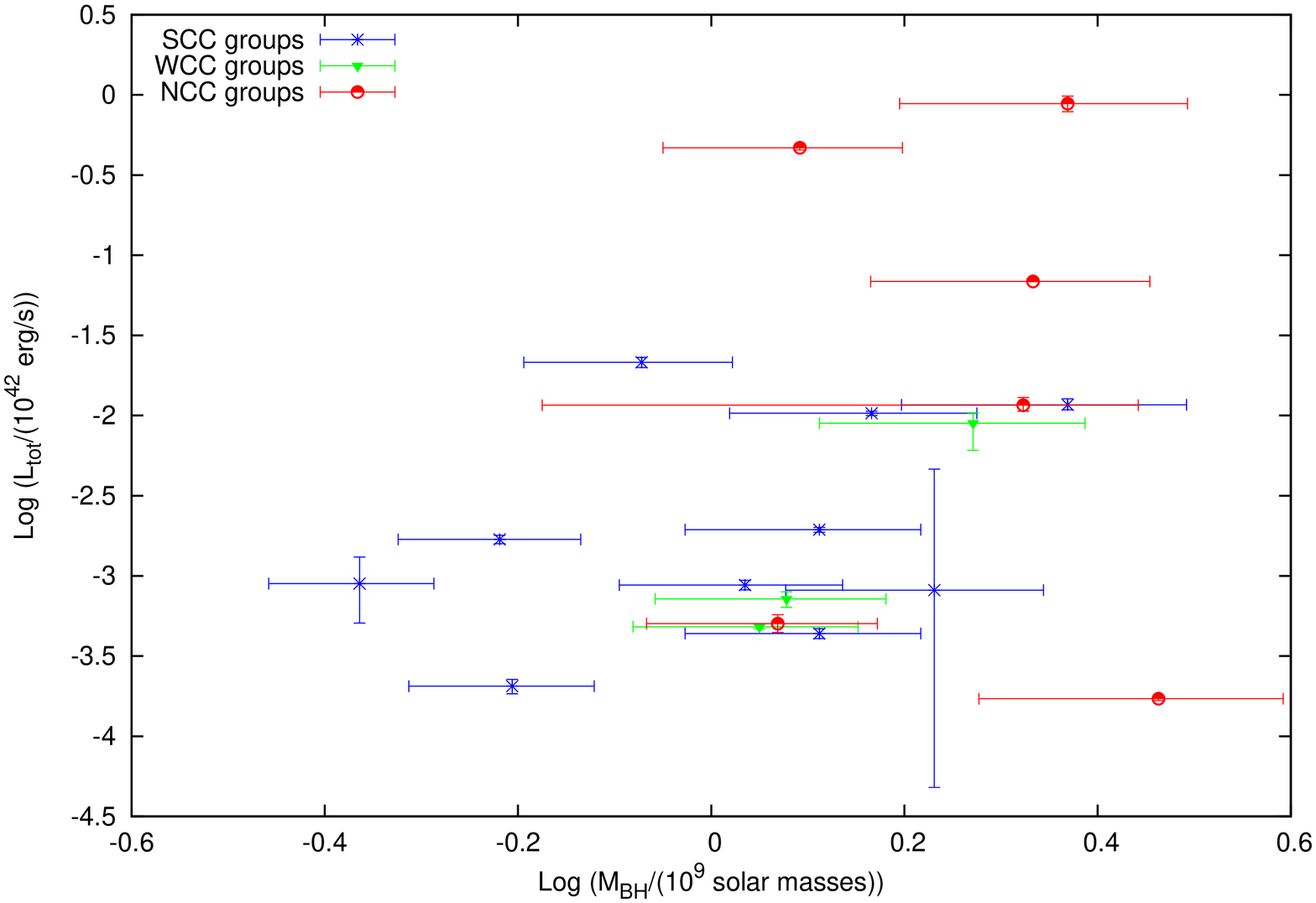} 
\caption{Left: Total radio luminosity vs. CCT with best-fit line for
  \textit{clusters} from \cite{ 2009A&A...501..835M}. Right: Total radio luminosity vs. mass of SMBH. The blue asterisks represent SCC groups, the green triangles represent WCC groups and the red circles represent NCC groups.}
\label{fig:Rtot}
\end{figure*}

Figure~\ref{fig:Rtot} shows the same plot for groups. We do not find
indications of a trend between the two quantities. Here we show the
best fit obtained for the CC clusters from \cite{
  2009A&A...501..835M} to highlight the difference between clusters
and groups. The power-law fit for the CC clusters from \cite{
  2009A&A...501..835M} is given by:

\begin{equation}
 L_{\mathrm{tot}} = (0.041\pm0.016)\times (t_{\mathrm{cool}})^{-3.16\pm0.38}~.
\end{equation}

It is interesting to note that \textit{all} SCC groups and most of the
WCC groups show a much lower radio output than the best fit for
clusters. This was first alluded to in \cite{ 2009A&A...501..835M},
where the groups in that sample (all SCC) were clear outliers and
we confirm this for the first time with a large sample of groups. We
discuss this in detail in Sect.~\ref{AGNact}.

\subsubsection{Total radio luminosity vs.~SMBH mass}

The total radio luminosity shows no trend with the SMBH mass
(Fig.~\ref{fig:Rtot}). Classifying the sample as SCC, WCC, and NCC also
does not yield any discernible correlations. This is in contrast with the HIFLUGCS sample, which shows a
weak correlation for the SCC clusters (correlation coefficient of
0.46). We calculate a correlation coefficient of 0.29 for all groups and 0.20
for only SCC groups. In Table~\ref{meanSMBHmassLrad} we present the mean SMBH masses and radio luminosities for the different CC types along with their standard errors to investigate whether different CC groups have systematically different masses and/or luminosities. We observe that the NCC groups have a systematically higher SMBH mass and radio luminosity than the SCC and WCC groups.

\begin{table*}
 \begin{center}
  \caption{Mean SMBH masses and radio luminosities for different CC groups.}\label{meanSMBHmassLrad}
  \begin{tabular}{|c|c|c|}
   \hline
   Group & Mean SMBH mass ($10^{9}$ solar masses) & Mean radio luminosity ($10^{42}$ erg/s) \\ \hline
   SCC &1.17$\pm$0.18&0.0050$\pm$0.0023\\
   WCC &1.39$\pm$0.24&0.0034$\pm$0.0028\\
   NCC &1.98$\pm$0.27&0.24$\pm$0.15\\ \hline
  \end{tabular}

 \end{center}

\end{table*}

\subsection{BCG properties}
\begin{figure*}
\centering
\includegraphics[scale=0.50, angle=270]{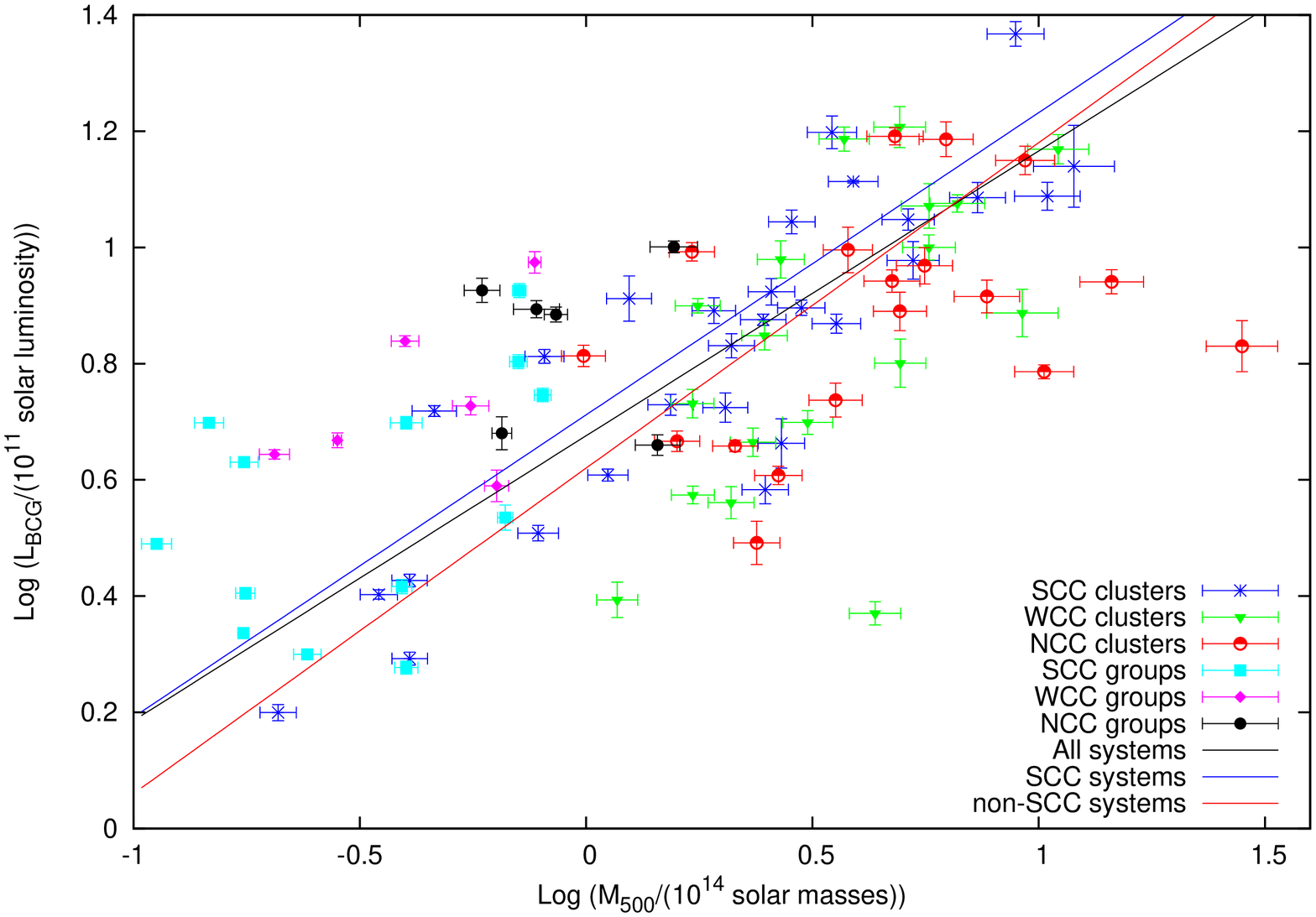} 
\caption{BCG luminosity vs.~$ M_{500}$. The black line is the best fit
  for all systems, the blue line for only SCC systems, and the red line
  for non-SCC systems.}
\label{fig:M500Lbcg}
\end{figure*}

We present the scaling relation between the BCG and the cluster/group X-ray luminosity and mass in Figs.~\ref{fig:LxLbcg} and \ref{fig:M500Lbcg}. The fits shown here are for a combined relation for groups and clusters. The details are summarized in Table~\ref{scat}. The derivation of the scatter is explained in Appendix~\ref{Scatcalc}. We observe that most group BCGs lie above the best-fit relations. Additionally, extending the scaling relations from clusters to groups leads to a higher intrinsic scatter in most cases.
 \subsubsection{BCG luminosity vs.~X-ray luminosity}

\begin{figure*}
\centering
\includegraphics[scale=0.50, angle=270]{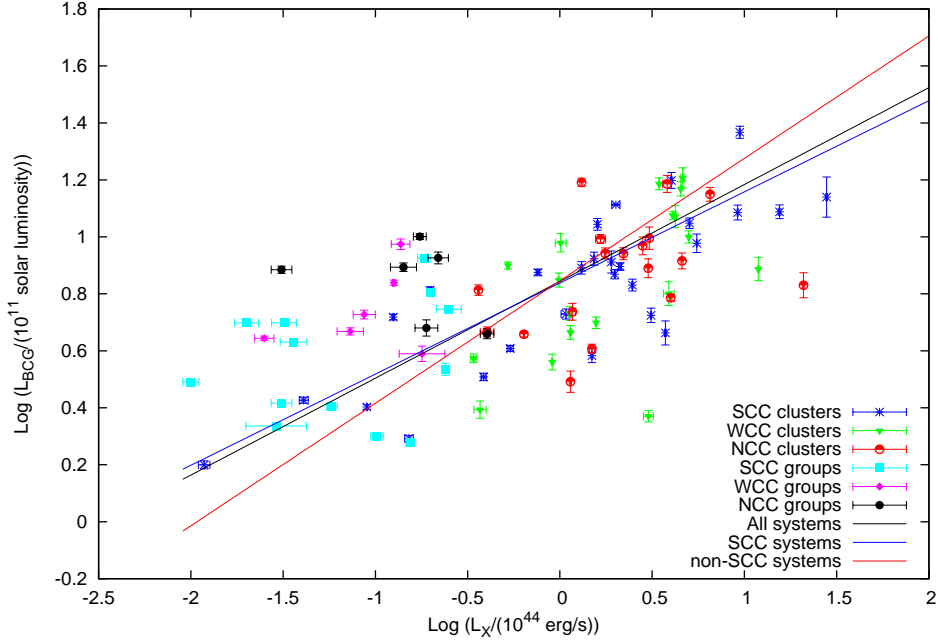} 
\caption{BCG luminosity vs.~X-ray luminosity. The black line is the
  best fit for all systems, the blue line for only SCC systems, and the
  red line for non-SCC systems.}
\label{fig:LxLbcg}
\end{figure*}

 Figure~\ref{fig:LxLbcg} shows the relation between the two quantities
 for both the HIFLUGCS clusters and the groups, with fits for all the
 members, SCC members, and non-SCC members. The correlation
 coefficients are 0.65, 0.79, and 0.46 for all, SCC and non-SCC systems,
 respectively. 

 It is seen that most group BCGs have a
 higher luminosity than expected from the derived scaling relation. The best-fit
 power-law relation is given by
\begin{equation}
 \left ( \frac{L_{\mathrm{BCG}}}{10^{11}L_{\odot}} \right )  = c\times \left ( \frac{L_{\mathrm{X}}}{10^{44} \mathrm{ergs/s}} \right )^{m}~,
\end{equation}
where $m = 0.34\pm0.03$ and $c = 6.98\pm0.16$ for all systems, $m
= 0.32\pm0.03$ and $c = 6.89\pm0.20$ for SCC systems, and $m =
0.43\pm0.09$ and $c = 7.01\pm0.30$ for non-SCC systems. We observe that when the relations are extended to the group regime, there is no significant difference in slopes and normalisations for the subsets.

\subsubsection{BCG luminosity vs.~\texorpdfstring{$ M_{500}$}{M500}}

Figure~\ref{fig:M500Lbcg} shows that the luminosity of the BCG grows
with the cluster/group mass. The correlation coefficients are 0.65,
0.86 and 0.44 for all, SCC and non-SCC systems respectively. The best
fit powerlaw relation is given by:
\begin{equation}
 \left ( \frac{L_{\mathrm{BCG}}}{10^{11}L_{\odot}} \right )  = c \times \left (\frac{M_{500}}{10^{14} M_{\odot}} \right)^{m}
\end{equation}
where $m = 0.49\pm0.03$ and $c = 4.74\pm0.12$ for all systems, $m
= 0.52\pm0.04$ and $c = 5.15\pm0.14$ for SCC systems, and $m =
0.56\pm0.08$ and $c = 4.17\pm0.03$ for non-SCC systems. We see that the normalisations for the SCC systems are around 23\% higher than those for non-SCC systems.

The total mass $M_{500}$ was calculated from the virial temperature
(taken from \citealt{ 2011A&A...532A.133M} and \citealt{ Eckmiller} for the clusters and groups, respectively) through the scaling relation given in 
\cite{2001A&A...368..749F}:
\begin{equation}
 \left ( \frac{M_{500}}{10^{13} h^{-1}_{71} M_{\odot}}  \right ) = (2.5\pm0.2)\left ( \frac{kT_{\mathrm{vir}}}{1 \mathrm{keV}} \right )^{1.676\pm0.054}~.
\label{M500}
\end{equation}

\begin{table*}
 \centering
 \caption{Best-fit results and scatter for BCG-cluster scaling relations: ``stat'' refers to statistical scatter and ``int'' refers to intrinsic scatter.}
\begin{tabular}[]{| c c  c  c  c  c  c |}
\hline
 \textbf{Category} & \textbf{Slope} & \textbf{Normalisation} & $\sigma_{\mathrm{int},L_{\mathrm{X}}}$ & $\sigma_{\mathrm{stat},L_{\mathrm{X}}}$ & $\sigma_{\mathrm{int},L_{\mathrm{BCG}}}$ & $\sigma_{\mathrm{stat},L_{\mathrm{BCG}}}$ \\ \hline
  $L_{\mathrm{X}}-L_{\mathrm{BCG}}$ Clusters & 0.36$\pm$0.03 &4.54$\pm$0.34 & 0.67 & 0.07 & 0.24 & 0.03 \\
 $L_{\mathrm{X}}-L_{\mathrm{BCG}}$ Clusters+Groups &0.34$\pm$0.03&6.98$\pm$0.16& 0.85 & 0.08 & 0.21 & 0.03 \\
  $L_{\mathrm{X}}-L_{\mathrm{BCG}}$ Clusters (SCC) &0.32$\pm$0.03&5.15$\pm$0.38& 0.64 & 0.08 & 0.21 & 0.02 \\
  $L_{\mathrm{X}}-L_{\mathrm{BCG}}$ Clusters+Groups (SCC) &0.32$\pm$0.03&6.89$\pm$0.20& 0.57 & 0.08 & 0.18 & 0.02 \\
  $L_{\mathrm{X}}-L_{\mathrm{BCG}}$ Clusters (NSCC) &0.50$\pm$0.07&3.49$\pm$5.09&0.61 & 0.06 & 0.21 & 0.03 \\
  $L_{\mathrm{X}}-L_{\mathrm{BCG}}$ Clusters+Groups (NSCC) &0.43$\pm$0.09&7.01$\pm$0.30& 0.64 & 0.07 & 0.29 & 0.03 \\ \hline
  &&&$\sigma_{\mathrm{int},M_{500}}$ & $\sigma_{\mathrm{stat},M_{500}}$ & $\sigma_{\mathrm{int},L_{\mathrm{BCG}}}$ & $\sigma_{\mathrm{stat},L_{\mathrm{BCG}}}$ \\ \hline
  $M_{500}-L_{\mathrm{BCG}}$ Clusters &0.62$\pm$0.05&3.52$\pm$0.28& 0.30 & 0.07 & 0.19 & 0.04 \\
  $M_{500}-L_{\mathrm{BCG}}$ Clusters+Groups &0.49$\pm$0.03&4.74$\pm$0.12& 0.40 & 0.07 & 0.20 & 0.03 \\
  $M_{500}-L_{\mathrm{BCG}}$ Clusters (SCC) &0.62$\pm$0.01&4.30$\pm$0.29& 0.21 & 0.07 & 0.13 & 0.04 \\
  $M_{500}-L_{\mathrm{BCG}}$ Clusters+Groups (SCC) &0.52$\pm$0.04&5.15$\pm$0.14& 0.31 & 0.06 & 0.16 & 0.03 \\
  $M_{500}-L_{\mathrm{BCG}}$ Clusters (NSCC) &0.75$\pm$0.09&2.55$\pm$0.36& 0.31 & 0.07 & 0.23 & 0.05 \\
  $M_{500}-L_{\mathrm{BCG}}$ Clusters+Groups (NSCC) &0.56$\pm$0.08&4.17$\pm$0.03& 0.53 & 0.07 & 0.29 & 0.03 \\ \hline
\end{tabular}
\label{scat}
\end{table*}

\section{Discussion of results}\label{Discussion}

\subsection{Cool-core fraction and physical properties}

A comparison between the group sample and the HIFLUGCS sample with
respect to cool-core fractions is presented in Table~\ref{CCfrac}.
\begin{table*}
\centering
\caption{Comparison between observed HIFLUGCS and group sample fractions.}
\begin{tabular}{| c  c  c |}
\hline
\textbf{Point of Distinction} & \textbf{Group sample} & \textbf{HIFLUGCS} \\ \hline
\% of CC systems (SCC+WCC) & 77& 72 \\ 
\% of SCC systems & 50 & 44 \\ 
\% of WCC systems & 27 & 28 \\ 
\% of NCC systems & 23 & 28 \\ \hline  
\end{tabular}
\label{CCfrac}
\end{table*}

We notice that the observed fraction of CC groups is similar to that
of clusters. It is worth recalling that the Malmquist bias results in higher observed CC cluster fractions and this should
qualitatively extend to the group sample. Simulations have shown that
SCC systems are selected preferentially because of their higher luminosity
at a given temperature and correcting this bias reduces the fraction
of SCC systems by about 25\% \citep{ 2010A&A...513A..37H,
  2011A&A...532A.133M}. \cite{2011A&A...526A..79E} show that the CC
bias due to steeper surface brightness profiles increases for less
luminous systems such as groups. Additionally, as this group sample is compiled from
Chandra archives, it is statistically incomplete like most other group samples, and therefore in this case could also suffer from an archival bias,
possibly resulting in a preferential selection of CC objects. These
biases will be quantified with access to a complete sample (Lovisari
et al.~in prep.). Here, we conclude that there is no significant
difference in SCC, WCC, and NCC fractions between clusters and groups.

The strong correlation between CCT and $K_{0}$ highlights that the
central entropy may also serve as a good proxy to quantify the CC/NCC
nature. Studies like \cite{2008ApJ...681L...5V}, \cite{2009ApJS..182...12C} and
\cite{2011arXiv1106.4563R} make use of the central entropy as the
defining parameter for a CC system. This tight relation also ensures
that systems defined by either parameter can be compared reasonably
accurately for both clusters and groups.

\subsection{Temperature profiles}\label{Tprofdiscuss}

A central temperature drop in the HIFLUGCS sample was a clear
indication of cool gas in the centre, corroborated with short cooling
times and high surface-brightnesses, particularly for the SCC clusters. We, find, however that there are some SCC groups that do not have a
central temperature drop, indicating the absence of cool gas; despite having
centrally peaked surface brightnesses and very short cooling times. We note that adding an additional power law to the spectral fit to account for possible low-mass X-ray binary emission does not improve the quality of the fit and also does not change this behaviour of the temperature profile. 

\begin{figure*}
\centering
\includegraphics[scale=0.33,angle=-90]{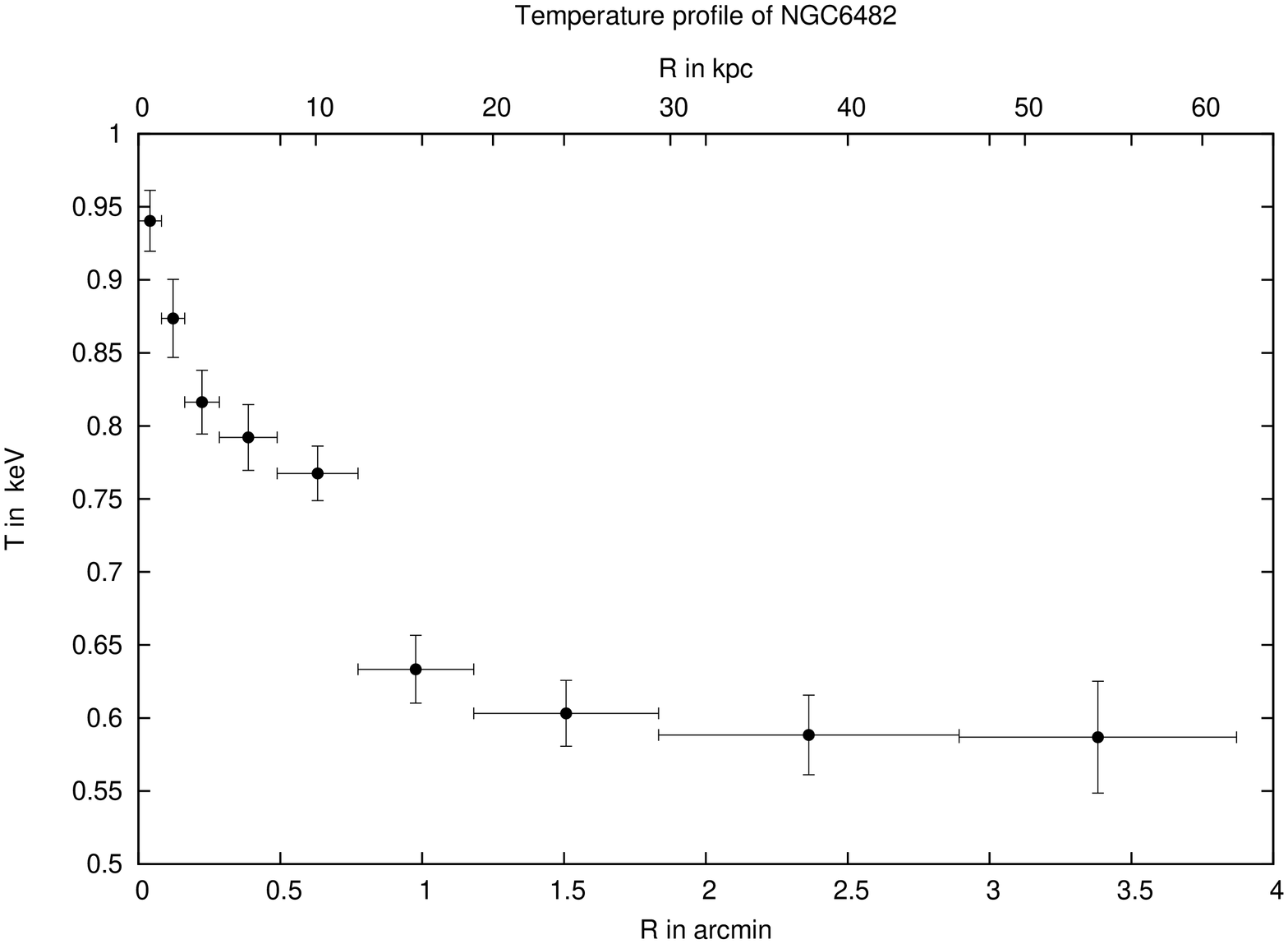}
\includegraphics[scale=0.33,angle=-90]{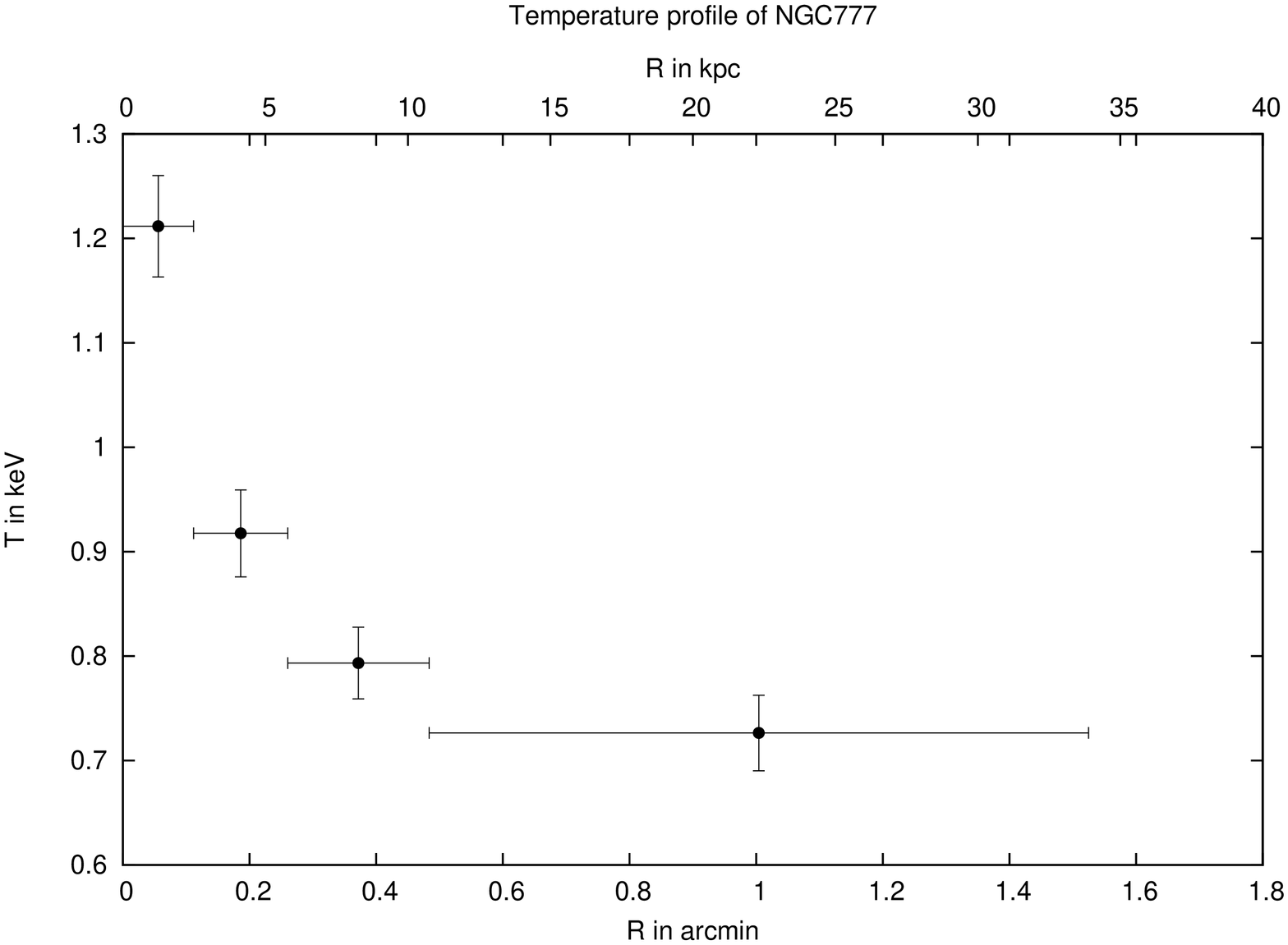}
\includegraphics[scale=0.33,angle=-90]{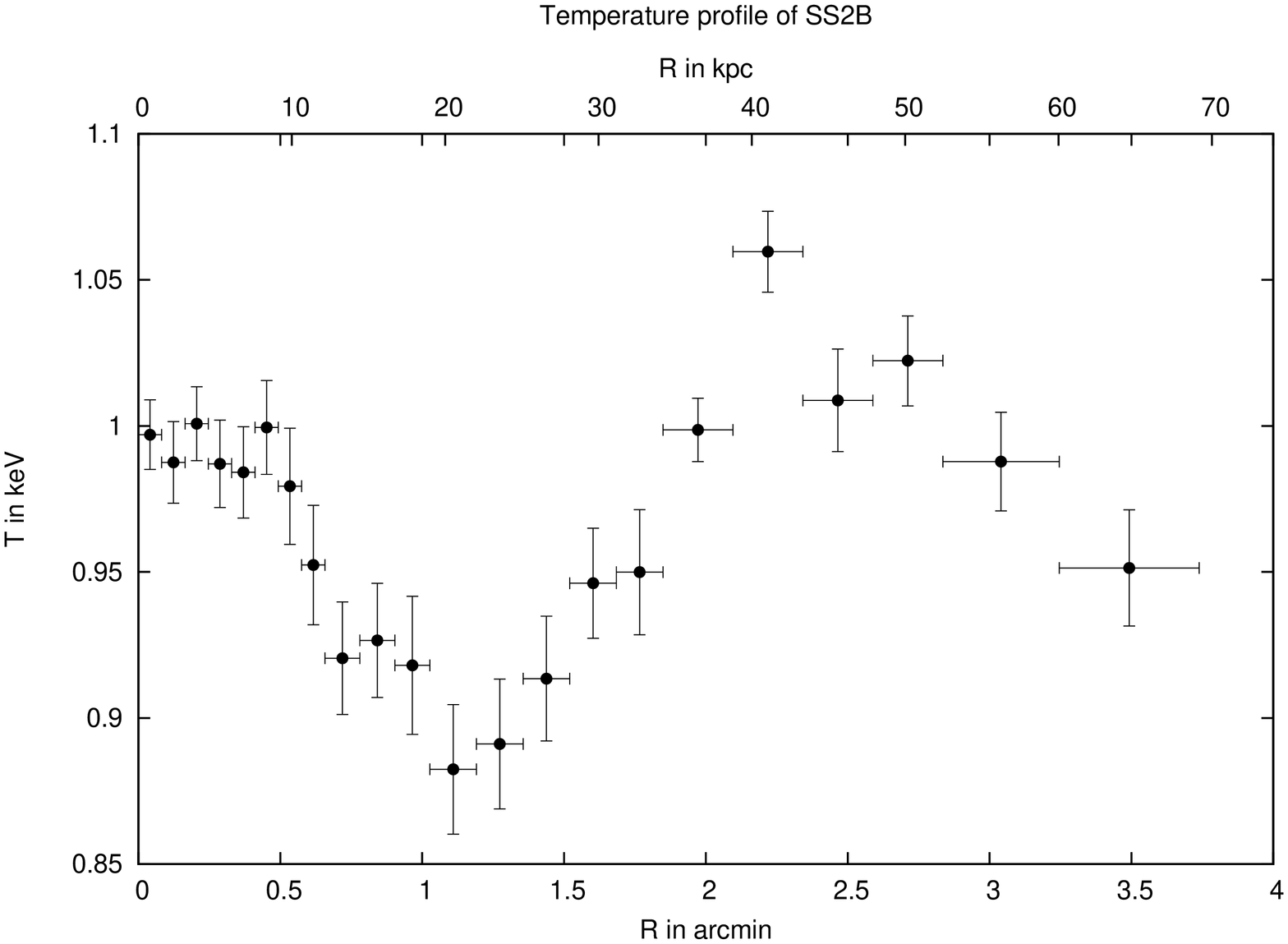}
\includegraphics[scale=0.33,angle=-90]{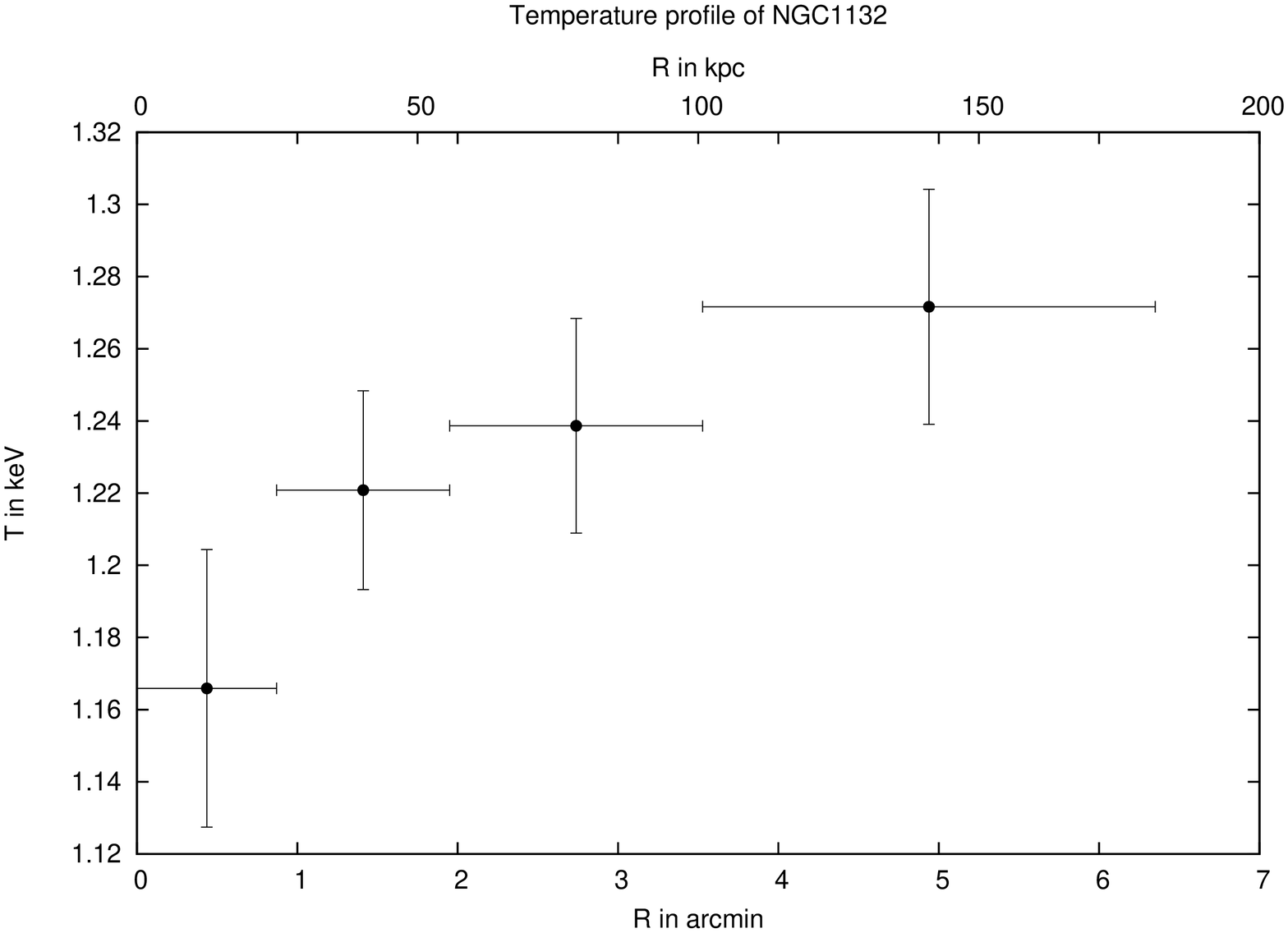}
\caption{Temperature profiles for NGC~6482, NGC~777 (top, both SCC), SS2B153, and NGC~1132 (bottom, SCC and WCC respectively).}
\label{fig:tempspecial}
\end{figure*}

The temperature profiles of three groups that show this feature are shown in Fig.~\ref{fig:tempspecial}.
The group NGC~6482 has been confirmed as a fossil group by
\cite{2004MNRAS.349.1240K}, who argue the shape of the temperature
profile could be obtained through a steady state cooling flow solution
\citep{1984Natur.310..733F}, although not ruling out heating
mechanisms such as supernovae or AGN, despite the lack of a clear
radio source. \cite{2007ApJ...658..299O} argue that SS2B is a group
whose core has been partially reheated by a currently quiescent AGN in
the past, and essentially that the system is currently being observed at a
  stage when the effects of both heating and cooling are visible. in principle, this could also be the explanation for NGC~6482 and NGC~777.

\begin{figure*}
\centering
\includegraphics[scale=0.33]{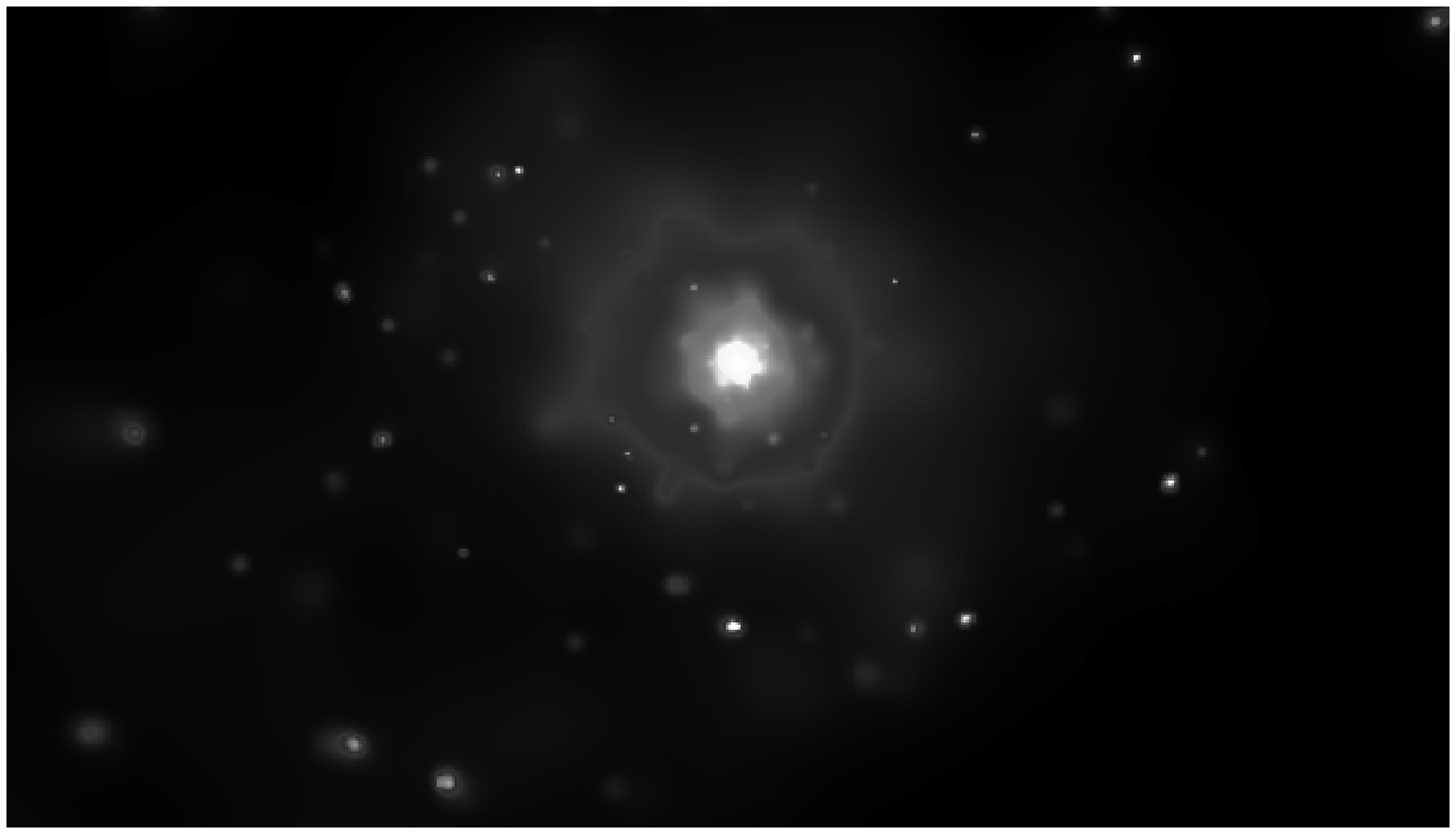}
\includegraphics[scale=0.33]{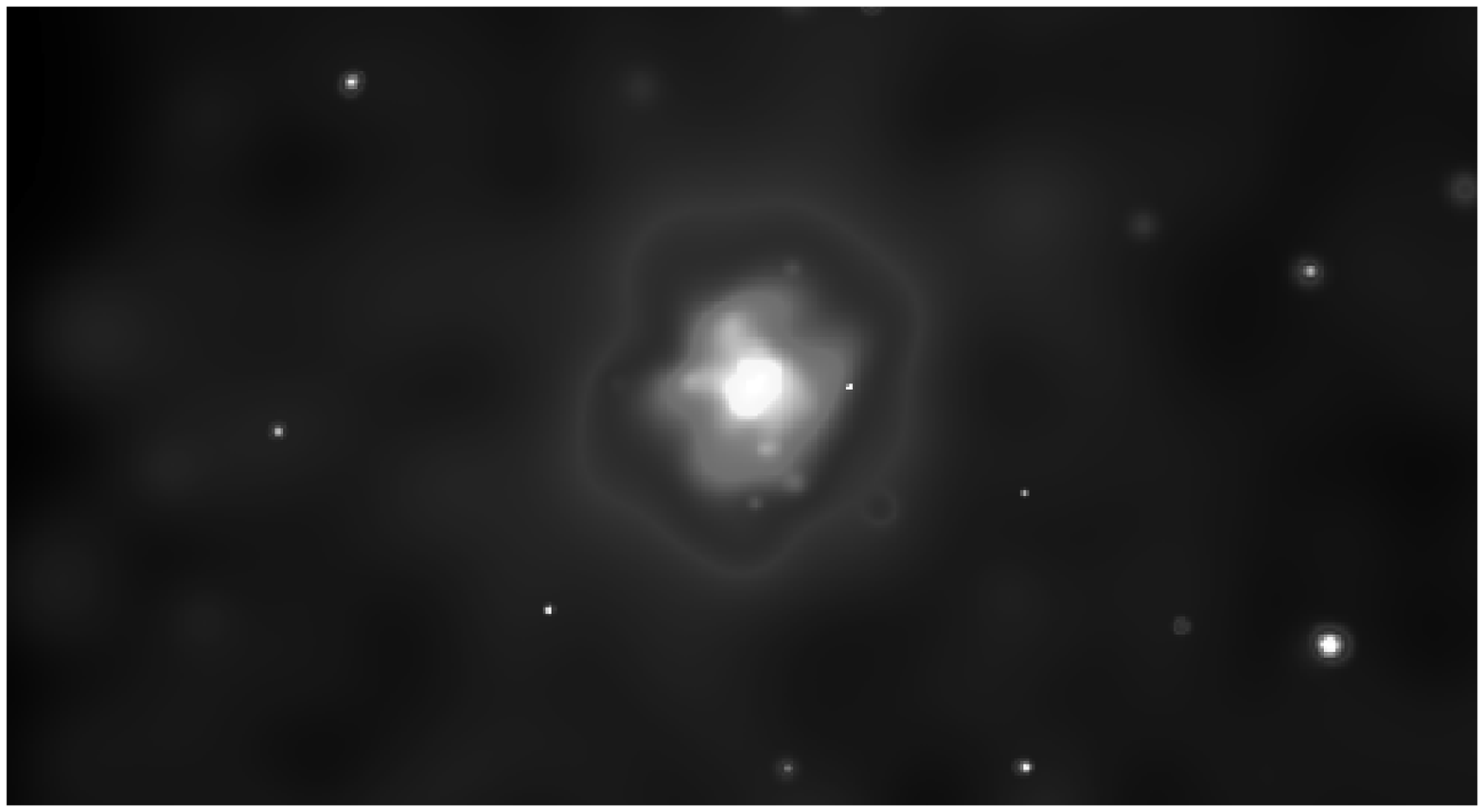}

\includegraphics[scale=0.33]{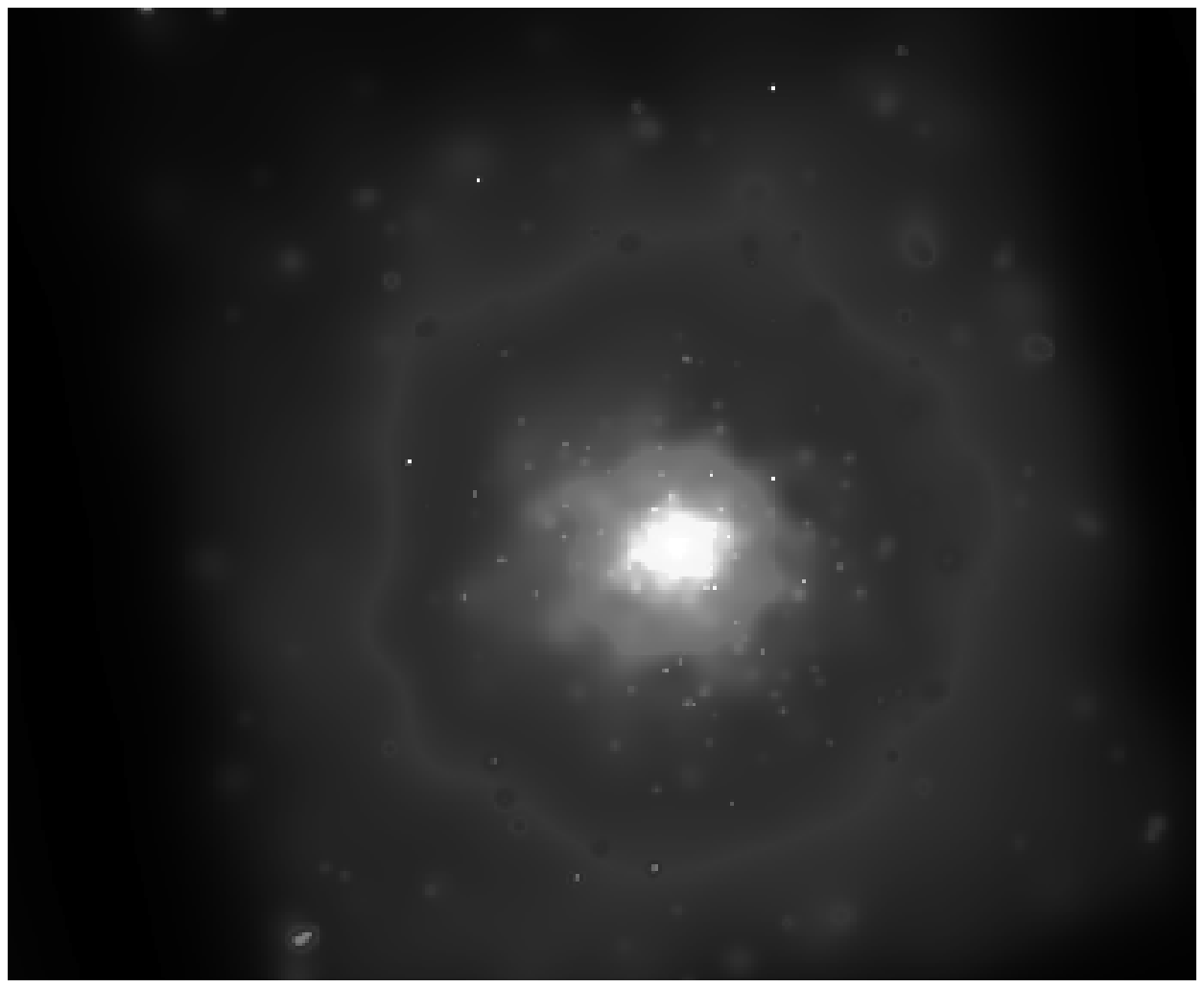}

\caption{X-ray images of NGC~6482 (top left), NGC~777 (top right), and SS2B (bottom). The images are adaptively smoothed and exposure corrected in energy bands of 0.5-2.0 keV.}
\label{fig:Xrayimg}
\end{figure*}

We notice that by X-ray morphology, NGC~777 and SS2B are quite similar
in appearance to NGC~6482, the fossil group
(Fig.~\ref{fig:Xrayimg}). In the optical band, all three systems are
dominated by a bright elliptical galaxy and few other galaxies. Thus,
NGC 777 and SS2B could also be classified as fossil groups, but we are
unable to confirm this because of the lack of magnitude information. The
next question that arises is if fossil groups might be a
special class of systems. In the current sample, we also have NGC~1132, another confirmed fossil group \citep{1999ApJ...514..133M} with a CCT of 1.08~Gyr (making it a WCC, but it could be a SCC within
errors), a CRS, and an almost flat temperature profile with a barely
significant central temperature drop
(Fig.~\ref{fig:tempspecial}). Thus, we speculate that fossils in
general could have the features described here, i.e. a partially
heated cool core. There is at least one fossil group for which this
has been shown to be the case \citep{2004ApJ...612..805S}. It is also
worth stating here that \cite{2012AJ....144...48H} show that 67\% of
fossil groups from their sample have CRSs. However, \cite{2010A&A...517A..52D} present fossil groups that do not show a central temperature rise. These are intriguing results that we are currently investigating with a sample of fossil groups.

Some studies, such as \cite{2006MNRAS.372.1496S} and
\cite{2008ApJ...675.1125B}, define a CC cluster through a central
temperature drop. This has never been a problem until now as it was
corroborated with short CCTs and low central entropies. We
have shown, however, that this may not always be the case and caution must be
exercised while using the central temperature drop as a CC diagnostic,
particularly on the group regime.

\subsection{AGN activity}\label{AGNact}

One of the most important conclusions of the study of the HIFLUGCS
sample by \cite{ 2009A&A...501..835M} was that as the CCT decreases,
the likelihood of a cluster hosting a CRS increases. \textit{All} the
SCC clusters contain a CRS and this fraction drops to 45\% for the NCC
clusters. The fraction of groups and clusters with a CRS, classified
on the basis of the CCT, is shown in Table~\ref{fraction}. We see
that, unlike clusters, the CRS fraction in groups does not scale with decreasing CCT.

\begin{figure*}
\centering
 \includegraphics[scale=0.45,angle=-90]{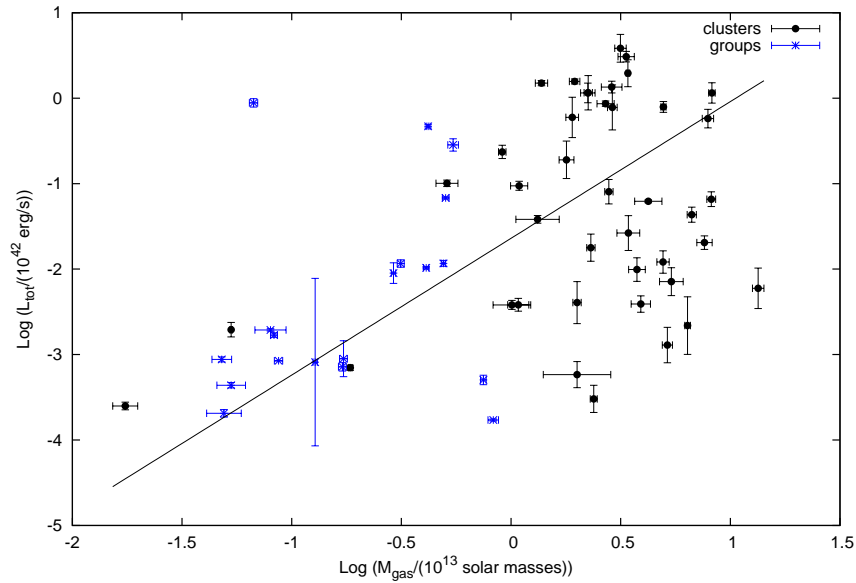}
\caption{Total radio luminosity vs. gas mass. The gas masses presented here are within $r_{500}$ and are taken from
\cite{2011A&A...526A.105Z} and \cite{ Eckmiller} for clusters and groups, respectively.}
\label{fig:Ltotall}
\end{figure*}

Figure~\ref{fig:Rtot} shows that almost all CC groups have a
very low radio luminosity compared to CC clusters. Quantitatively, we
find that the median of the radio luminosity of the CC groups is
$0.169 \cdot 10^{40}$ erg/s compared to $6.21 \cdot 10^{40}$ erg/s
for CC clusters. This is more than an order of magnitude difference
between clusters and groups. Assuming that gas cooling is responsible
for the radio output of the AGN, such a low value implies that not
enough gas is being accreted onto the SMBH. This could simply be due
to groups having less gas than clusters to accrete onto
the SMBH. We note that the gas mass for clusters is also an order of magnitude higher than that for groups (Fig.~\ref{fig:Ltotall}), making it possible that the reason for a lower radio output for groups is simply a lack of enough cooling gas, but, we also note that the correlation between the gas mass and the radio luminosity is weak (0.43 for a combined sample of clusters and groups), raising the question whether this simple explanation is sufficient to explain the low radio output. What role does star formation play here? We discuss this in Section~\ref{Star}.

The correlation coefficient between SMBH mass and radio output is very low, leading one to suspect that there is no correlation between these two quantities. This correlation has always been contentious, with studies
leading to conflicting results. \cite{1998MNRAS.297..817F} and
\cite{2001ApJ...551L..17L} show a correlation between $
\mathrm{L}_{\mathrm{5 GHz}} $ and the mass of the SMBH, whereas
\cite{2006ApJ...637..669L} show the absence of one. Interestingly, the NCC SMBH masses are systematically higher than the other SMBHs, raising the possibility that these objects might probably be suffering from stronger radio outbursts from larger SMBHs which might have destroyed their cool core. Though a tantalizing possibility, it could also simply be a selection effect and might have nothing to do with the CC nature of these objects.  

\begin{table*}
\centering
\caption{CRS fractions.}
\begin{tabular}{ | c | c | c |}
\hline
 Point of Distinction & \textbf{Group sample} & \textbf{HIFLUGCS sample} \\ \hline
\% of CC systems with CRS & 70  &75  \\ 
\% of NCC systems with CRS &100  &45  \\ 
\% of WCC systems with CRS & 57 & 67 \\ 
\% of SCC systems with CRS & 77 & 100 \\ \hline
\end{tabular}
\label{fraction}
\end{table*}

\subsection{BCG and cluster properties}

Figures~\ref{fig:LxLbcg} and~\ref{fig:M500Lbcg} show a clear trend
that the BCG luminosity increases with the system mass/X-ray luminosity, but with a
large amount of intrinsic scatter. Combined with the HIFLUGCS sample,
there are 85 clusters and groups spanning four orders of magnitude in
X-ray luminosity and three orders of magnitude in cluster mass, one of the
largest comparisons carried out with CC/NCC
distinction. Unlike \cite{2009A&A...501..835M}, who find a segregation beween SCC clusters and non-SCC clusters in the $M_{500}-L_{\mathrm{BCG}}$ relation with different slopes and normalisations, a combined fit to groups and clusters does not yield stastically significant different slopes (although it does yield different observed normalisations). The higher normalisations and flatter slopes obtained by extending these relations suggest that groups
in general have more luminous BCGs than clusters relative to $L_{\mathrm{X}}$ or $M_{500}$. Quantifying this further, we observe that $\left \langle \mathrm{log} L_{\mathrm{BCG}} - \mathrm{log} L_{\mathrm{X}} \right \rangle$ ($\left \langle \mathrm{log} L_{\mathrm{BCG}} - \mathrm{log} M_{500} \right \rangle$) is 1.7334 for groups (1.0252) vs. 0.6375 for clusters (0.3740).

\begin{figure*}
 \includegraphics[scale=0.33,angle=270]{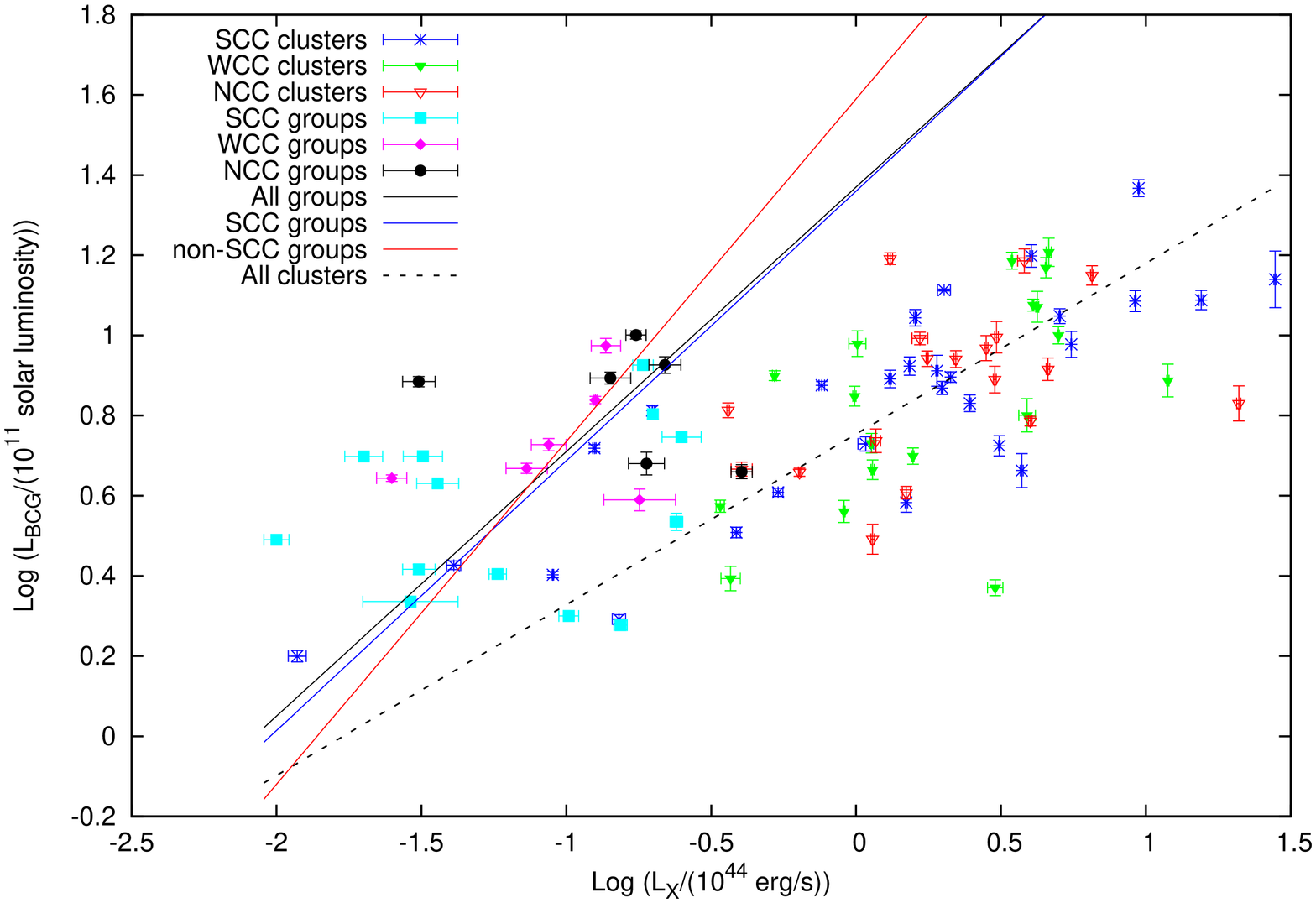}
\includegraphics[scale=0.33,angle=270]{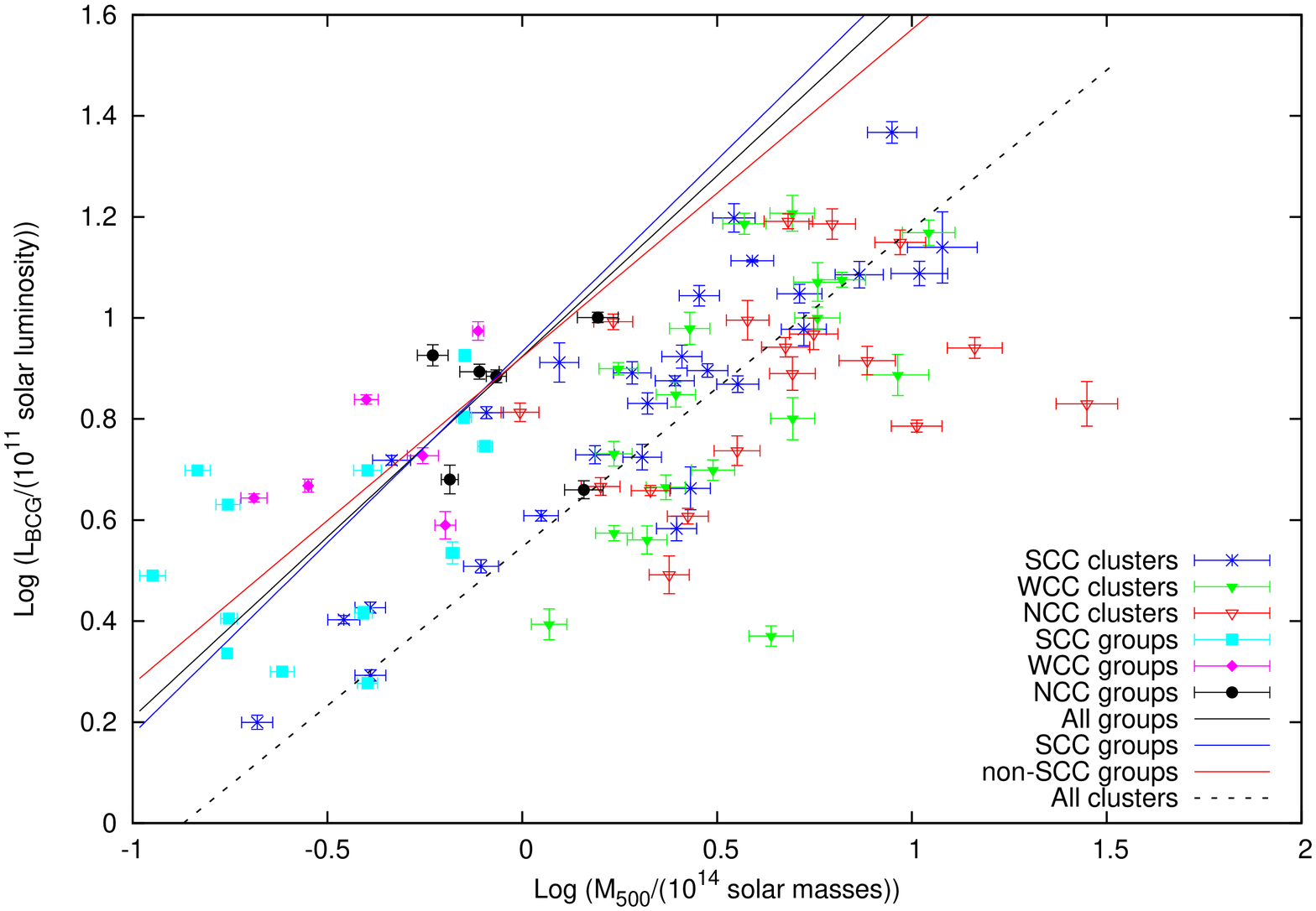}
\caption{$L_{\mathrm{X}}-L_{\mathrm{BCG}}$ and
  $M_{500}-L_{\mathrm{BCG}}$ relations. The black line is the best fit
  for all groups, the blue line for SCC groups and the red line for non-SCC groups. The dotted line shows the fit for all clusters.}
\label{fig:BCGonlyG}
\end{figure*}

To illustrate this point further, the same relations are presented again with only the fits for only groups in Fig.~\ref{fig:BCGonlyG}. For the complete group sample, the normalisation for the $L_{\mathrm{X}}-L_{\mathrm{BCG}}$ relation increases by more than a factor of 5 ($23.4\pm15.2$ vs $4.54\pm0.34$) and by more than a factor of 2 for the $M_{500}-L_{\mathrm{BCG}}$ relation ($8.39\pm0.80$ vs $3.52\pm0.28$) compared to that for clusters only. There are also indications that the slope is steeper, but not significantly so ($0.66\pm0.27$ for the $L_{\mathrm{X}}-L_{\mathrm{BCG}}$ and $0.72\pm0.10$ for the $M_{500}-L_{\mathrm{BCG}}$) relation. Assuming that the NIR luminosity is a good stellar mass proxy (e.g.~\citealt{2001ApJ...550..212B}), all the above points hint that groups have relatively more stellar mass in their BCGs than clusters.

\subsection{The Role of star formation}\label{Star} 
When trying to understand the relation between gas cooling and AGN feedback in groups, one cannot ignore the role of star formation. As pointed out by authors like \cite{2011arXiv1108.3678L}, \cite{2003ApJ...591..749L}, and \cite{2008A&A...485..633L}, less massive cold  
clusters/groups are more prolific star forming environments. \cite{2008ApJ...687..899R} show that for groups and clusters, star formation kicks in when the central entropy is below 30 keV $\mathrm{cm}^{2}$, with the requirement that the X-ray and galaxy centroids are within 20 kpc. In our sample, all our CC BCGs are within 20 kpc of the X-ray EP, and all SCC groups and a few WCC groups have a central entropy well below the entropy limit, thus fulfilling these criteria for star formation. \cite{2010ApJ...719.1844H} use UV data from GALEX to show that there is a good correlation between gas cooling time and star formation rate (SFR) for CC cluster BCGs, such that the SFR increases with decreasing cooling time. Additionally, in most cases the classical mass deposition rates for our CC groups is not too high (a rough estimate yields a median of $< 10~\mathrm{M}_{\odot}$/yr calculated at a radius where the cooling time is 7.7 Gyr) which means that one cannot rule out the possibility that star formation is being 
fueled by most of the cooling gas. This is of course a hypothesis, and confirmation of this can be provided if stronger correlations between the cooling times and star formation rates were seen for groups than for clusters. Therefore, we plan to acquire $\mathrm{H}\alpha$ data for the group BCGs to constrain SFRs.

\section{Summary and conclusions}\label{Summary}

With a sample of 26 Chandra galaxy groups we have peformed a study of
the ICM cooling, AGN feedback and the BCG properties on the galaxy
group scale. The major results of our study are as follows.
\begin{itemize}
\item The group sample has similar SCC, WCC, and NCC fractions to the HIFLUGCS cluster sample.
\item We find that 23\% of the groups that have CCT $\leq$ 1 Gyr do not show a
  central temperature drop. We speculate that this could be due
  to a partial reheating of the cool core in the past. Additionally, we also speculate that this might be a characteristic feature of fossil groups.
\item An increase in the CRS fraction with decreasing CCT is
  \textit{not} seen, unlike for the HIFLUGCS sample. This is the first
  indication of differences between clusters and groups in the AGN
  heating/ICM cooling paradigm.
\item There is no correlation seen between the CCT and the integrated
  radio luminosity of the CRS. We notice that CRSs for the SCC groups,
  in particular, have a much lower radio luminosity than clusters.
\item We extend the scaling relations between $L_{\mathrm{BCG}} $ and
  global cluster properties ($L_{\mathrm{X}} $ and $M_{500}$) into
  the group regime. Most group BCGs have a BCG luminosity above the
  best fit and we think this may be due to a higher stellar mass content in group
  BCGs than in cluster BCGs, for a given $L_{\mathrm{X}}$ and
  $M_{500}$.
\item We have speculated that star formation is a possible, effective answer
  to the fate of the cool gas in groups where, because of fueling star formation,
  not enough gas (low as it is) is being fed to the SMBH and hence the radio output
  of the CRSs is not as high. This could also explain why some SCC
  groups do not show a CRS.
\end{itemize}
In conclusion, we have demonstrated differences between clusters and
groups vis-a-vis ICM cooling, AGN feedback, and BCG properties. These
results lend support to the idea that groups are not simply scaled-down versions of clusters. In the future, it would be interesting to study the impact of these processes on scaling relations for galaxy groups.

\begin{acknowledgements} {The authors would like to thank the anonymous referee who provided useful comments which helped improve the quality of the work. VB would like to thank Lorenzo Lovisari for
    helpful discussions which helped improve the quality of the
    work. VB also acknowledges financial support from the Argelander
    Institut f\"ur Astronomie. T.~H.~R acknowledges support from the
    DFG through the Heisenberg research grant RE 1462/5. G.S.~acknowledges support from DFG research grant RE 1462/6. H.I. and H.J.E. acknowledge support from DFG grant 1462/4. H.I. additionally acknowledges support from STFC grant ST/K003305/1. The program for calculating the CCT was kindly provided by
Paul Nulsen, which is based on spline interpolation on a table of values for the APEC model assuming an optically thin plasma by R.~K.~Smith. This research has made use of the NASA/IPAC Extragalactic Database (NED) which is operated by the Jet Propulsion Laboratory, California Institute of Technology, under contract with the National Aeronautics and Space Administration.}

\end{acknowledgements}
\bibliographystyle{aa}
\bibliography{ref}

\begin{table*}
 \begin{center}
  \caption{Radio data for CRS.}
\label{radiodata}
  \begin{tabular}{| c | c | c | c | c |c|}
  \hline \hline
  & & & & &\\
  Cluster & $\nu$ (GHz) & Flux density (mJy) & ${L}_\mathrm{1.4 GHz}$\newline ($10^{32}$ erg/s/Hz)  & $L_\mathrm{tot} 10^{42}$ (erg/s) & SI of 1 \\\hline
  & & & & &\\
  A0160 & $1.4^{\dagger}$ & 1042.8$\pm$35.7 & $0.44847^{+0.0154}_{-0.0153}$ & $0.4676^{+0.0147}_{-0.0276}$ &\\    
        & $0.408^\mathrm{a}$ & 1660$ \pm$60 & & &\\ 
  & & & & &\\
  HCG62 & $1.4^{\dagger}$ & 4.9$ \pm$0.5 & $0.000195^{+0.00002}_{-0.00002}$ & $0.000205^{+0.00002}_{-0.000021}$& YES\\ 
  & & & & &\\
  IC1262 & $1.4^{\dagger}$ & 87.6$ \pm$6.3 & $0.021156^{+0.001521}_{-0.005625}$ & $0.0215^{+0.0015}_{-0.0016}$ & YES\\ 
  & & & & &\\
  IC1633 & $1.4^{\mathrm{b}}$ & 1.59$ \pm$0.039 & $0.00015891^{+0.000005}_{-0.000005}$ & $0.00017142^{+0.000005}_{-0.000005}$& YES\\ 
  & & & & &\\
  MKW4 &$1.4^{\mathrm{c}}$&2.40$\pm$0.5&$0.00213^{+0.000045}_{-0.000045}$& $0.000815^{+0.003810}_{-0.000767}$&\\  
       & $4.86^{\mathrm{c}}$ & 0.35$ \pm$0.1 &&&\\
  & & & & &\\
  MKW8 & $1.4^{\dagger}$ & 2.54$ \pm$0.1 & $0.000402^{+0.000025}_{-0.000025}$ & $0.000504^{+0.000088}_{-0.000031}$&\\
       & $4.86^{\mathrm{c}}$ & 2.09$ \pm$0.15 & & &\\ 
  & & & & &\\
  NGC 326 & $1.4^{\dagger}$ & 1802$ \pm$59 & ${0.85239}^{+0.0279}_{-0.0279}$ & $ {0.880}^{+0.095}_{-0.101} $ &\\
	  & $0.174^{\spadesuit}$ & 5100$ \pm$510 & & &\\
	  & $0.074^{\ddagger}$ & 12320$ \pm$1440 & & &\\ 
  & & & & &\\
  NGC 507 & $1.4^{\dagger}$ & 120.5$ \pm$6.0 & $ {0.007150}^{+0.000354}_{-0.000354}$ & $ {0.008970}^{+0.001430}_{-0.002890}$ &\\
	  & $0.074^{\ddagger}$ & 3250.0$ \pm$490.0 & & &\\ 
  & & & & &\\
  NGC 533 & $1.4^{\dagger}$ & 28.6$ \pm$1.0 & $ {0.00183}^{+0.00006}_{-0.00006}$ & $ {0.00194}^{+0.000069}_{-0.000066}$& YES\\ 
  & & & & &\\
  NGC 777 & $1.4^{\dagger}$ & 7.0$ \pm$0.5 & $ {0.000413}^{+0.000029}_{-0.000029}$ & $ {0.000437}^{+0.000031}_{-0.000031}$& YES\\ 
  & & & & &\\
  NGC 1132 & $1.4^{\dagger}$ & 5.4$ \pm$0.6 & $ {0.000667}^{+0.00007}_{0.00007}$ & $ {0.000717}^{+0.00008}_{0.00008}$& YES\\ 
  & & & & &\\
  NGC 1550 & $1.4^{\dagger}$ & 16.6$ \pm$1.6 & $ {0.000547}^{+0.000053}_{-0.000053}$ & $ {0.00089}^{+0.00015}_{-0.00004}$&\\
	   & $2.38^{\mathrm{d}}$ & 8.0$ \pm$3.0 & & &\\
	   & $0.074^{\ddagger}$ & 670.0$ \pm$177.0 & & &\\ 
  & & & & &\\
  NGC 4936 & $1.4^{\dagger}$ & 39.8$ \pm$1.6 & $ {0.00117}^{+0.00005}_{-0.00005}$ & $ {0.00048}^{+0.00001}_{-0.00002}$&\\
	   & $0.843^{\clubsuit}$ & 35.2$ \pm$1.9 & & &\\ 
  & & & & &\\
  NGC 5129 & $1.4^{\dagger}$ & 7.2$ \pm$0.5 & $ {0.00082}^{+0.00006}_{-0.00006}$ & $ {0.00088}^{+0.00006}_{-0.00006}$& YES\\ 
  & & & & &\\
  NGC 5419 & $1.4^{\dagger}$ & 349.2$ \pm$12.2 & $ {0.01536}^{+0.00064}_{-0.00064}$ & $ {0.01162}^{+0.00097}_{-0.00114}$&\\
	   & $0.843^{\clubsuit}$ & 529.3$ \pm$16 & & &\\
	   & $0.074^{\ddagger}$ & 1580$ \pm$240 & & &\\ 
  & & & & &\\
  NGC 6269 & $1.4^{\dagger}$ & 50$ \pm$1.9 & $ {0.012532}^{+0.00048}_{-0.00048}$ & $ {0.01166}^{+0.00079}_{-0.00104}$ &\\
	   & $0.074^{\ddagger}$ & 560$\pm$100 & & &\\ 
  & & & & &\\
  NGC 6338 & $1.4^{\dagger}$ & 57$ \pm$1.8 & $ {0.0095806}^{+0.00030}_{-0.00030}$ & $ {0.01034}^{+0.00033}_{-0.00033}$& YES\\ 
  & & & & &\\
  RXCJ2214 &$1.4^{\dagger}$ &1604.7$\pm$50.7&$0.24691^{+0.00780}_{-0.00780}$&$0.2530^{+0.0051}_{-0.0051}$& YES\\
  & & & & &\\
  S0463	   & $0.843^{\clubsuit}$ & 314$ \pm$9.6 &-& $ {0.0687}^{+0.0003}_{-0.0003}$&YES\\ 
  & & & & &\\
  SS2B	   & $1.4^{\dagger}$ & 31.9$ \pm$1.4 & $ {0.00160}^{+0.00009}_{-0.00009}$ & $ {0.00169}^{+0.00010}_{-0.00010}$& YES\\ 
\hline \hline
  \end{tabular}
 
 \end{center}
$\dagger$ NVSS-\cite{ 1998AJ....115.1693C}
\hspace{2cm}
$\ddagger$ VLSS-\cite{ 2007AJ....134.1245C}
\newline
$\spadesuit$ 4C catalog-\cite{ 1965MmRAS..69..183P}
\hspace{2cm}
$\clubsuit$ SUMSS-\cite{1999AJ....117.1578B}
\newline
a \cite{1990MNRAS.247..387R}
\hspace{3.5cm}
b \cite{2000ApJS..128..469H} \newline
c From \cite{ 2009A&A...501..835M}, based on data from the VLA (Very Large Array) Archive \newline
d \cite{1978ApJS...36...53D}
\label{raddata}
\end{table*}
\clearpage
 \appendix
\onecolumn

\section{Calculation of scatter}\label{Scatcalc}

We explain the calculation of scatter given in Table \ref{scat}, which
is based on the weighted sample variance in the log-log plane \citep{2005A&A...441..893A,2011A&A...532A.133M}:
\begin{eqnarray*}
 \sigma_{\mathrm{raw},~L_{\mathrm{BCG}}}^2 & = & \mathrm{C}_{L_{\mathrm{BCG}}} \sum_{i=1}^N \frac{1}{\sigma_{i,~L_{\mathrm{BCG}}}^2}\left[\mathrm{lg}(L_{\mathrm{BCG_{i}}}) - (\lg c + m \mathrm{lg}(L_{\mathrm{X_{i}}})) \right]^2 \nonumber \\
  \mathrm{C}_{L_{\mathrm{BCG}}} & = & \frac{1}{(N-2)} \frac{N}{\sum_{i=1}^N (1/\sigma_{i,~L_{\mathrm{BCG}}}^2)} \, \\
  \sigma_{\mathrm{raw},~L_{\mathrm{X}}}^2 & = & {C_{L_{\mathrm{X}}}} \sum_{i=1}^N \frac{1}{\sigma_{i,~L_{\mathrm{X}}}^2}\left[\lg(L_{\mathrm{X_{i}}}) - (\lg(L_{\mathrm{BCG_{i}}}) - \lg c)/m \right]^2 \nonumber \\
  \mathrm{{C_{L_{\mathrm{X}}}}} & = & \frac{1}{(N-2)} \frac{N}{\sum_{i=1}^N (1/\sigma_{i,~L_{\mathrm{X}}}^2)} \,
\end{eqnarray*}
where ${\sigma}_{i,~L_{\mathrm{BCG}}}^2 = (\Delta
\mathrm{lg}(L_{\mathrm{BCG}_{i}}))^2 + {m}^2 (\Delta
\mathrm{lg}(L_{\mathrm{X}_{i}}))^2 $,
\\ \\
${\sigma}_{i,~L_{\mathrm{X}}}^2 = (\Delta \mathrm{lg}(L_{\mathrm{X}_{i}}))^2 + (\Delta \mathrm{lg}(L_{\mathrm{BCG}_{i}}))^2/{m}^2 $,\\ \\
Here, $N$ is the sample size, $c$ is the normalisation, $m$ is
the slope and the deltas represent the errors on the quantities. The statistical scatter, $\sigma_{\mathrm{stat}}$, is
estimated by calculating the root-mean-square of $\sigma_i$. The
intrinsic scatter is given as the difference between the raw and the
statistical scatters in quadrature. We simply replace $M_{500}$ in
place of $L_{\mathrm{X}}$ in the formula to calculate scatter for the
$M_{500}-L_{\mathrm{BCG}}$ relation.

\section{Temperature profiles}\label{Tempprofiles}

\begin{figure*}[h!]
\centering
\includegraphics[angle=-90,scale=0.5]{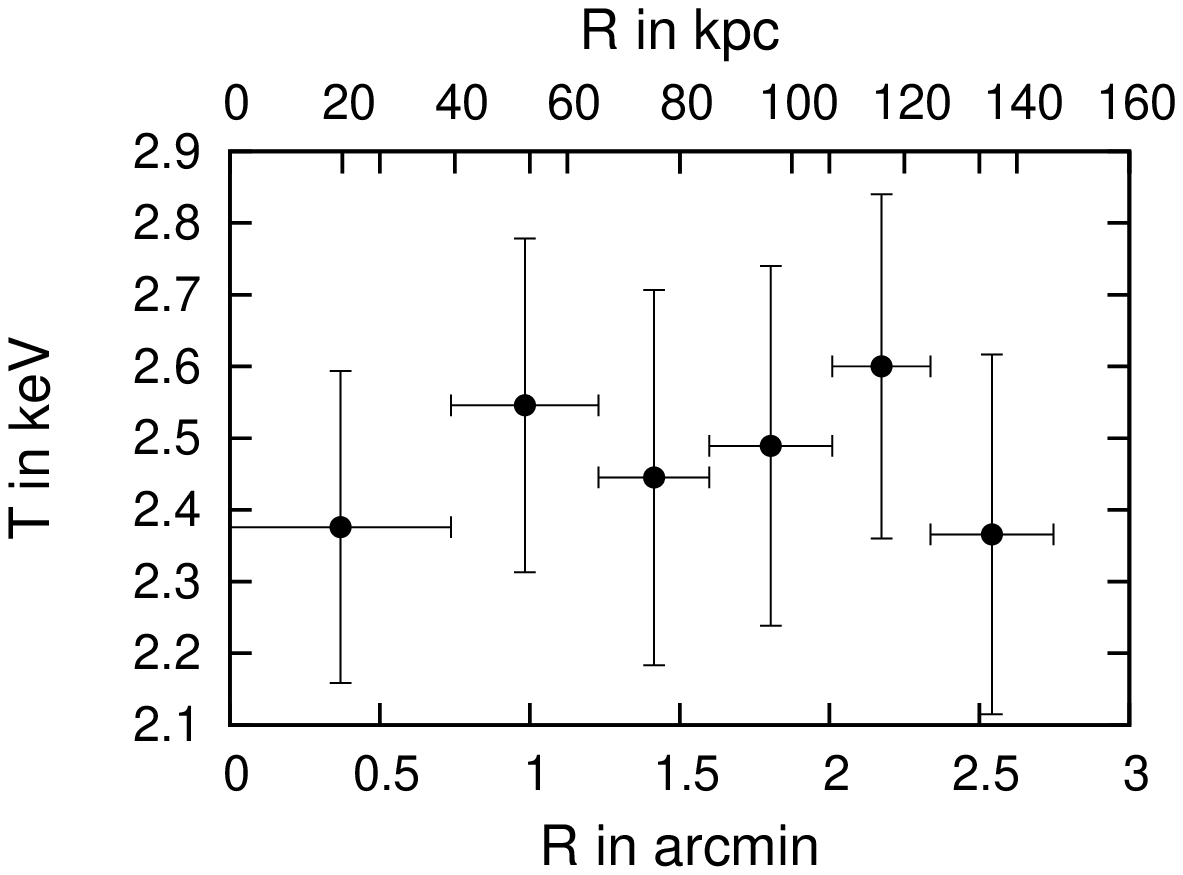}
\includegraphics[angle=-90,scale=0.50]{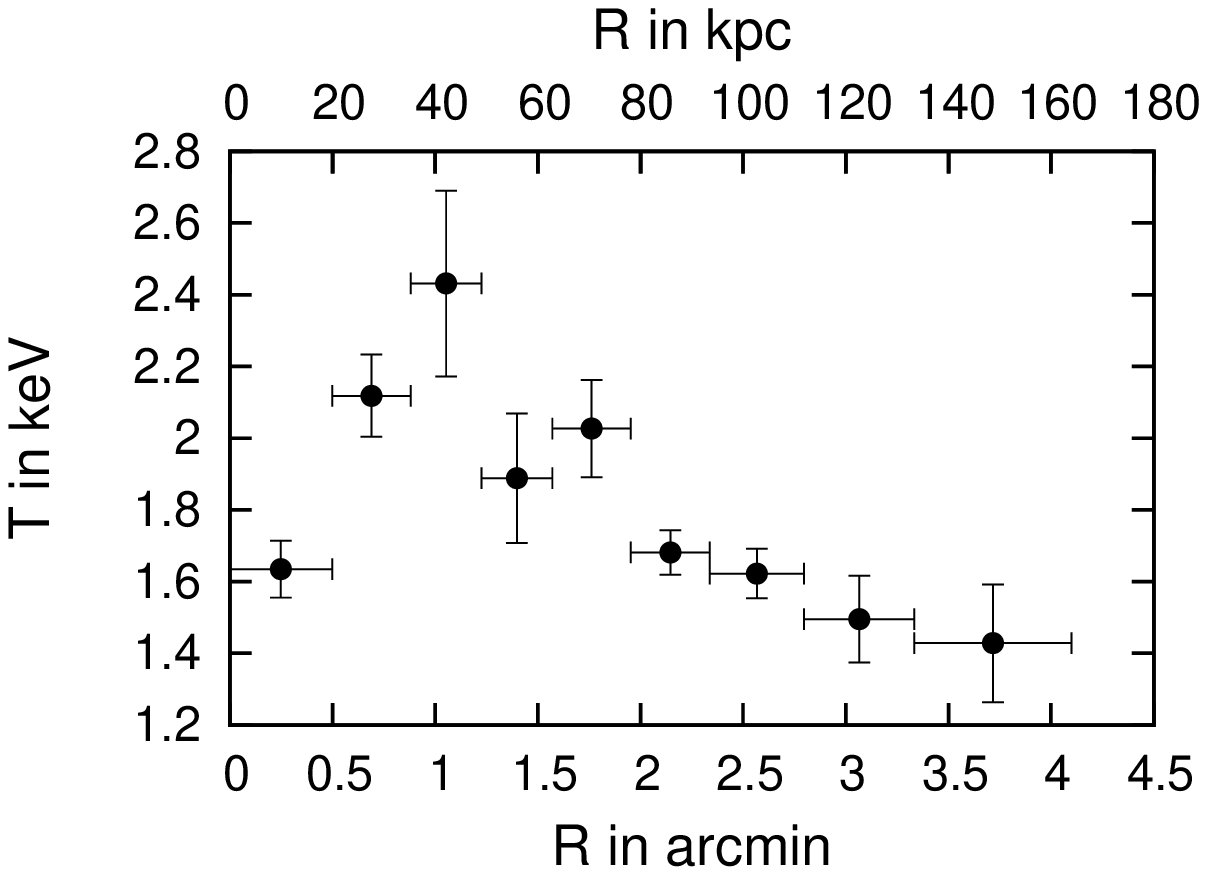}
\caption{Temperature profiles of A0160 (left) and A1177 (right).}
\end{figure*}
\begin{figure*}[h!]
\centering
\includegraphics[angle=-90,scale=0.50]{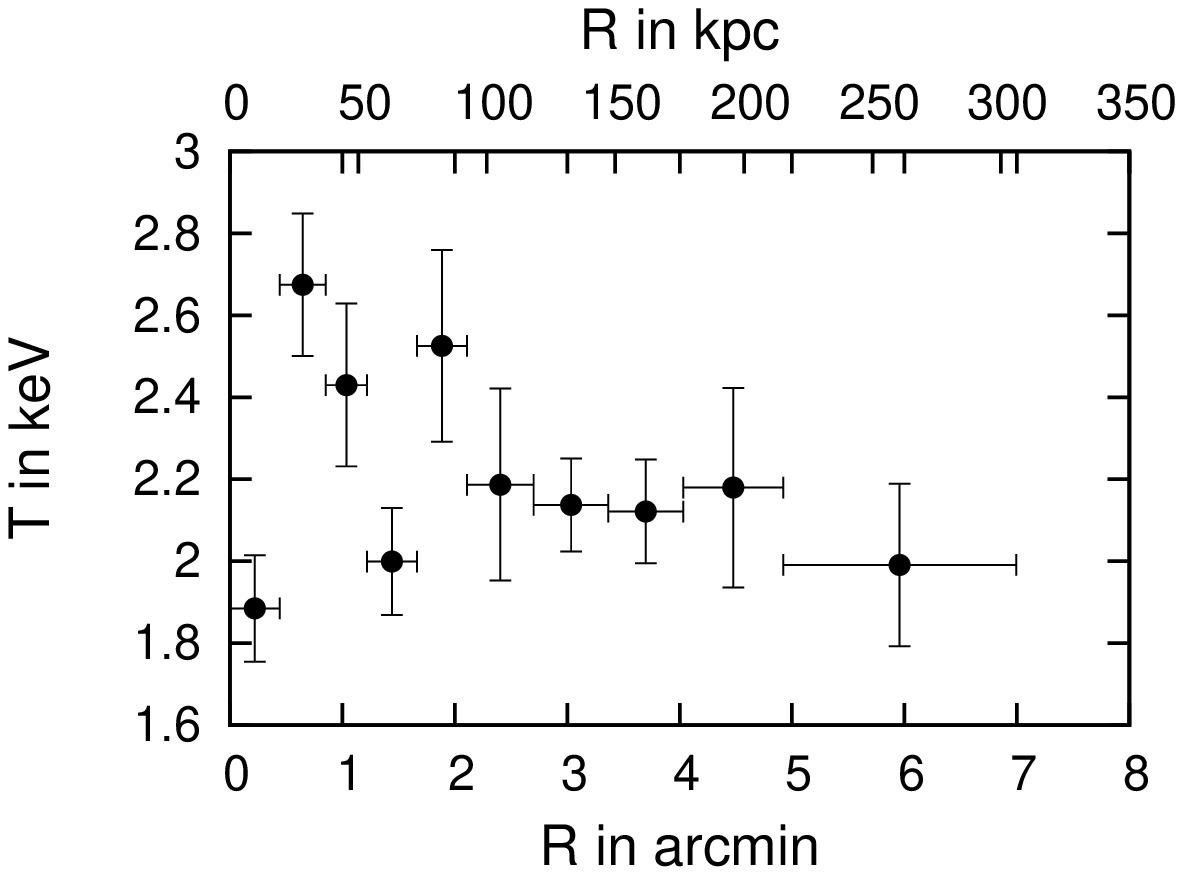}
\includegraphics[angle=-90,scale=0.50]{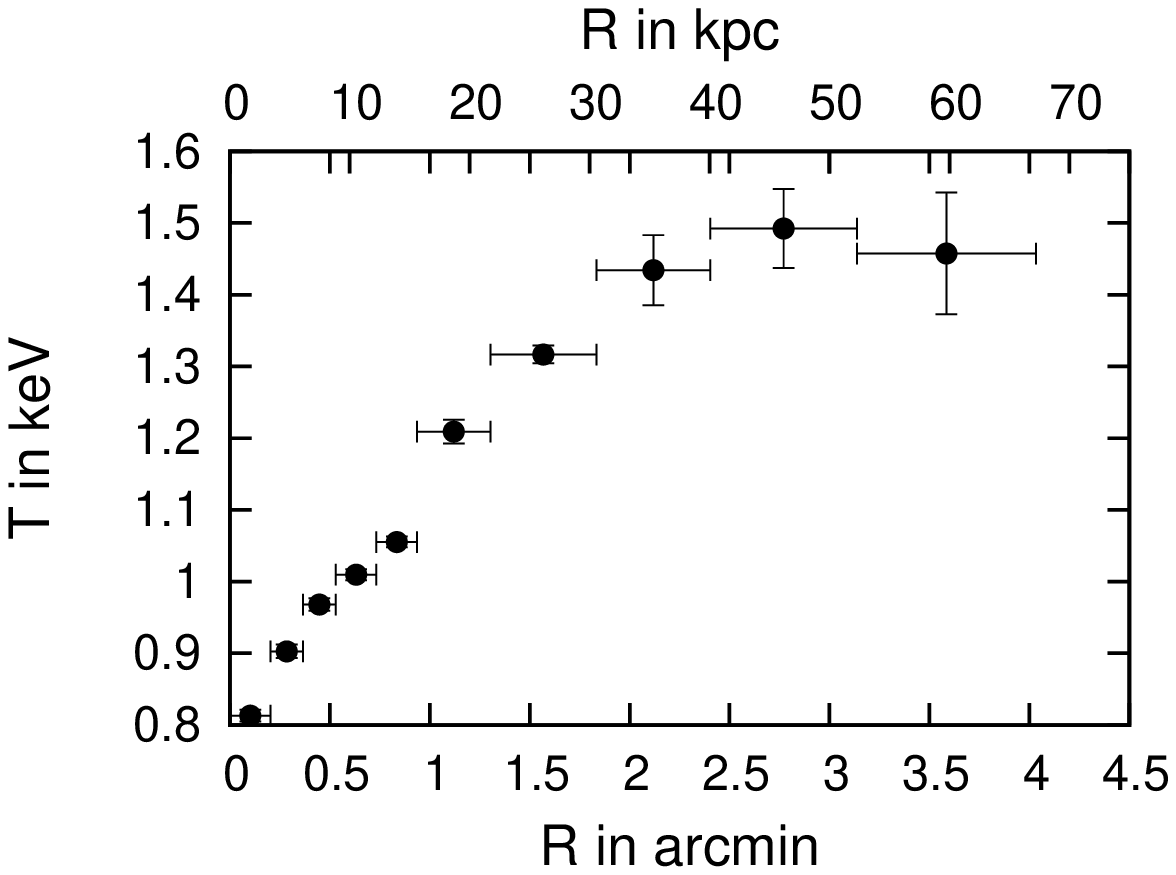}
\caption{Temperature profiles of ESO 55 (left) and HCG 62 (right).}
\end{figure*}
\begin{figure*}[h!]
\centering
\includegraphics[angle=-90,scale=0.50]{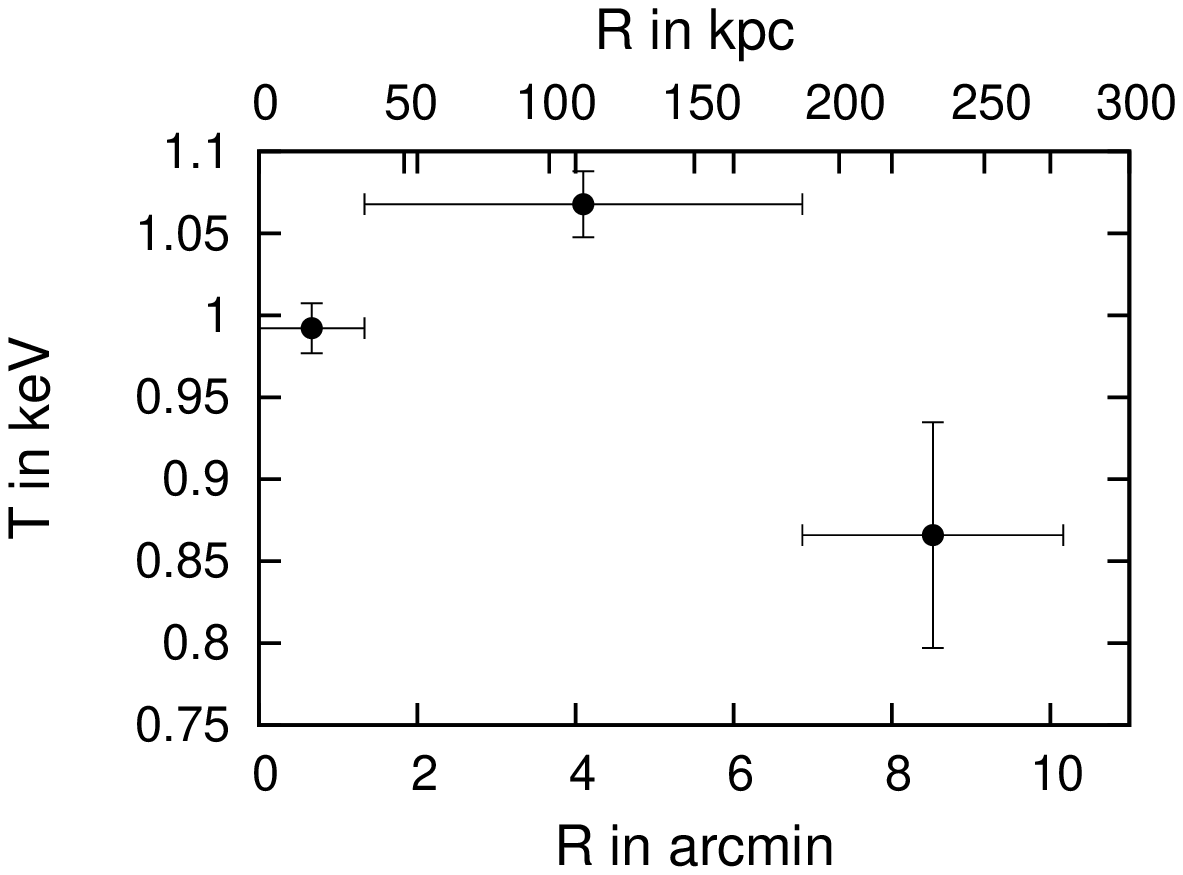}
\includegraphics[angle=-90,scale=0.50]{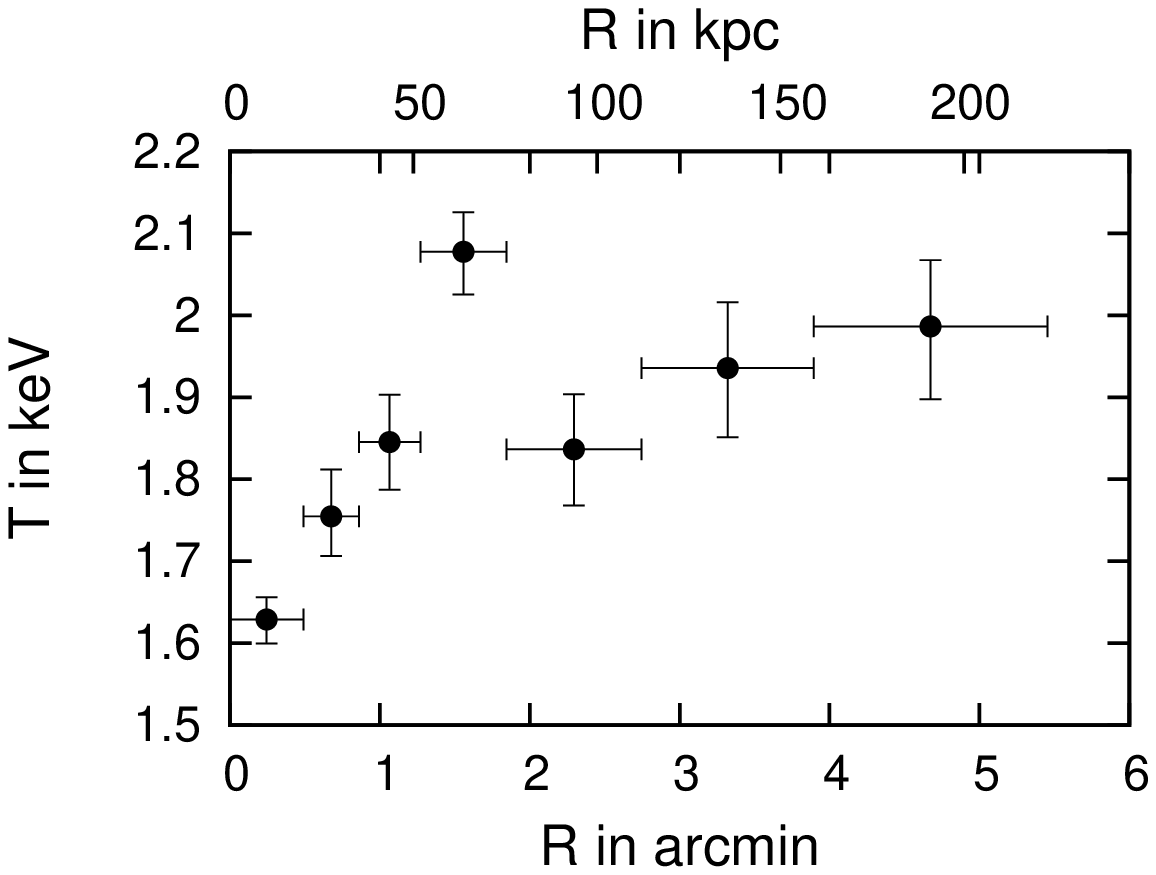}
\caption{Temperature profiles of HCG 97 (left) and IC 1262 (right).}
\end{figure*}
\begin{figure*}[h!]
\centering
\includegraphics[angle=-90,scale=0.50]{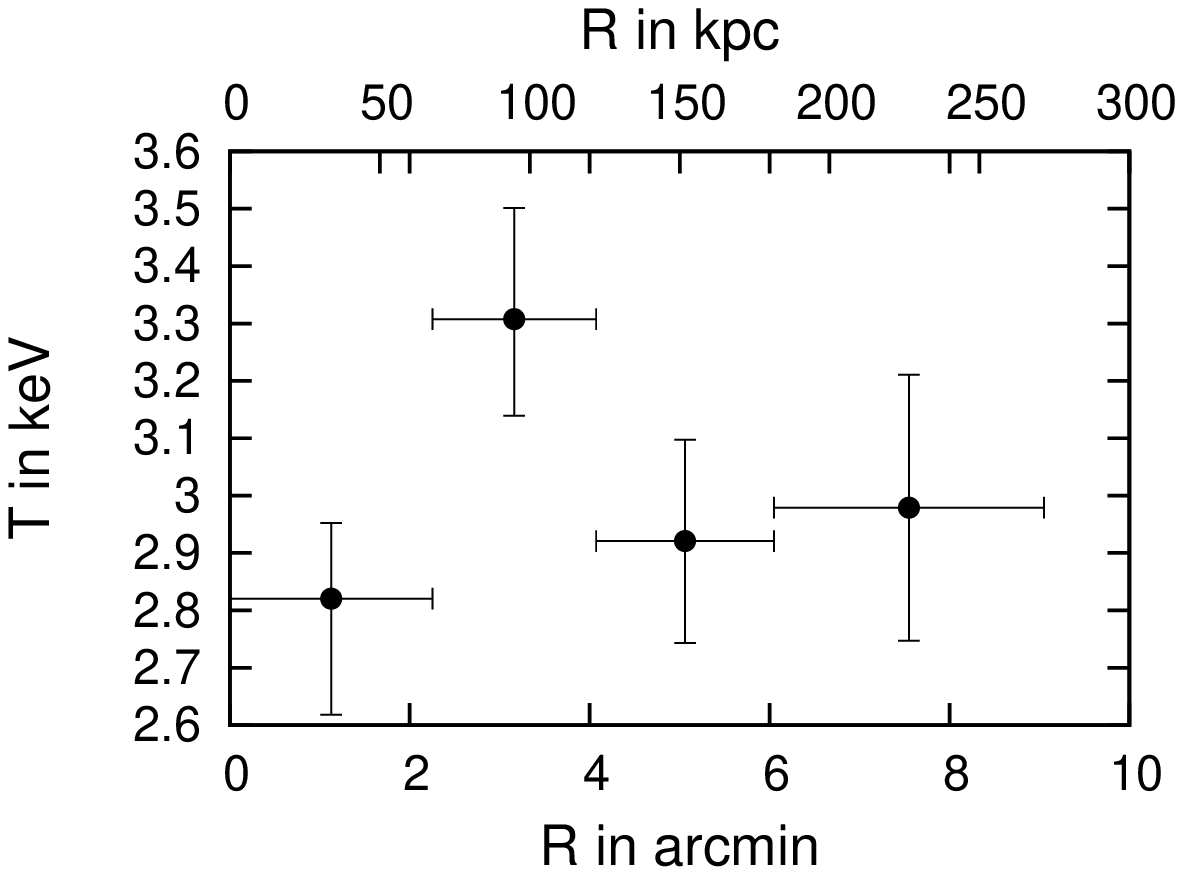}
\includegraphics[angle=-90,scale=0.50]{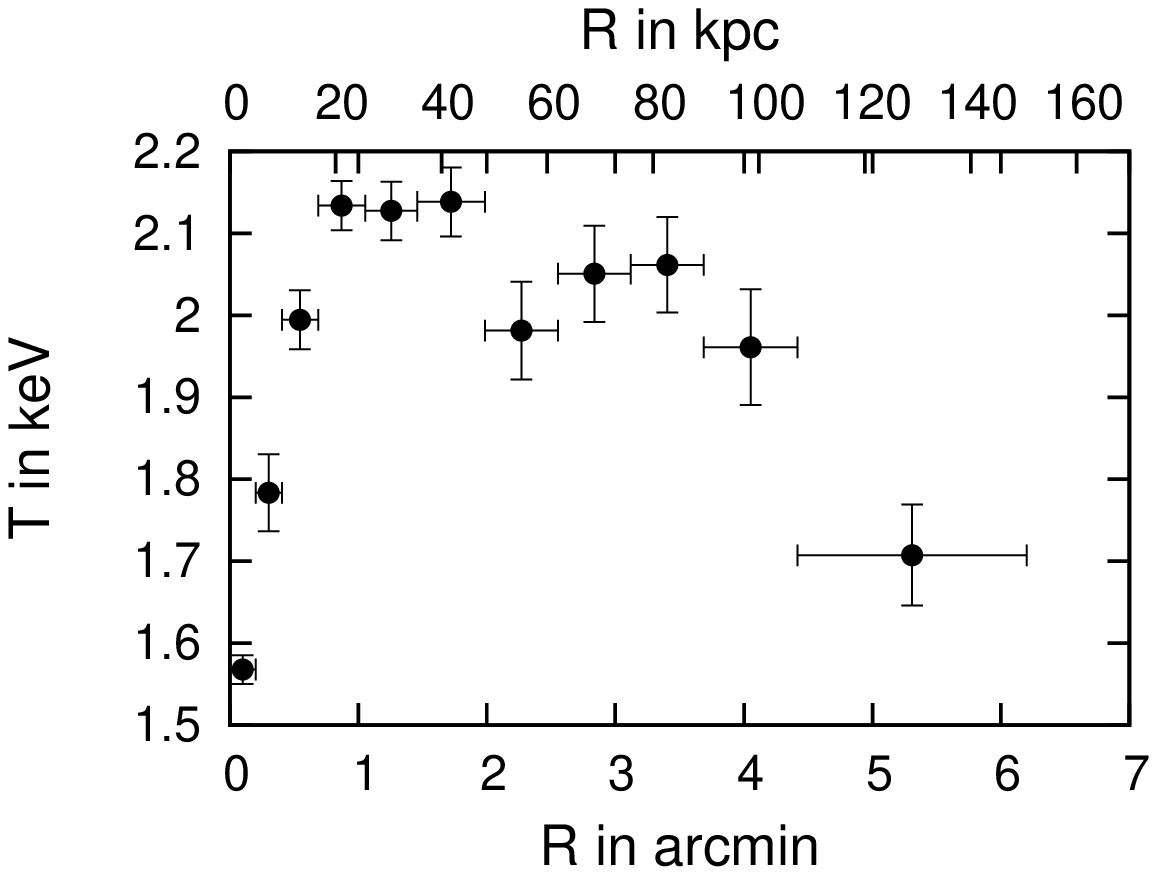}
\caption{Temperature profiles of IC 1633 (left) and MKW 4 (right).}
\end{figure*}
\begin{figure*}[h!]
\centering
\includegraphics[angle=-90,scale=0.50]{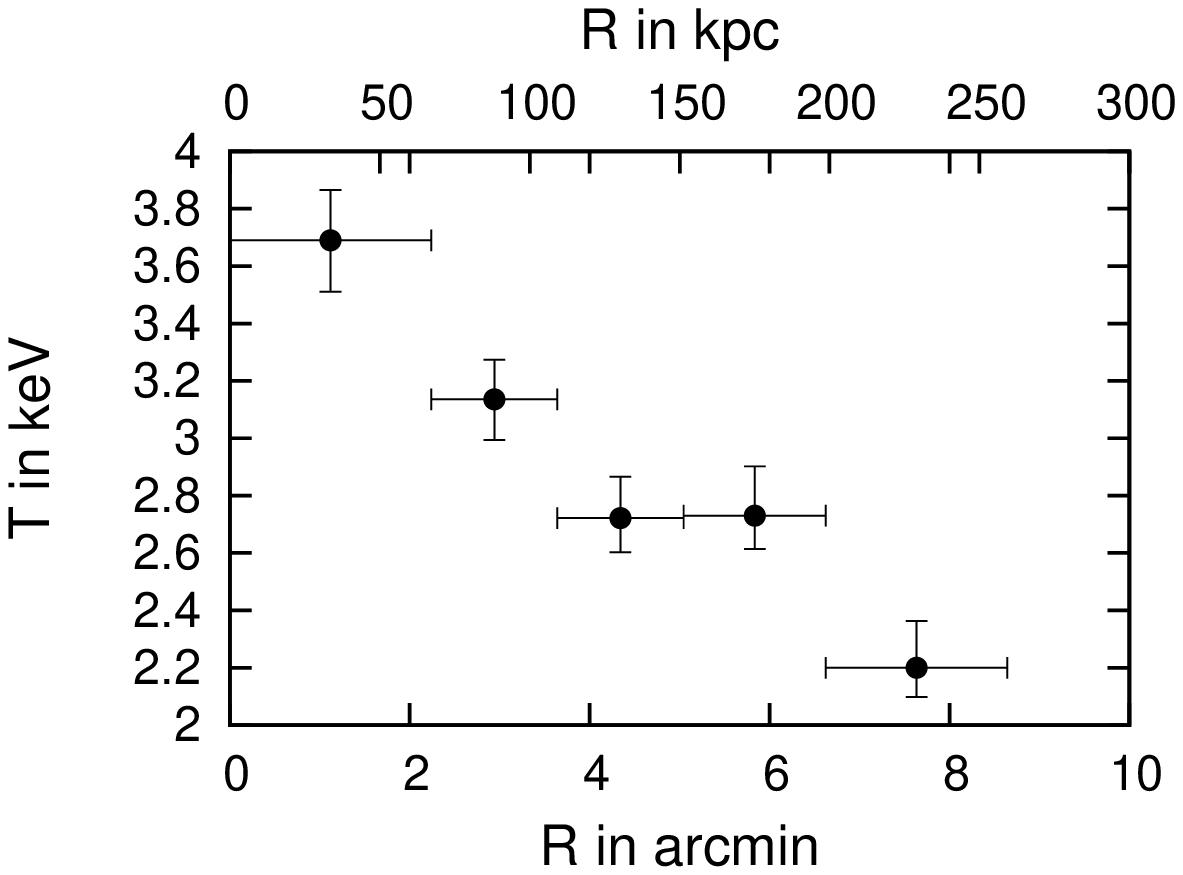}
\includegraphics[angle=-90,scale=0.50]{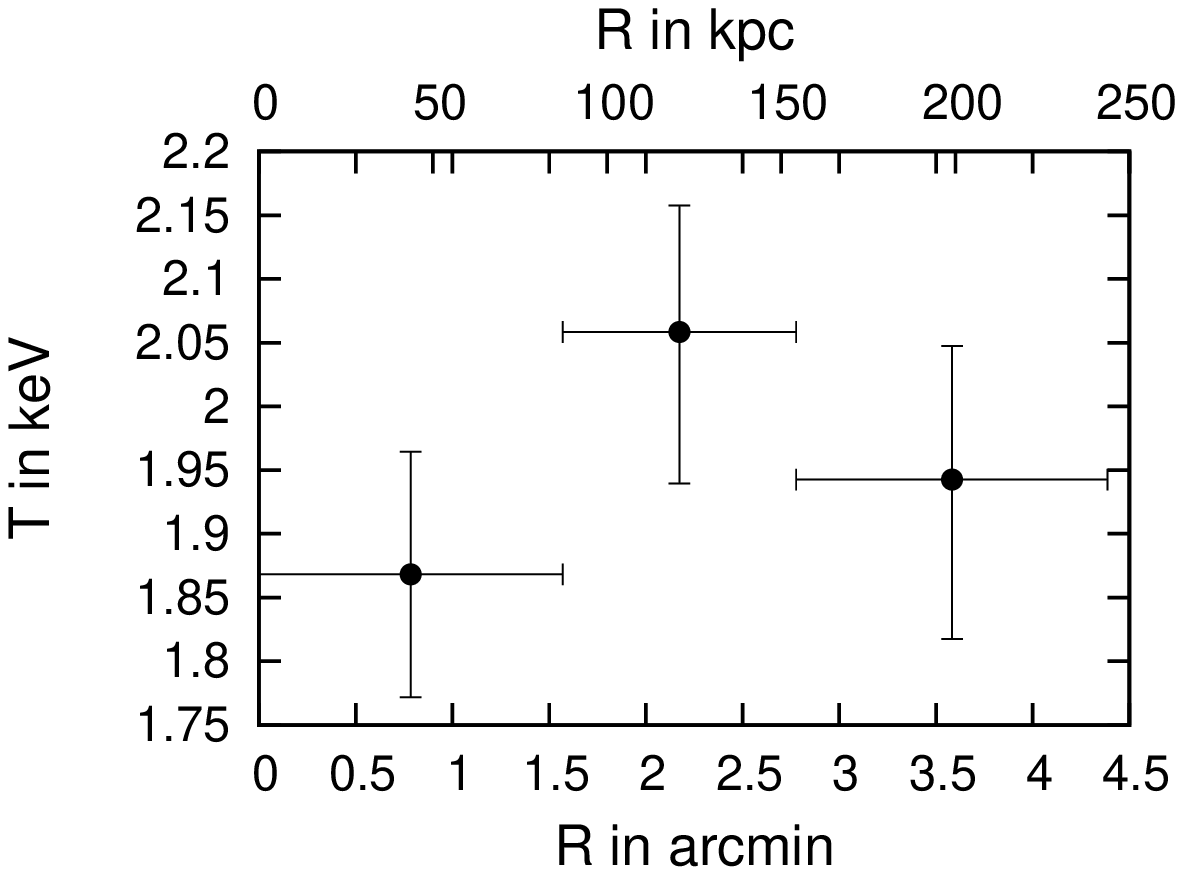}
\caption{Temperature profiles of MKW 8 (left) and NGC 326 (right).}
\end{figure*}
\begin{figure*}[h!]
\centering
\includegraphics[angle=-90,scale=0.50]{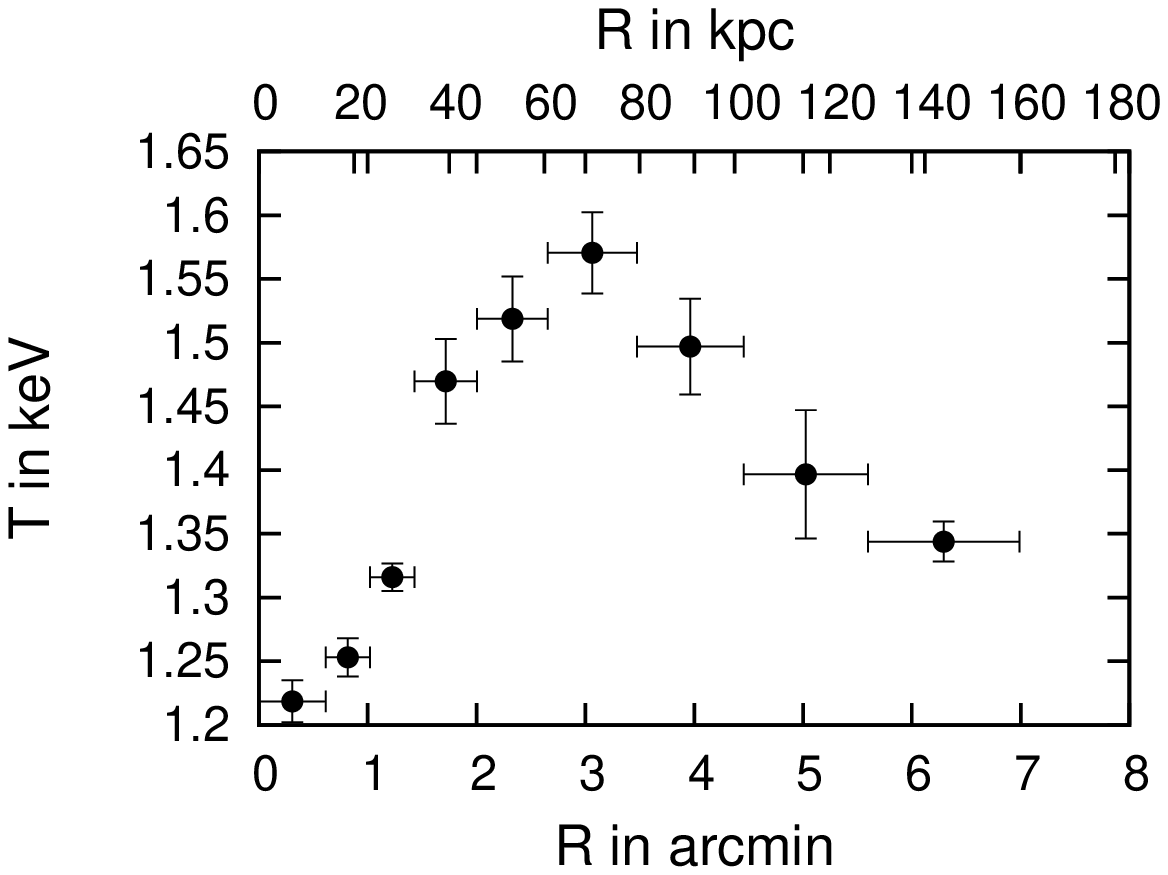}
\includegraphics[angle=-90,scale=0.50]{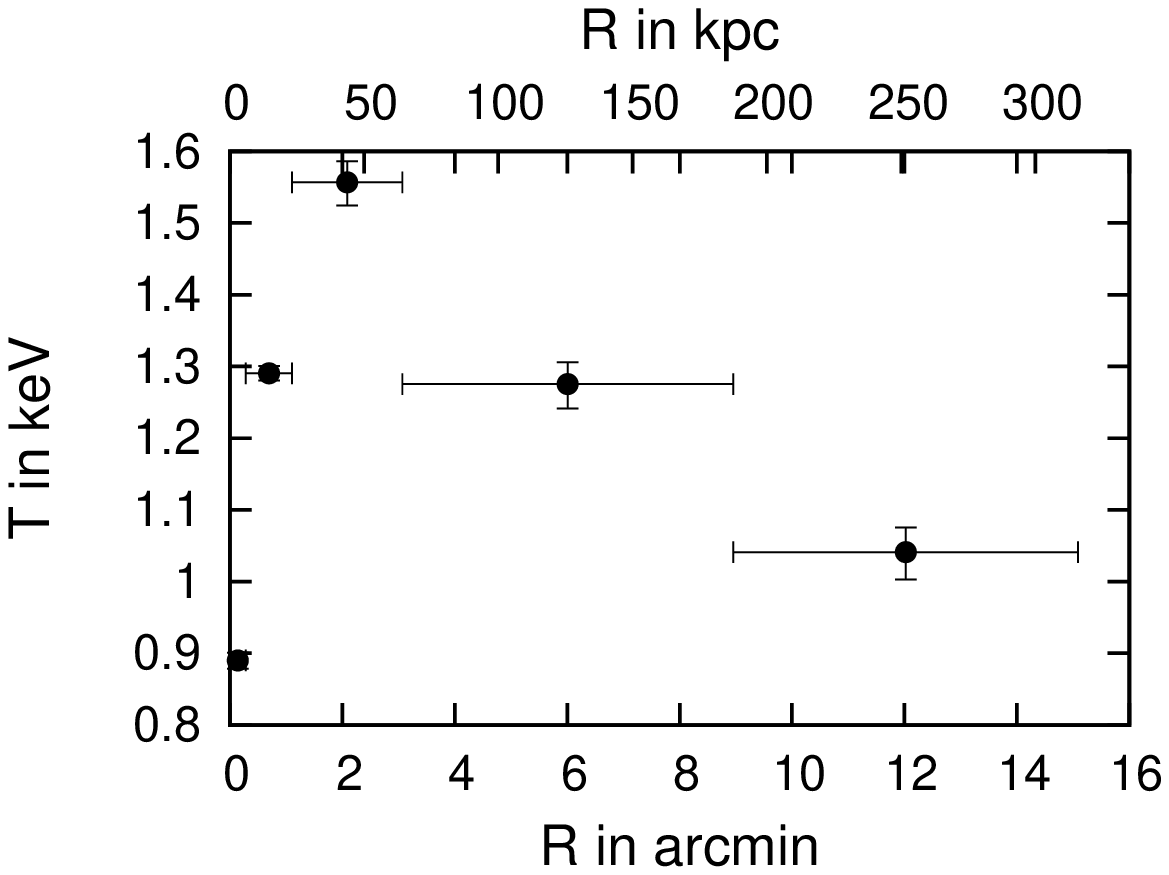}
\caption{Temperature profiles of NGC 507 (left) and NGC 533 (right).}
\end{figure*}
\begin{figure*}[h!]
\centering
\includegraphics[angle=-90,scale=0.50]{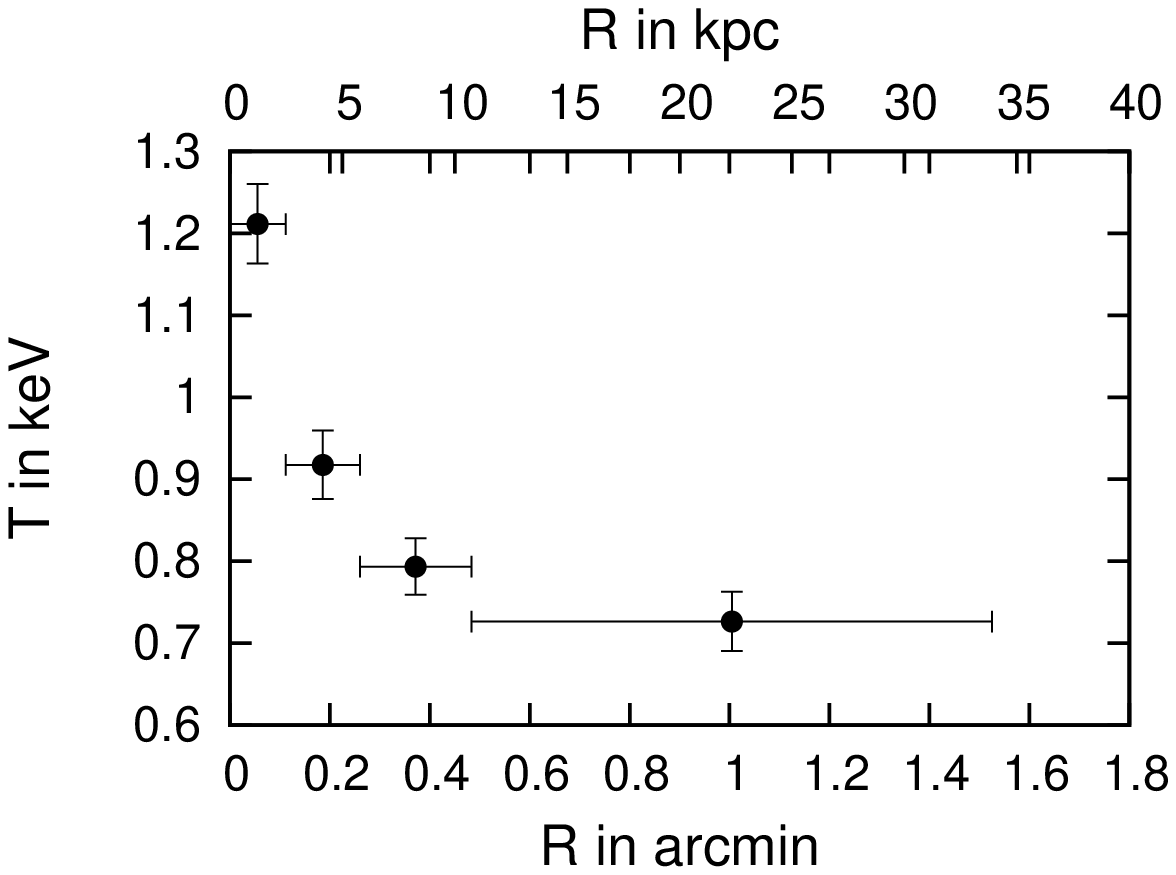}
\includegraphics[angle=-90,scale=0.50]{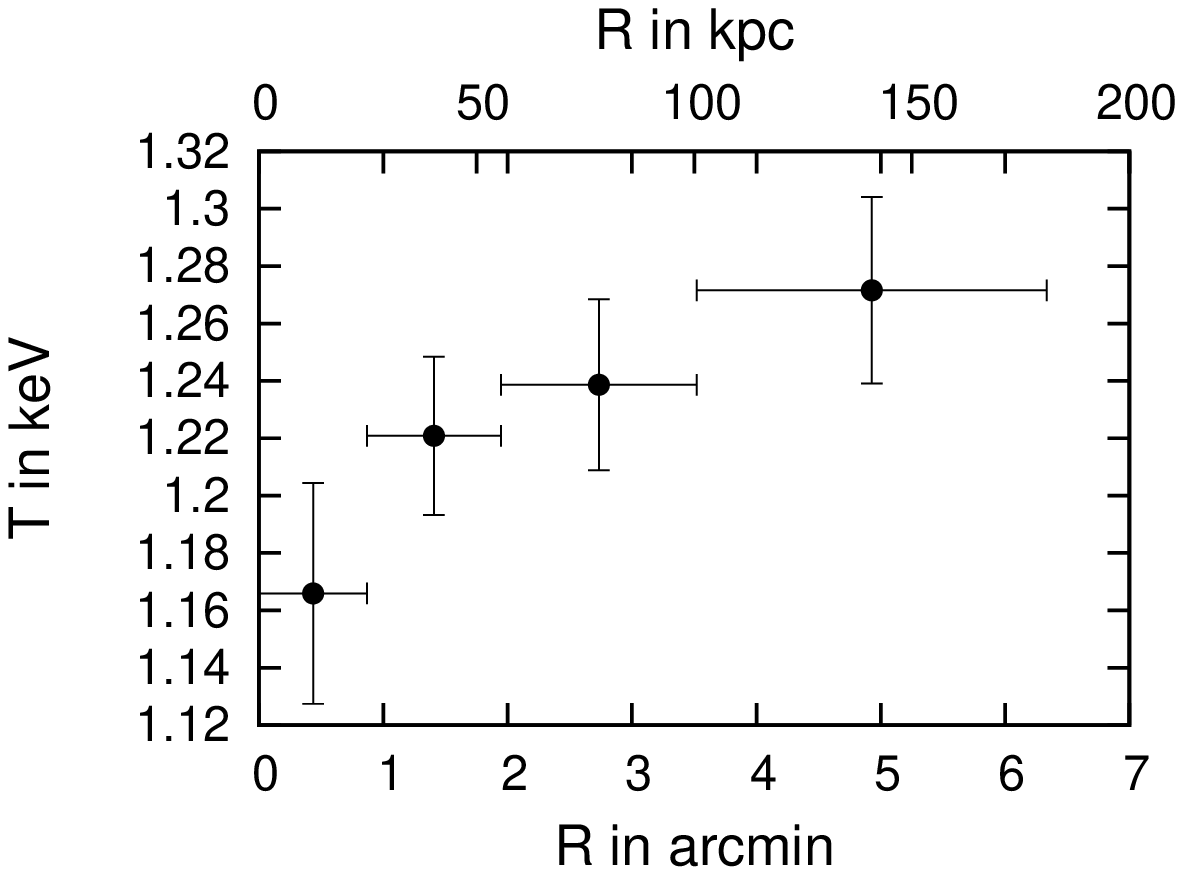}
\caption{Temperature profiles of NGC 777 (left) and NGC 1132 (right).}
\end{figure*}
\begin{figure*}[h!]
\centering
\includegraphics[angle=-90,scale=0.50]{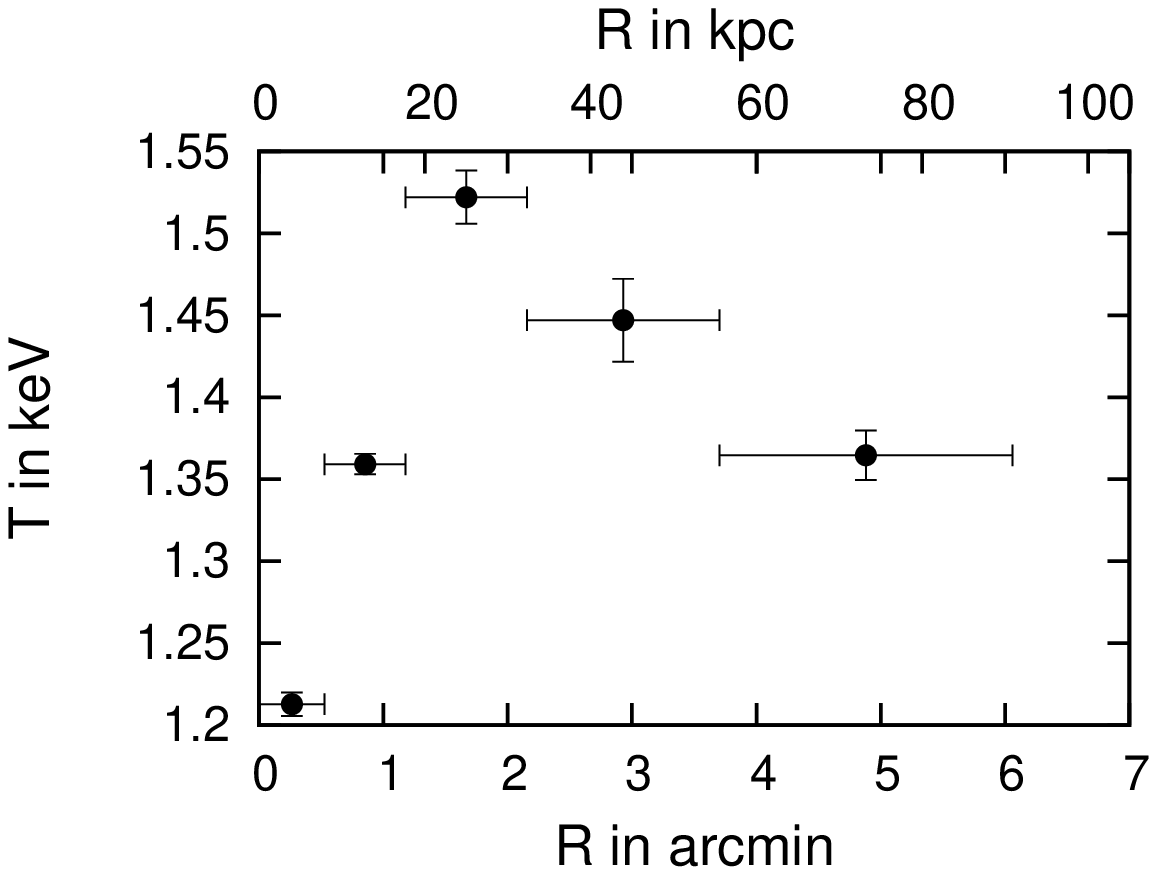}
\includegraphics[angle=-90,scale=0.50]{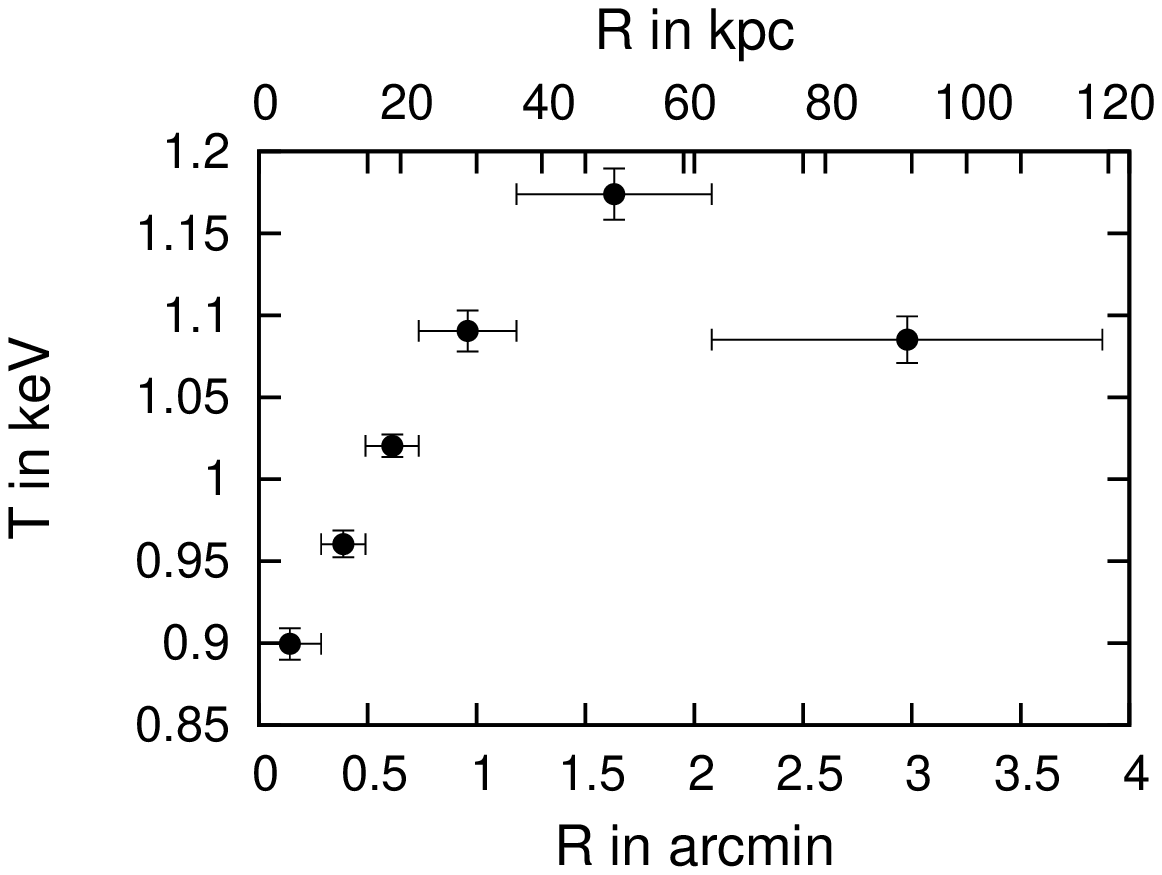}
\caption{Temperature profiles of NGC 1550 (left) and NGC 4325 (right).}
\end{figure*}
\begin{figure*}[h!]
\centering
\includegraphics[angle=-90,scale=0.50]{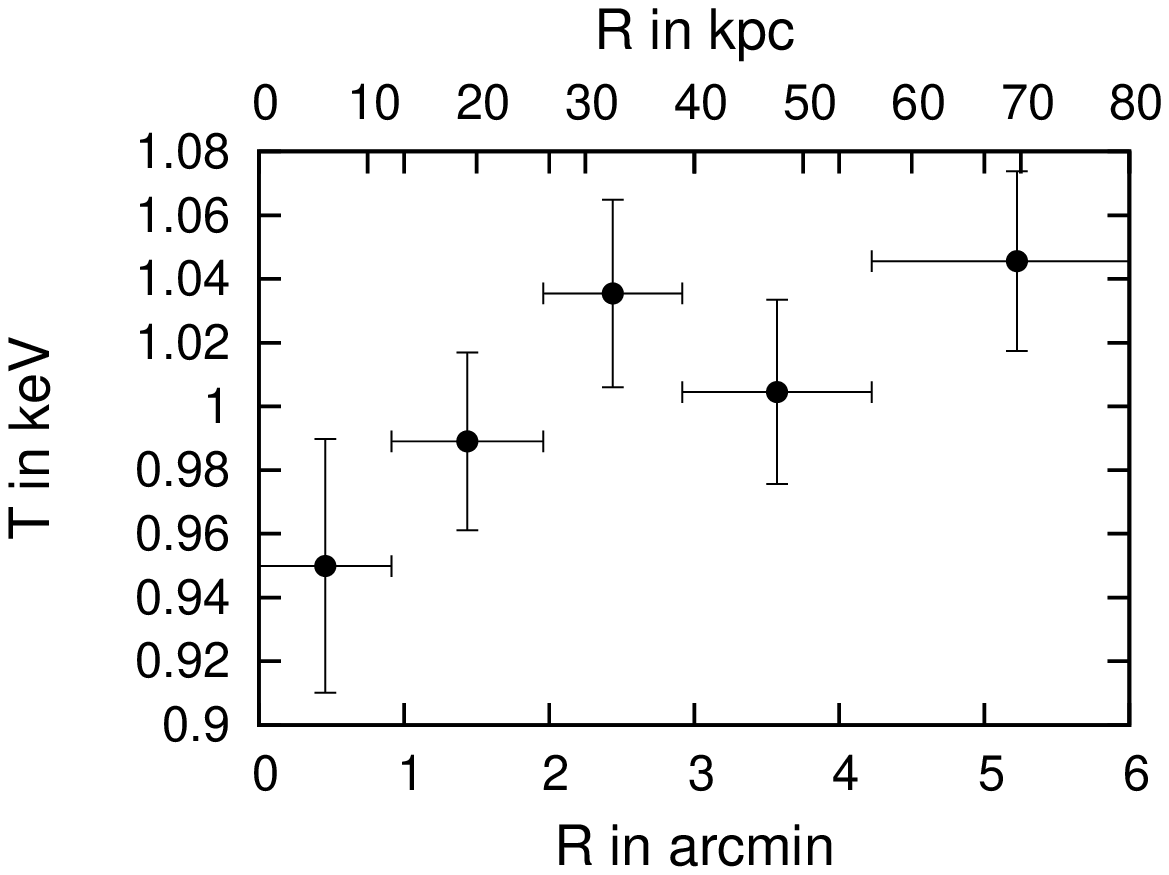}
\includegraphics[angle=-90,scale=0.50]{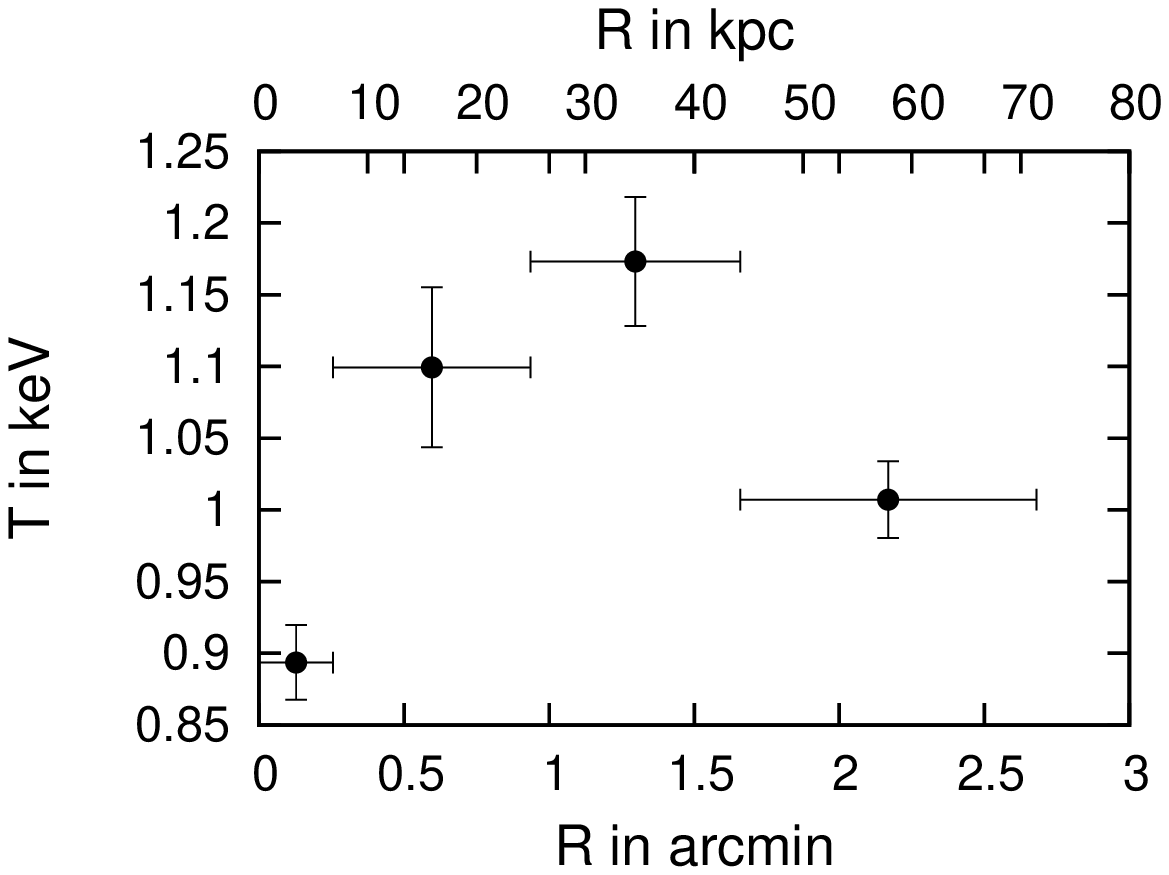}
\caption{Temperature profiles of NGC 4936 (left) and NGC 5129 (right).}
\end{figure*}
\begin{figure*}[h!]
\centering
\includegraphics[angle=-90,scale=0.50]{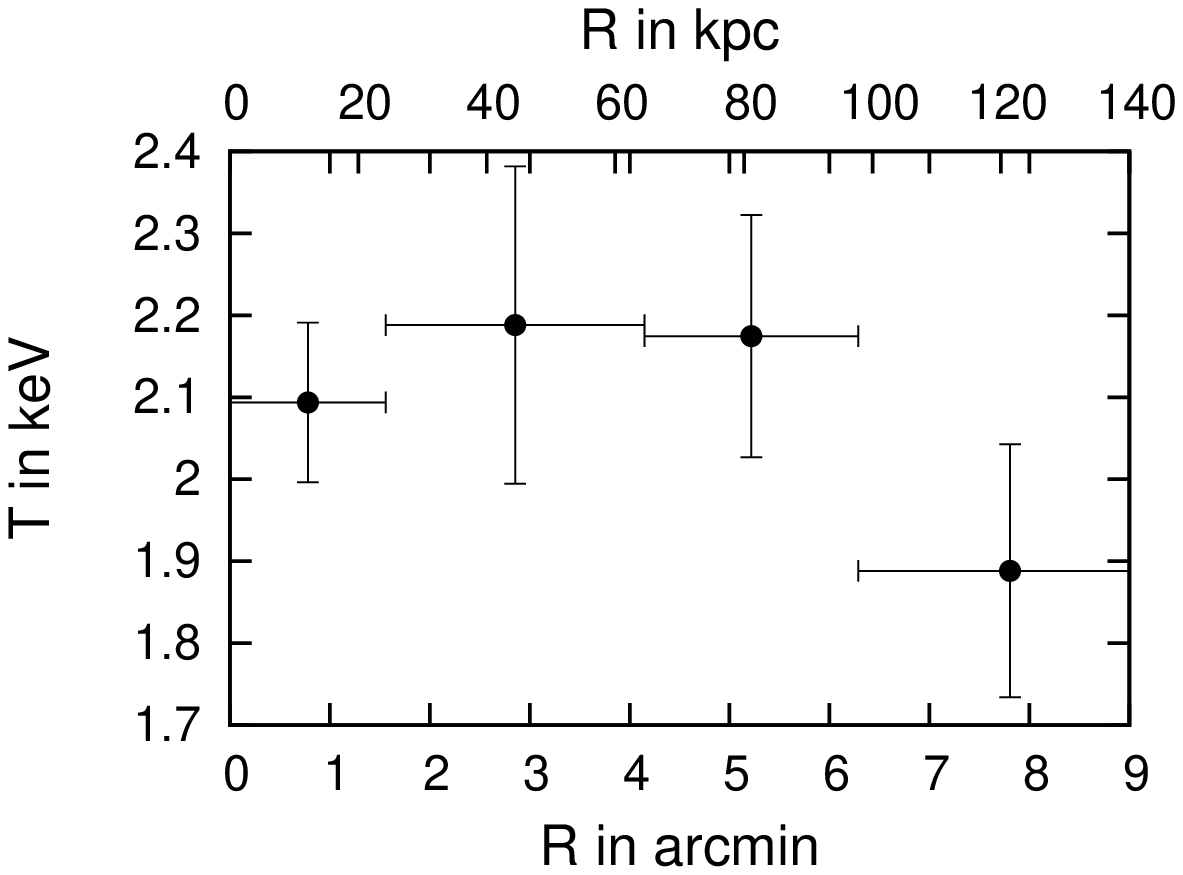}
\includegraphics[angle=-90,scale=0.50]{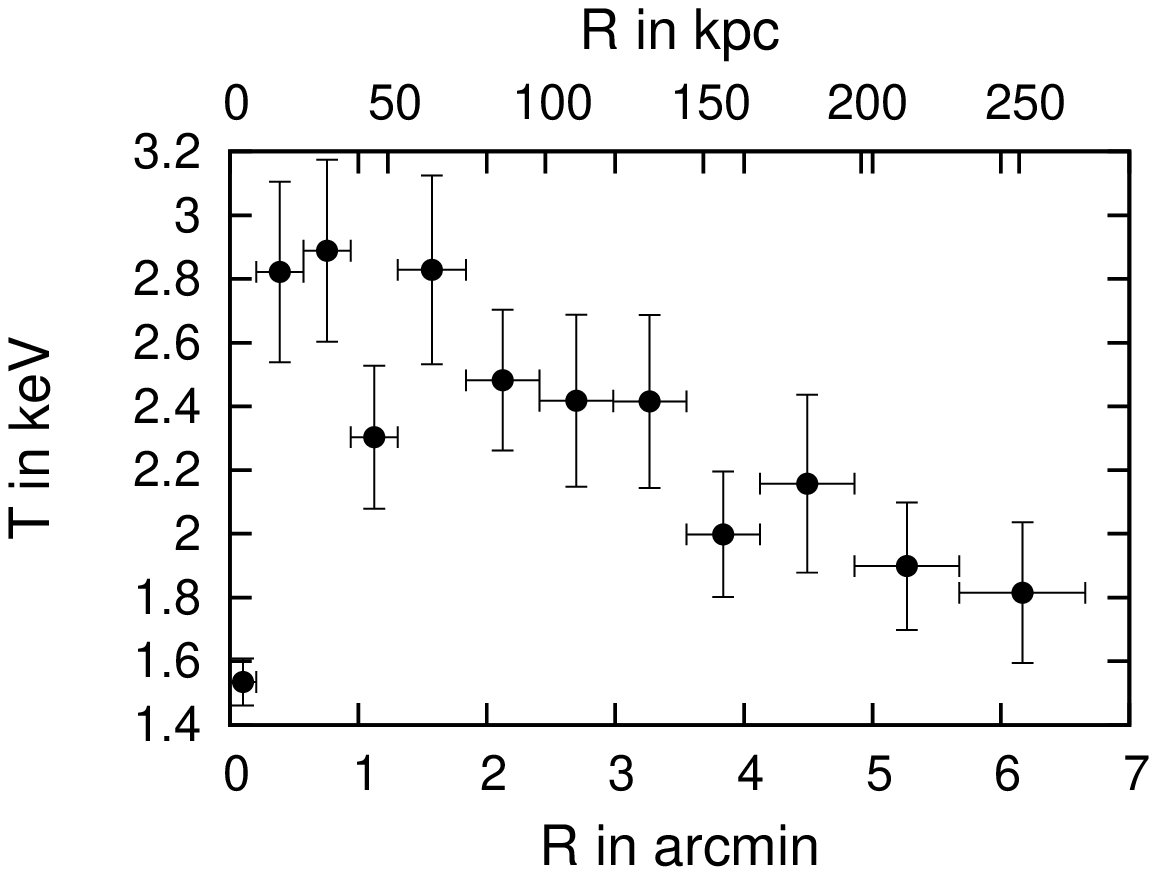}
\caption{Temperature profiles of NGC 5419 (left) and NGC 6269 (right).}
\end{figure*}
\begin{figure*}[h!]
\centering
\includegraphics[angle=-90,scale=0.50]{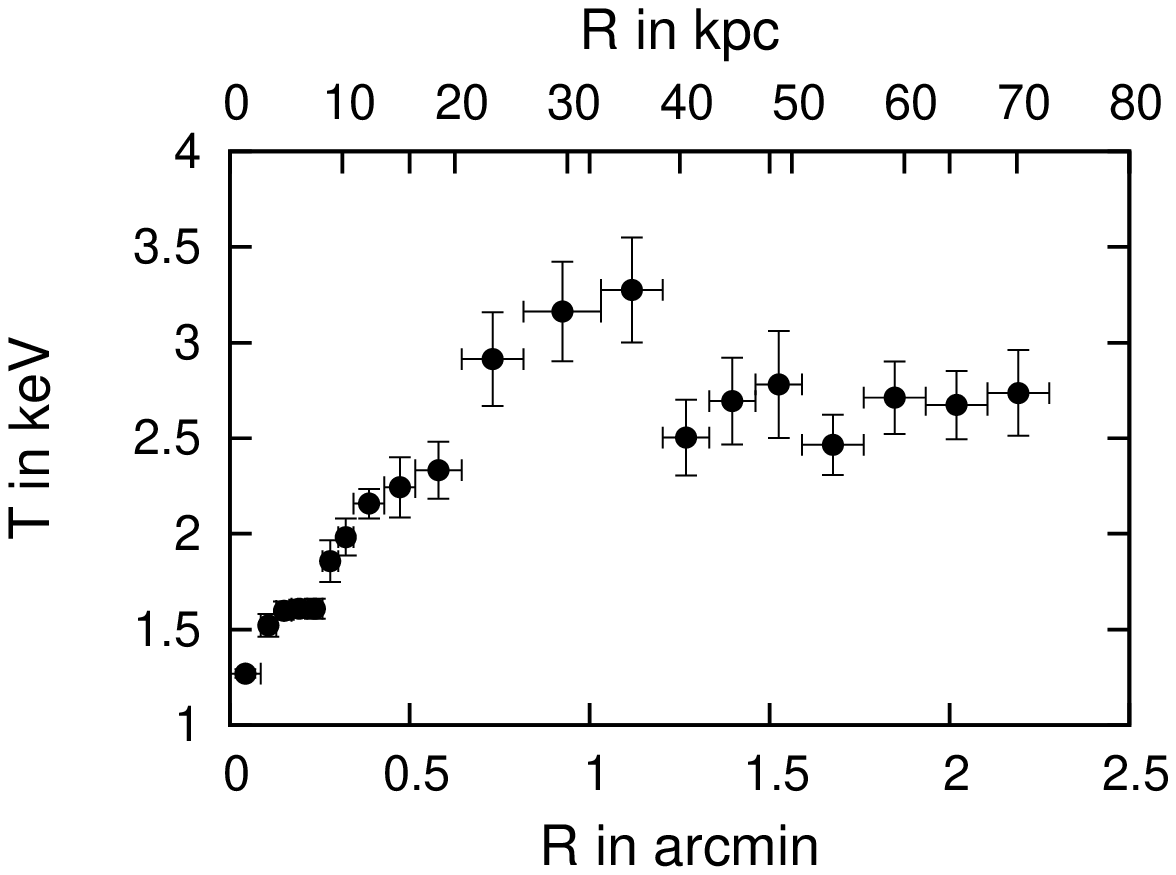}
\includegraphics[angle=-90,scale=0.50]{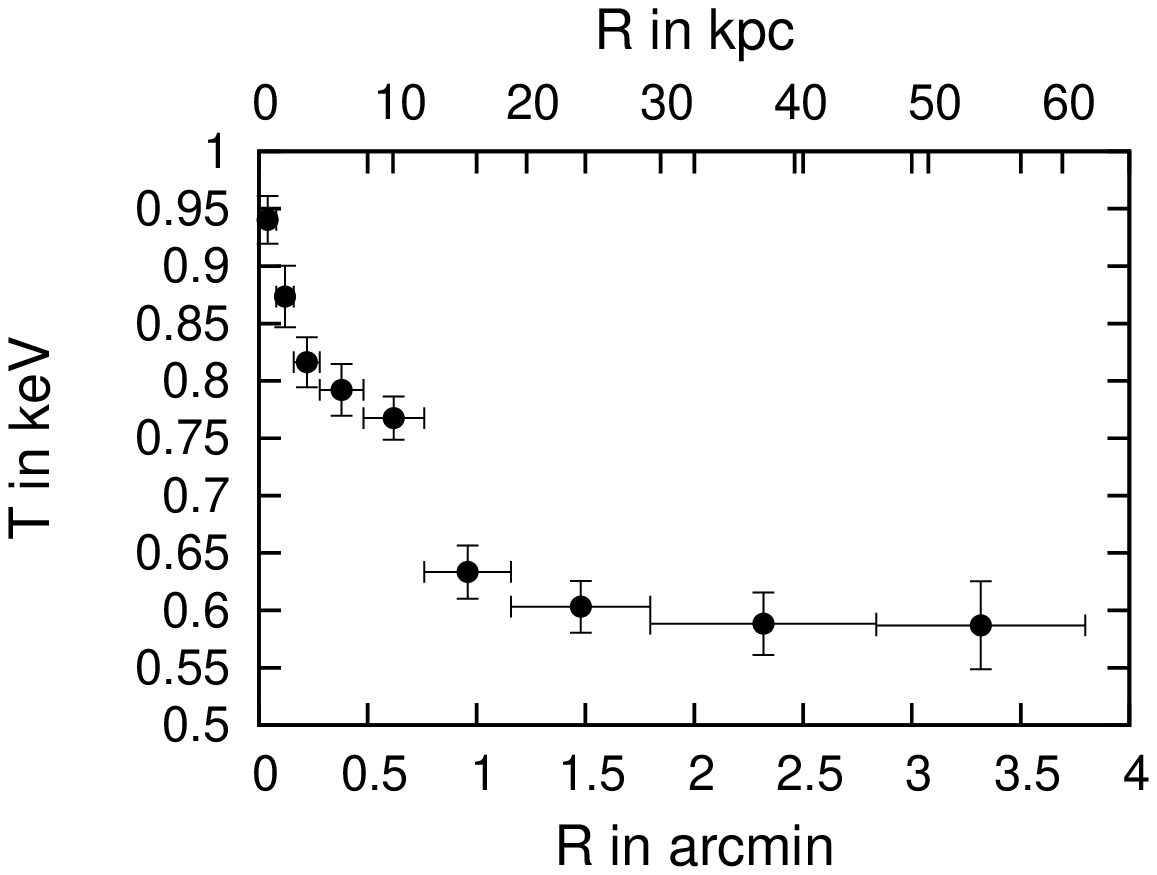}
\caption{Temperature profiles of NGC 6338 (left) and NGC 6482 (right).}
\end{figure*}
\begin{figure*}[h!]
\centering
\includegraphics[angle=-90,scale=0.50]{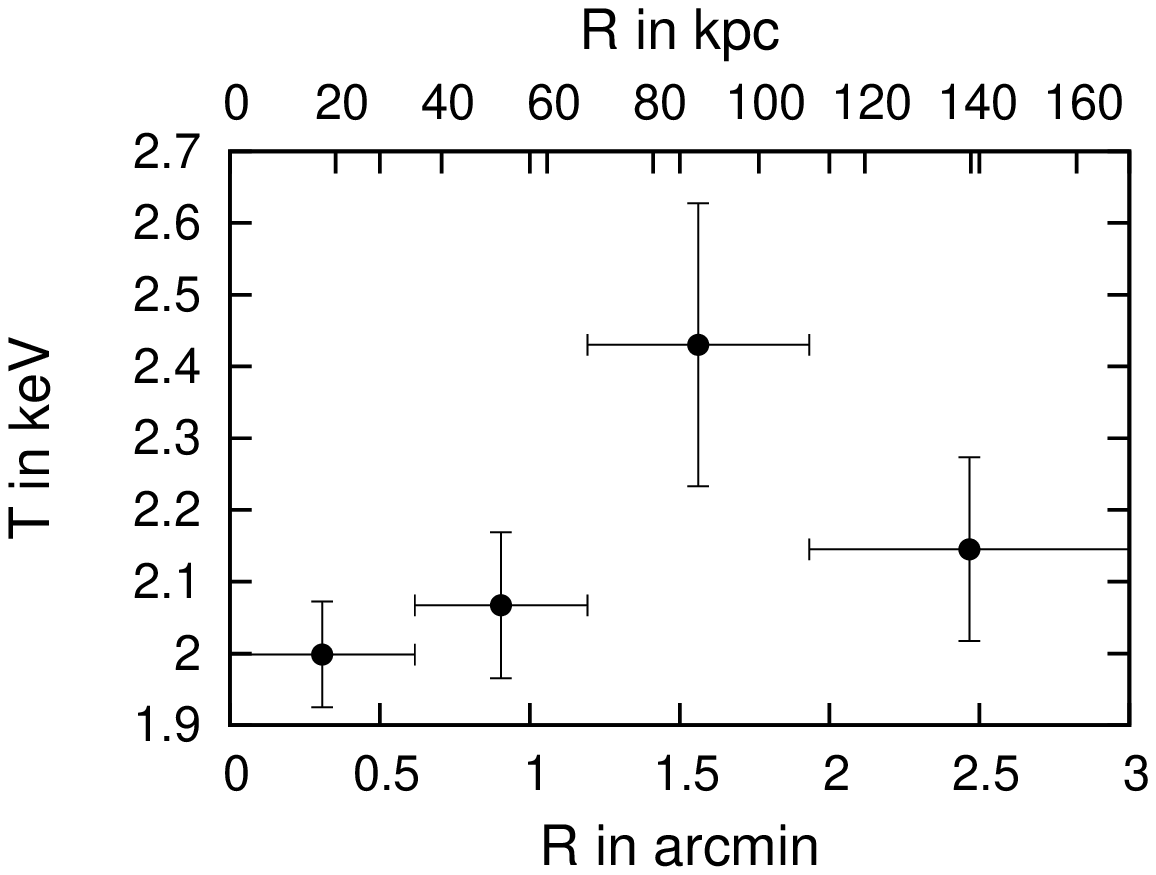}
\includegraphics[angle=-90,scale=0.50]{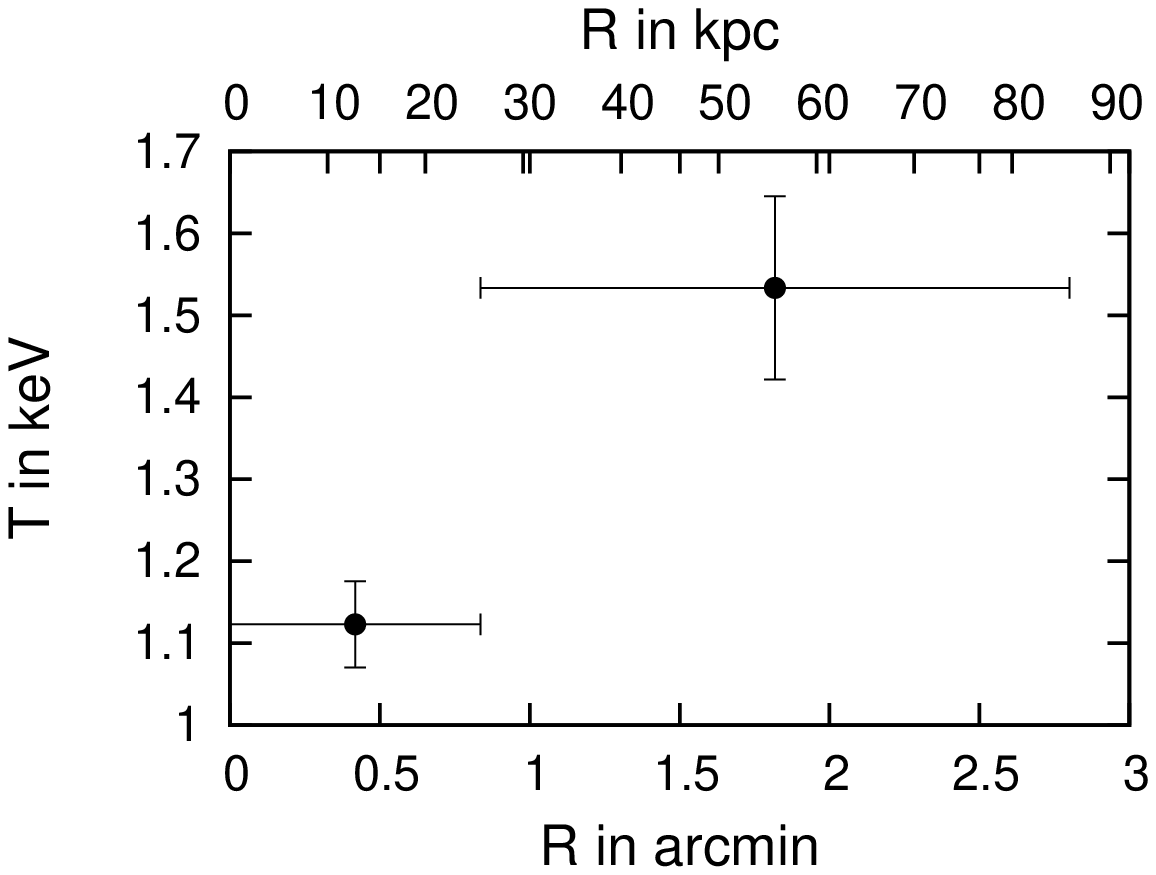}
\caption{Temperature profiles of RXCJ 1022 (left) and RXCJ 2214 (right).}
\end{figure*}
\begin{figure*}[h!]
\centering
\includegraphics[angle=-90,scale=0.50]{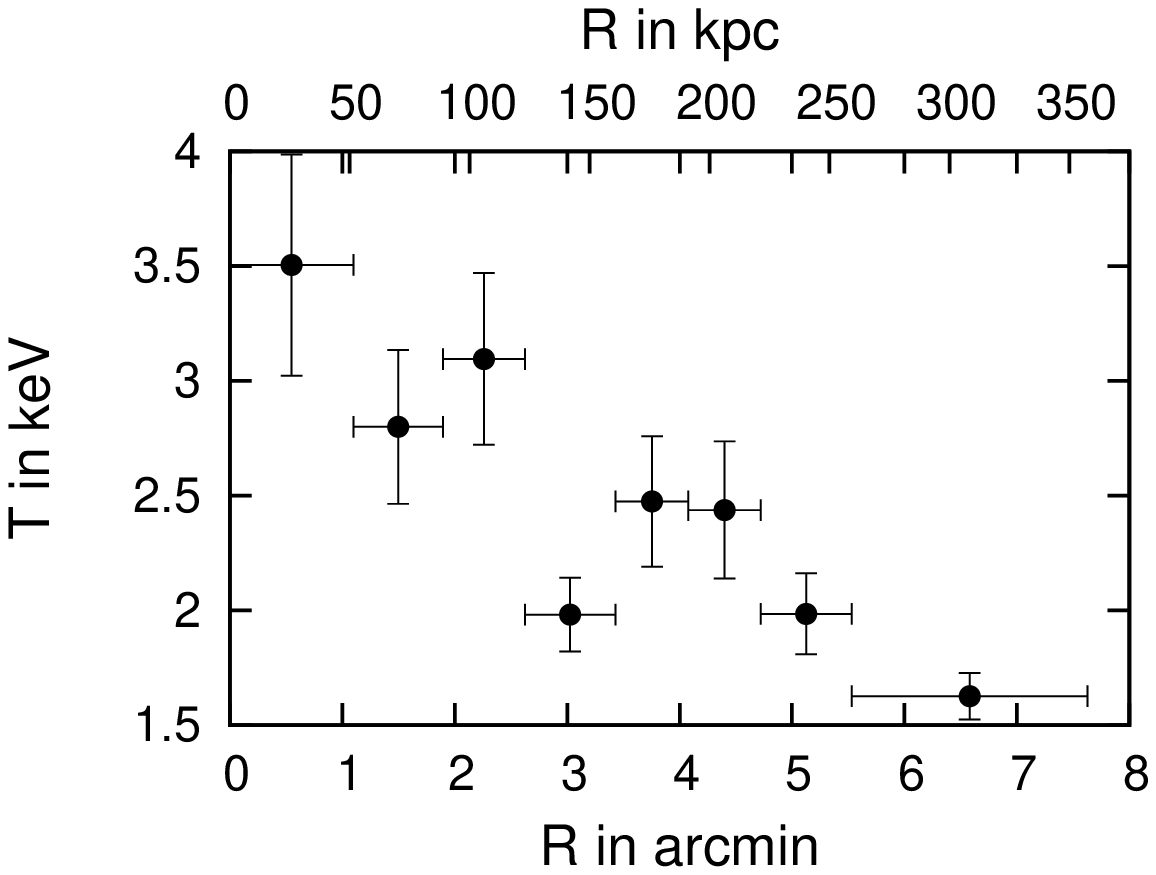}
\includegraphics[angle=-90,scale=0.50]{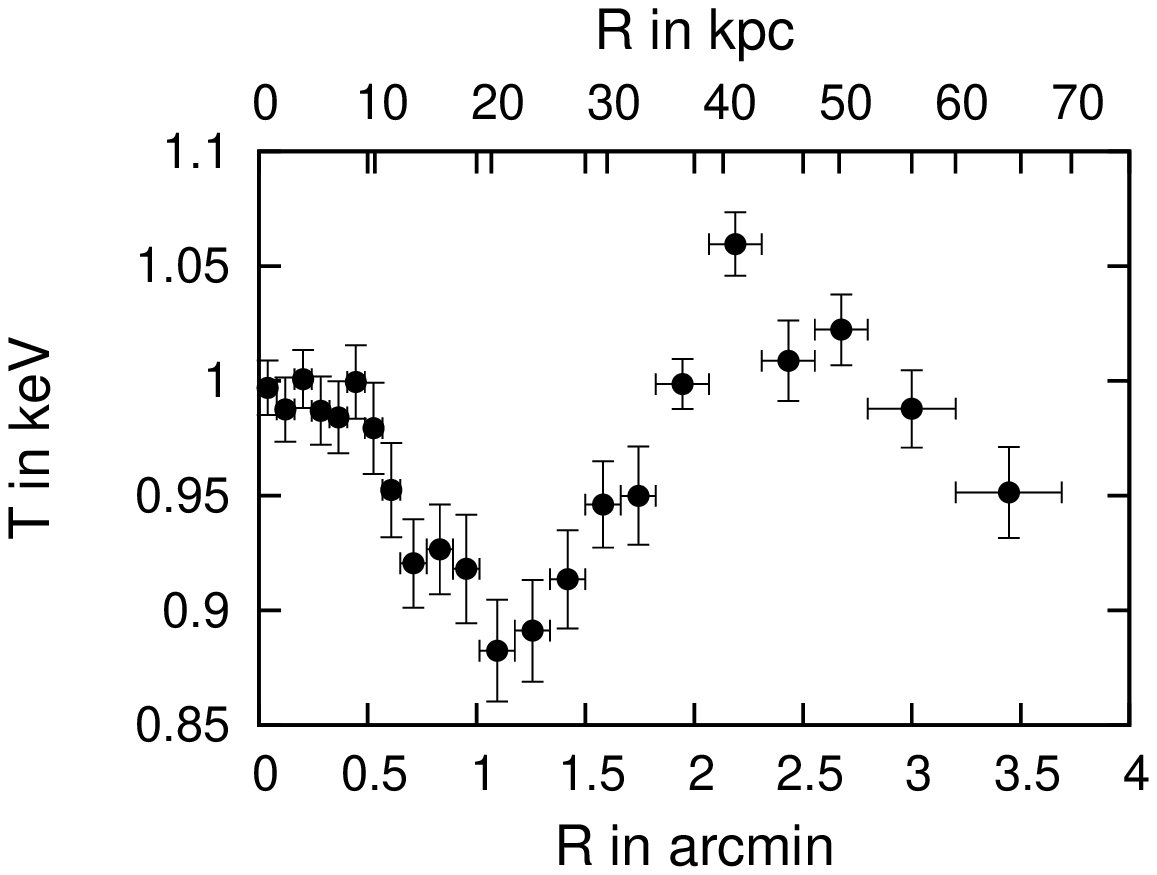}
\caption{Temperature profiles of S0463 (left) and SS2B (right).}
\end{figure*}
\clearpage

\section{NVSS radio contours on optical images for some groups}\label{radicont}

Here, we present 1.4 GHz NVSS radio contours (green) overlaid on
optical images from SAO-DSS with the EP marked (red X point). Co-ordinates are sexagesimal Right Ascension and Declination.
\begin{figure*}[h!]
\centering
\includegraphics[]{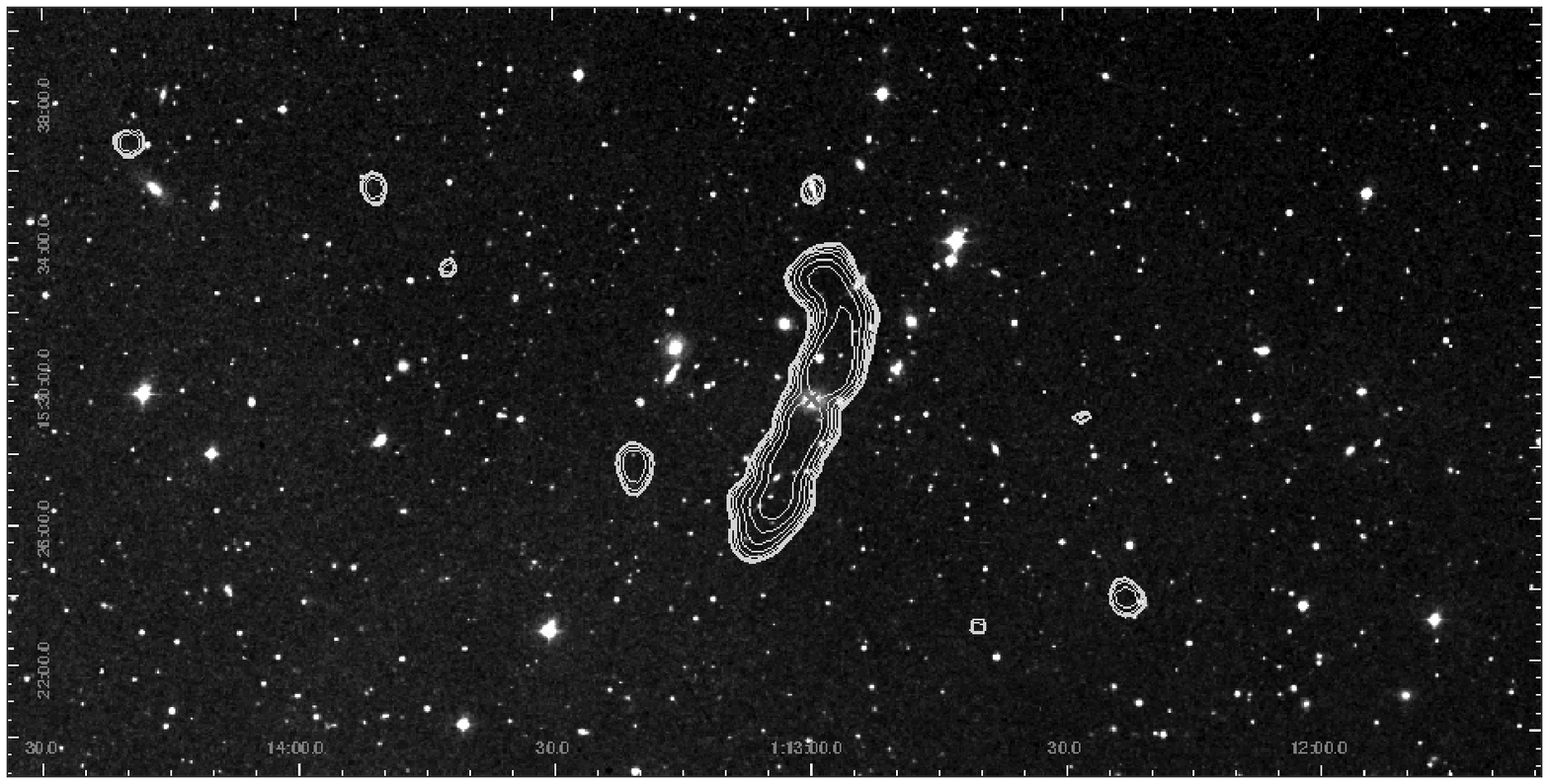}
 \includegraphics[]{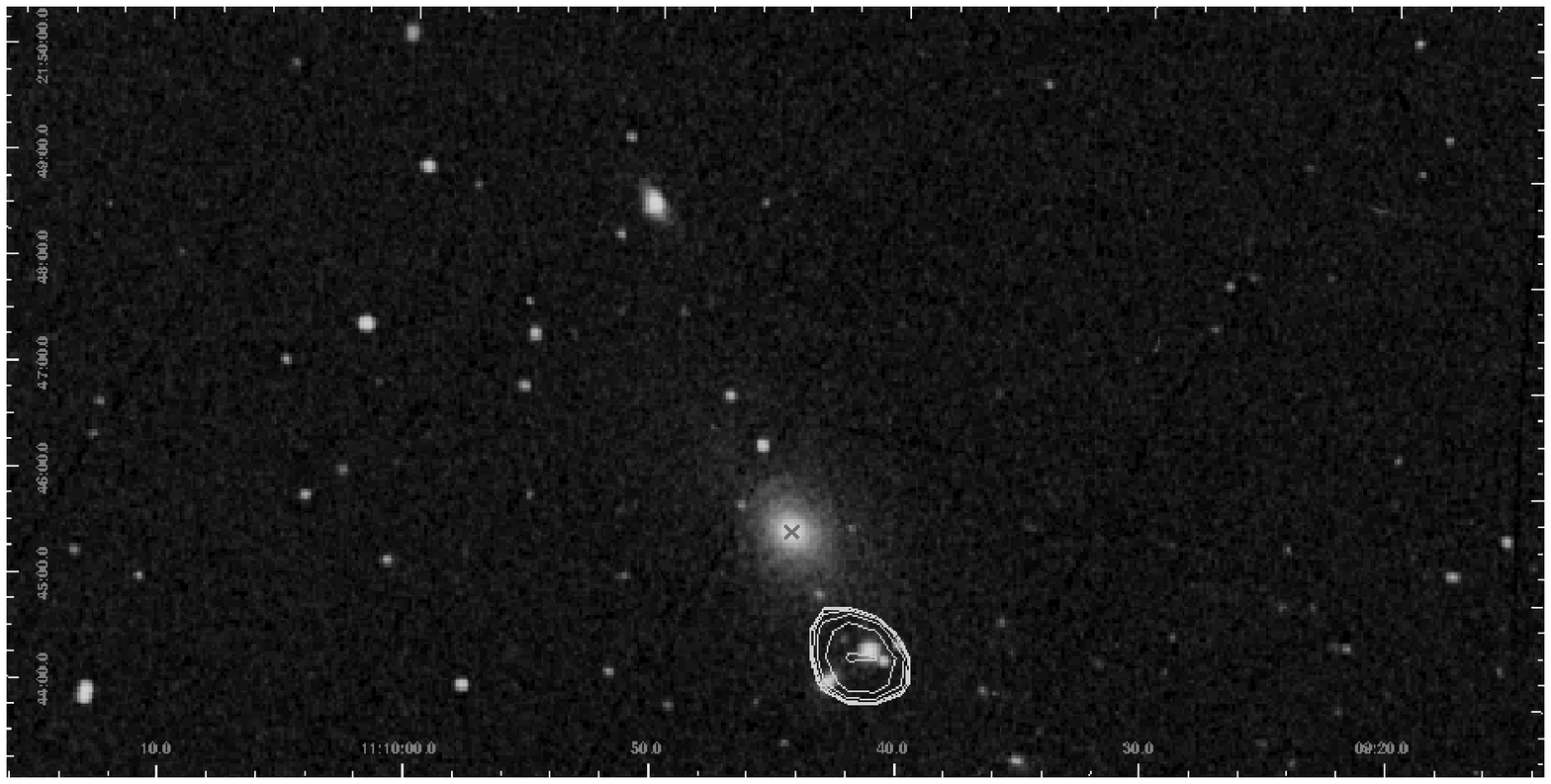}
\caption{A0160 and A1177.}
\end{figure*}
\begin{figure*}[h!]
 \includegraphics[]{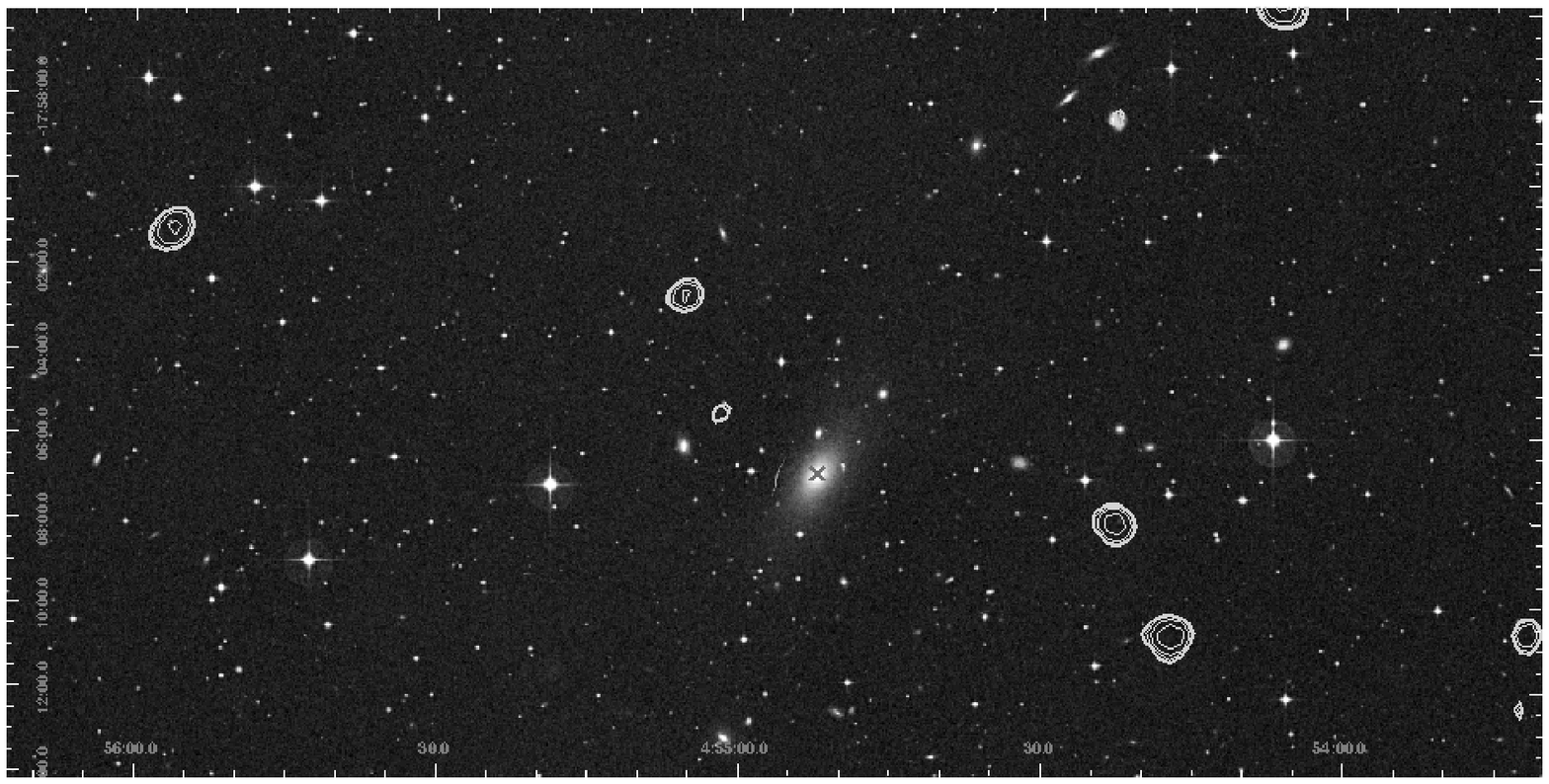}
\includegraphics{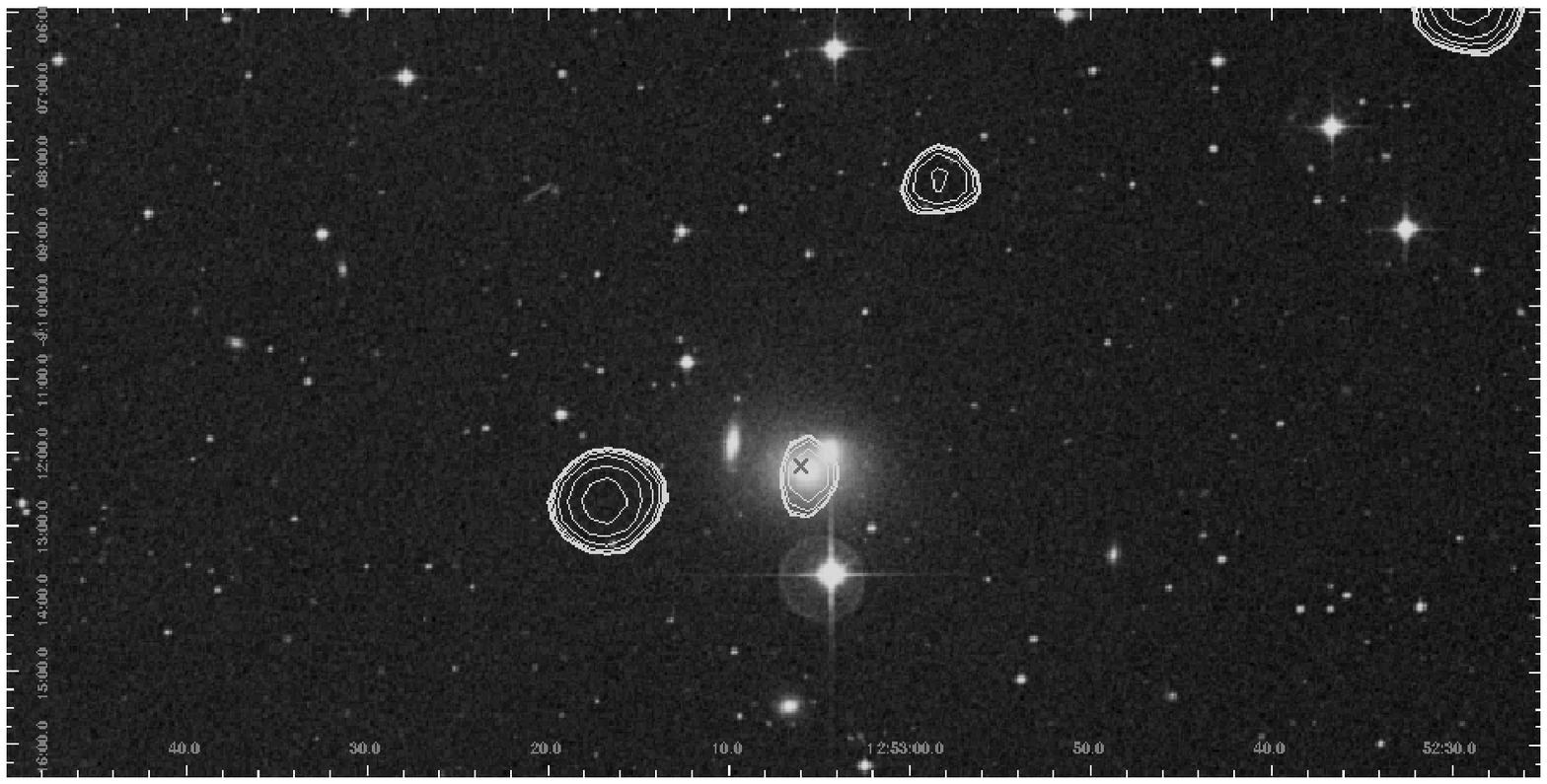}
\caption{ESO55 and HCG62.}
\end{figure*}
\begin{figure*}[h!]
 \includegraphics[]{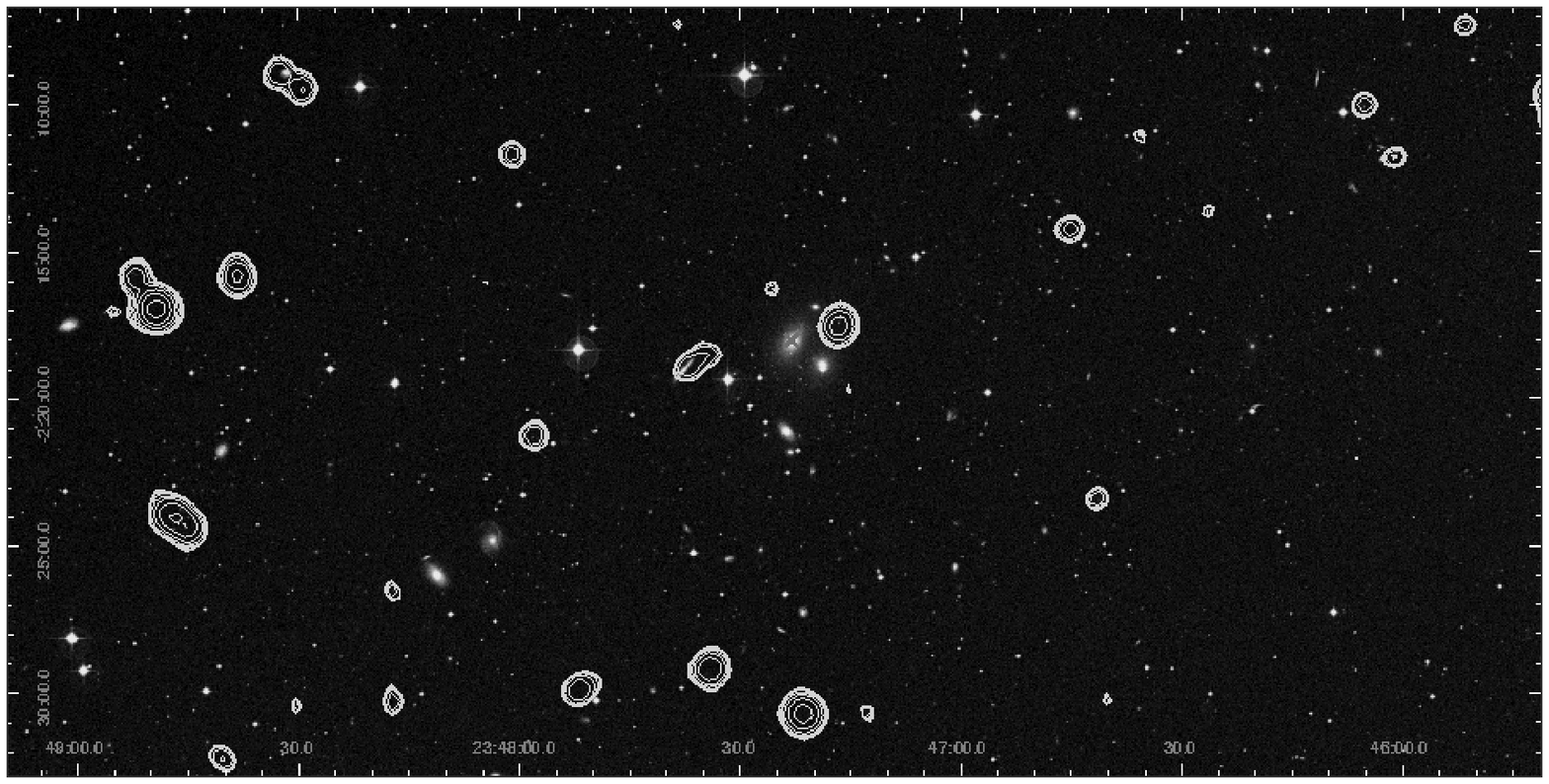}
\includegraphics{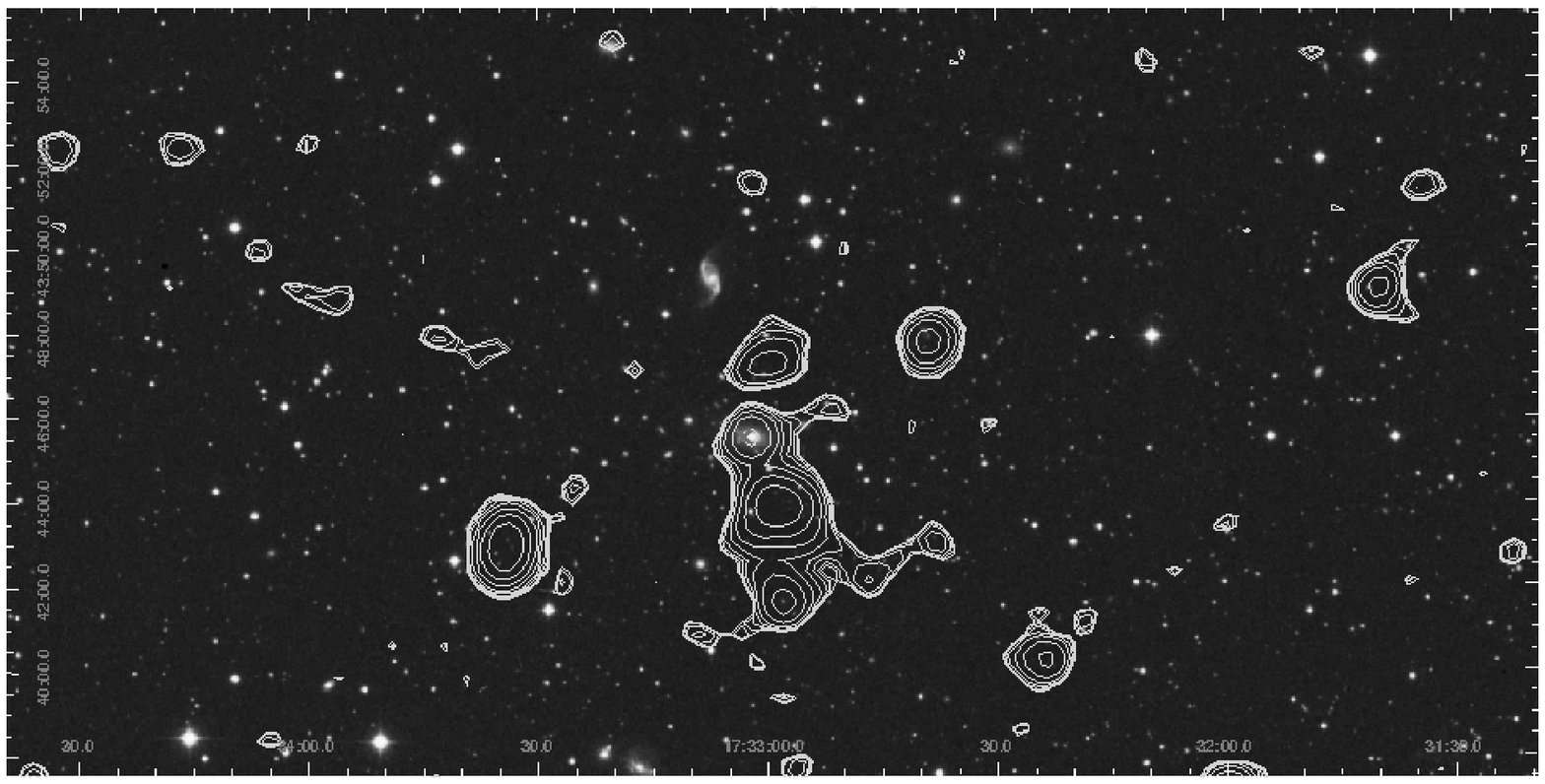}
\caption{HCG97 and IC1262.}
\end{figure*}

\begin{figure*}[h!]
\includegraphics{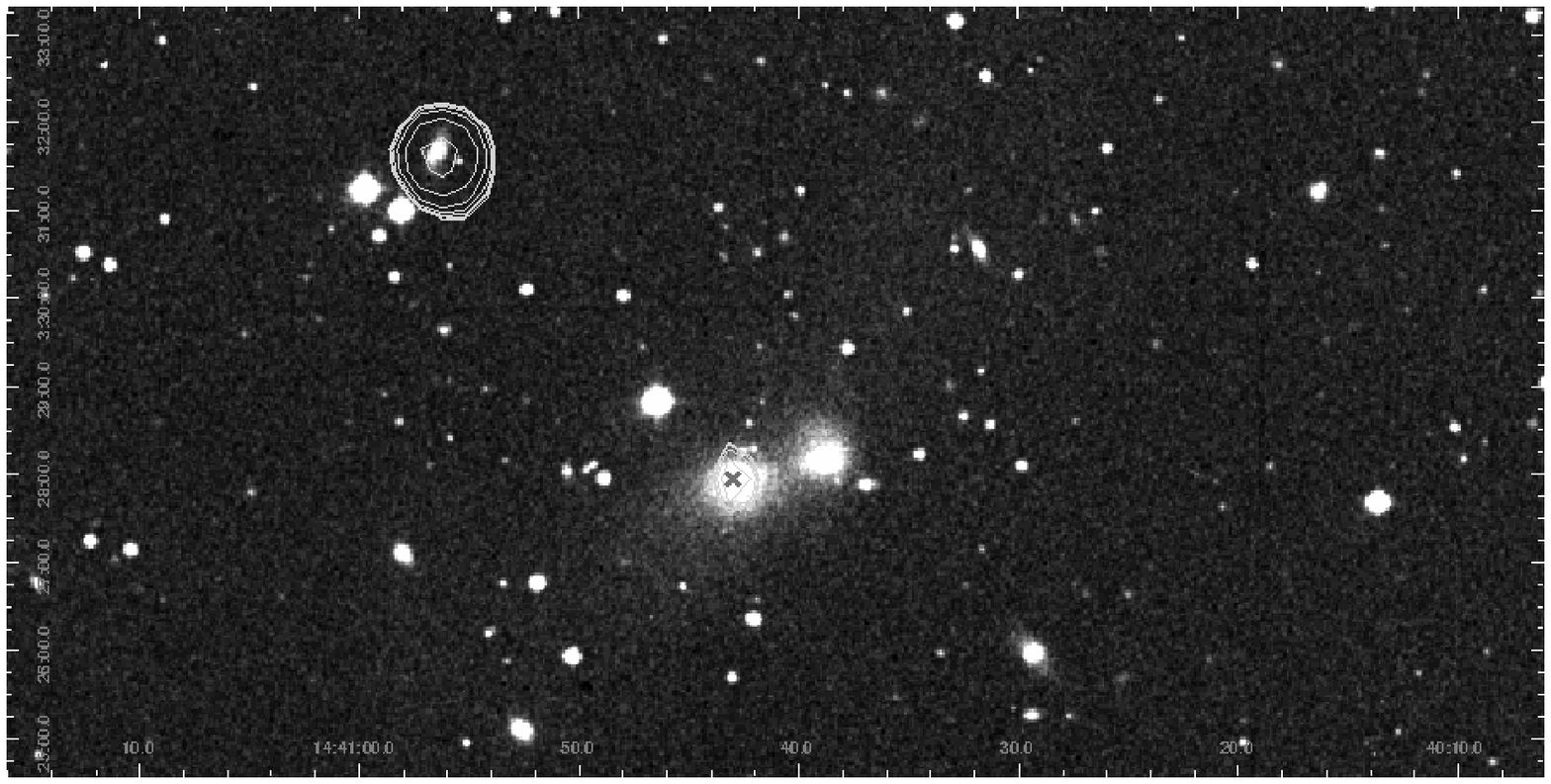}
\includegraphics{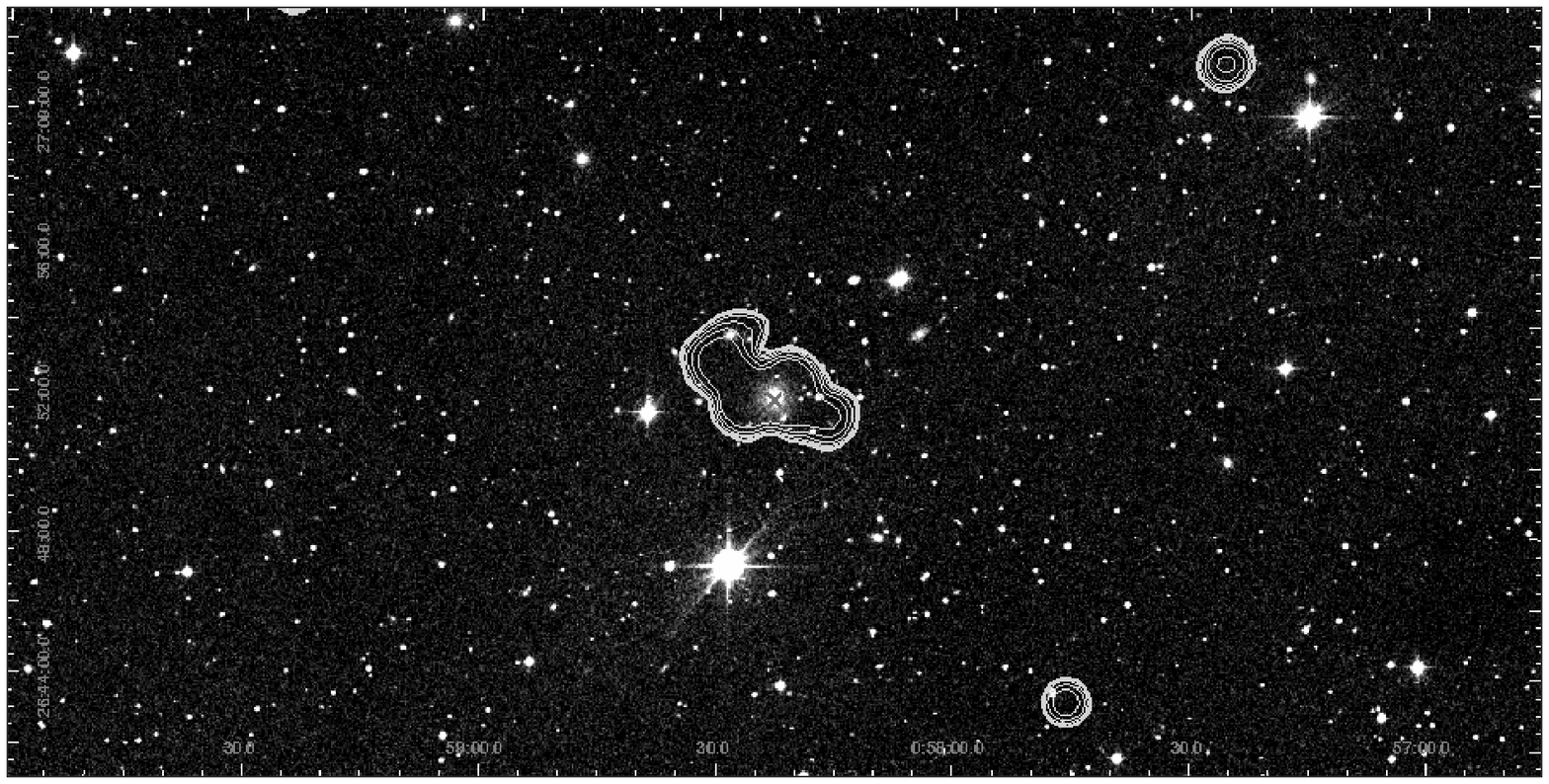}
\caption{MKW8 and NGC326.}
\end{figure*}

\begin{figure*}[h!]
\includegraphics{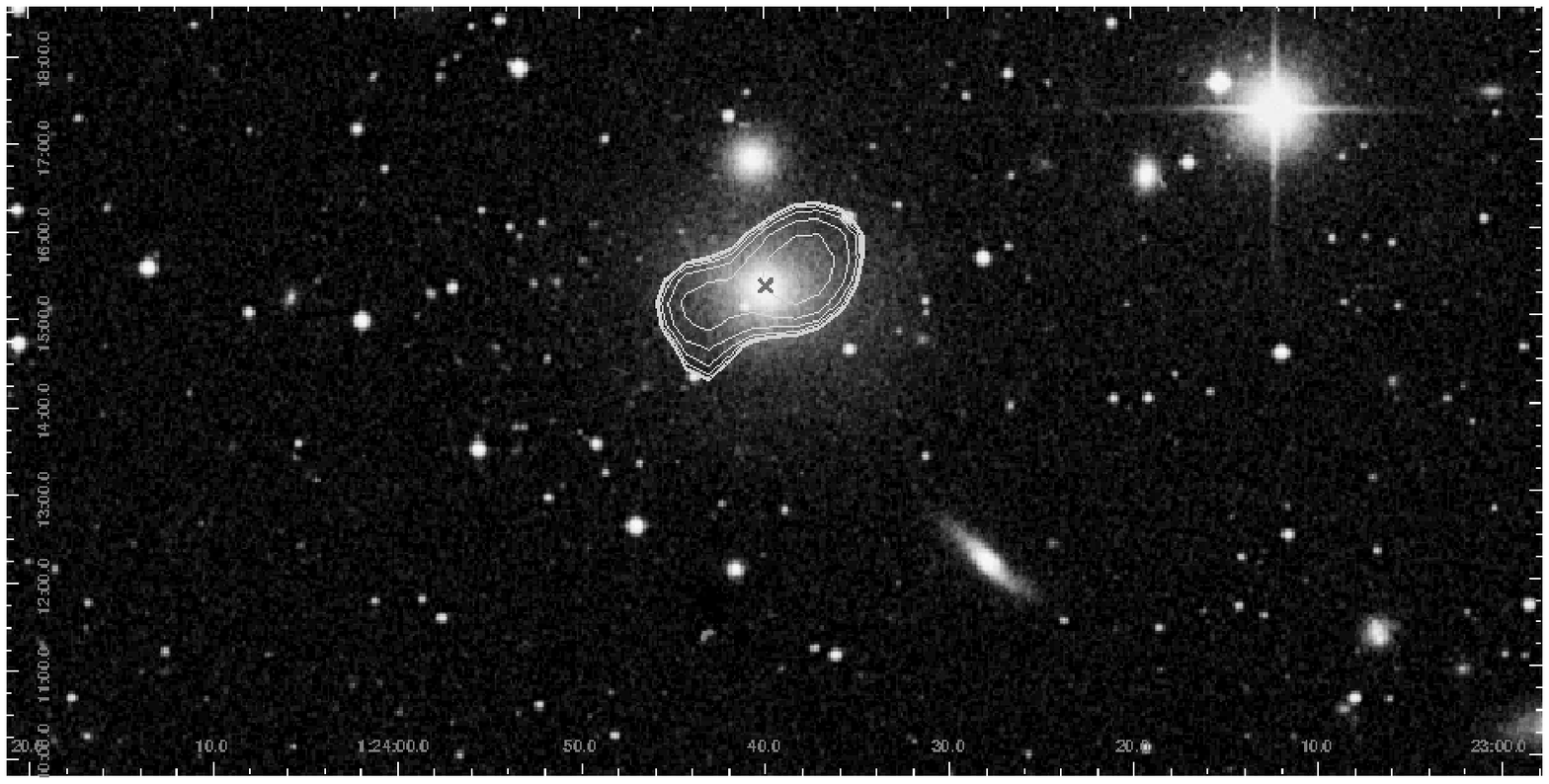}
\includegraphics{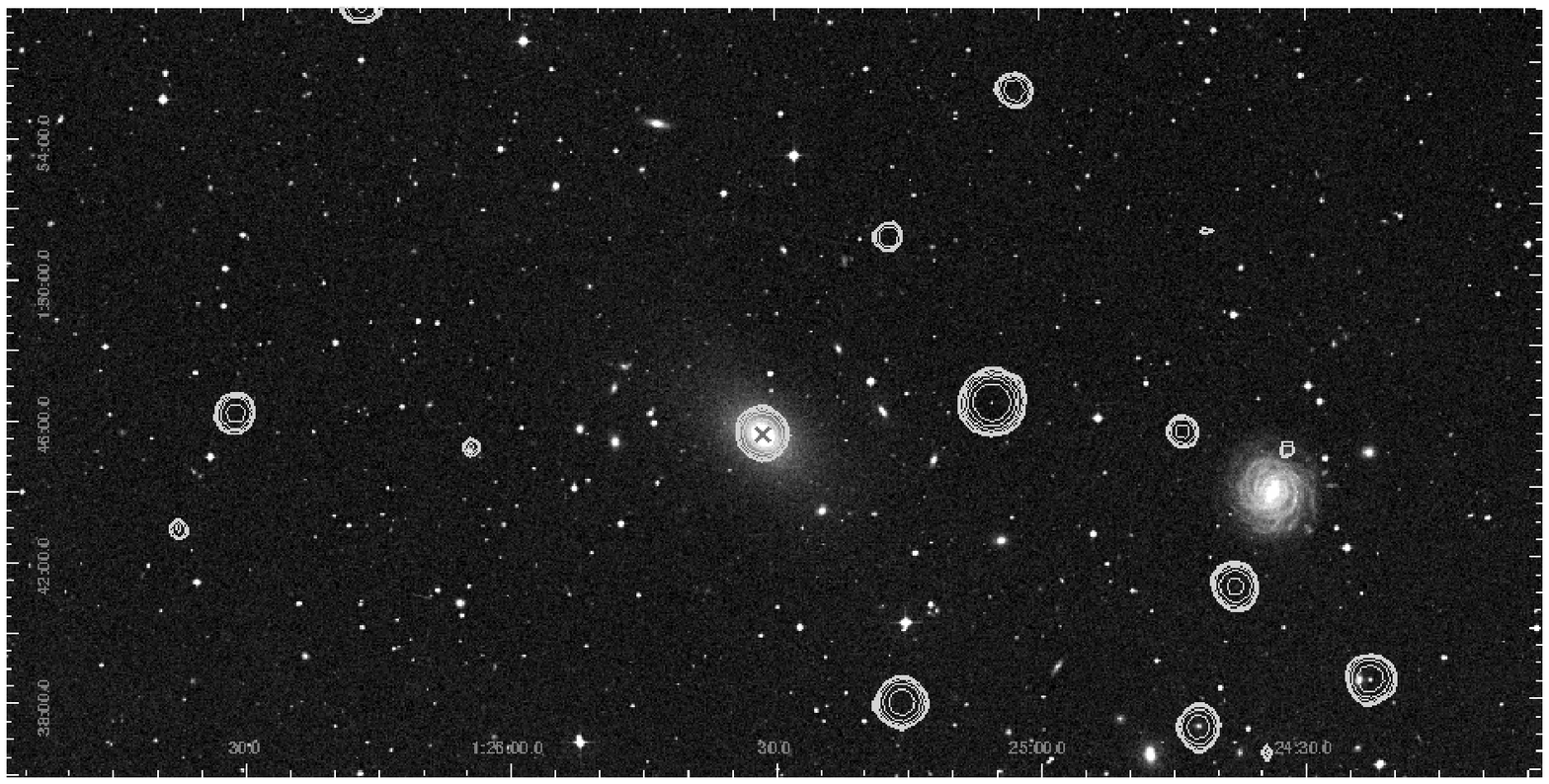}
\caption{NGC507 and NGC533.}
\end{figure*}

\begin{figure*}[h!]
\includegraphics{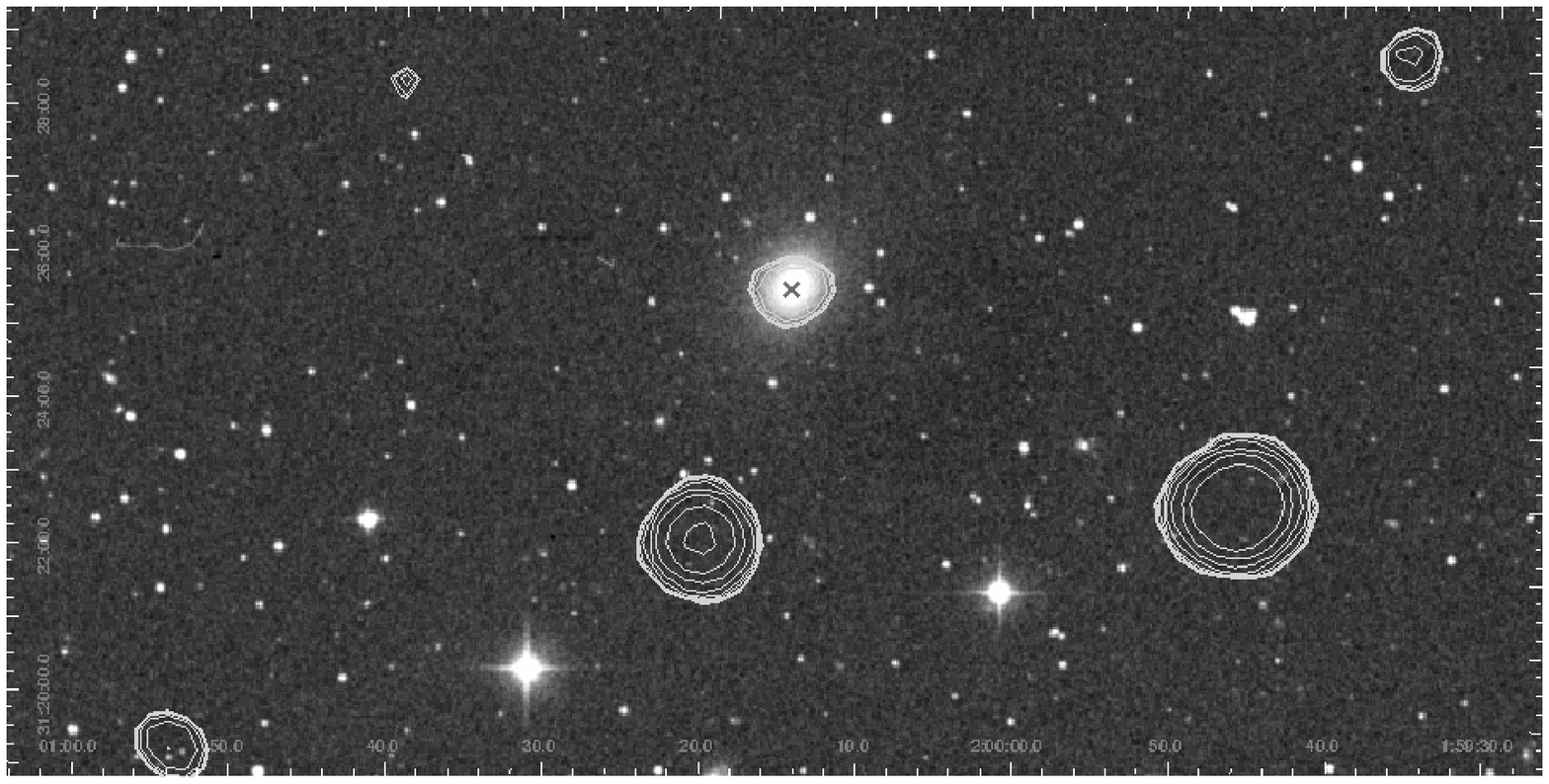}
\includegraphics{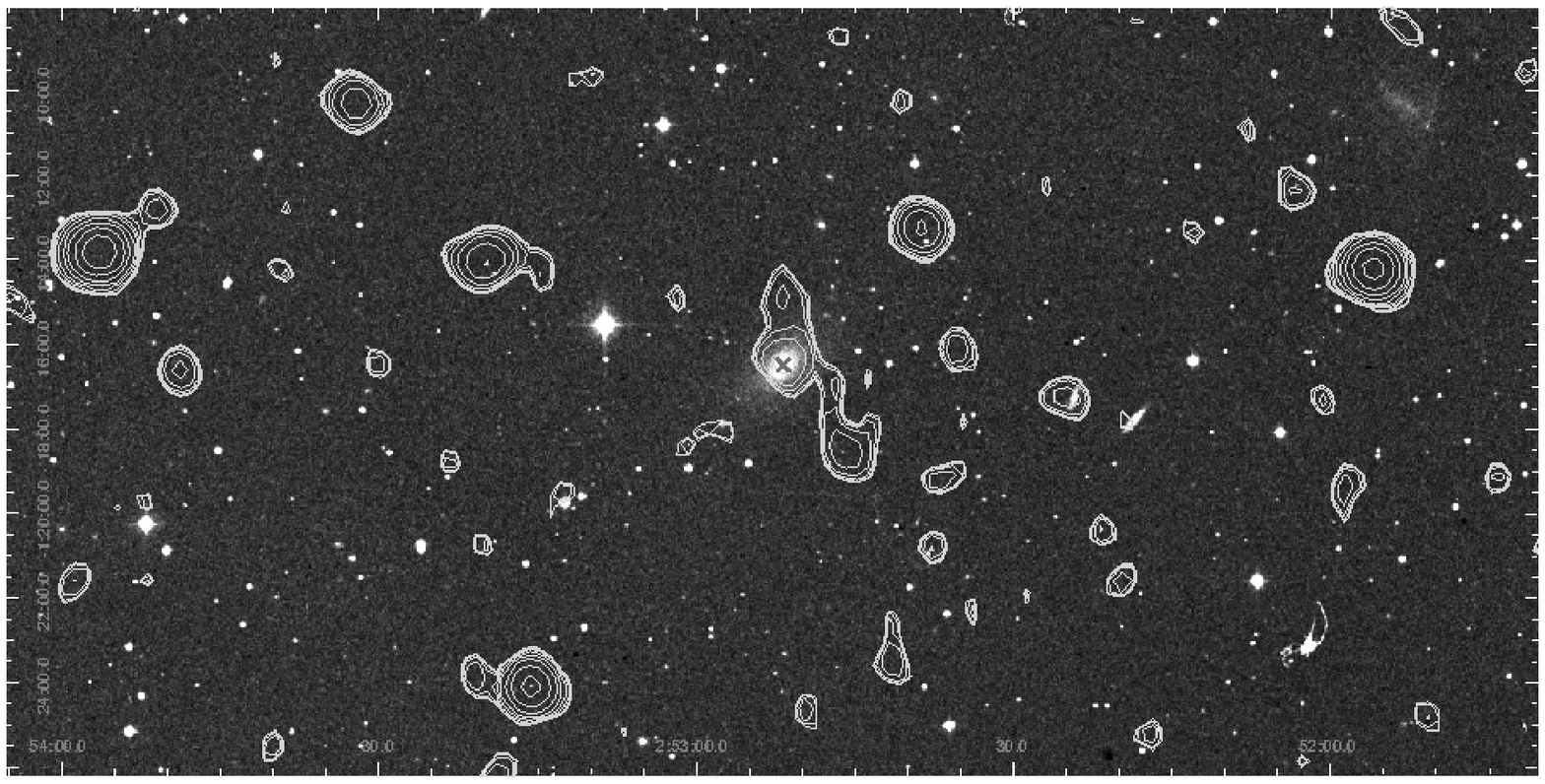}
\caption{NGC777 and NGC1132.}
\end{figure*}

\begin{figure*}[h!]
\includegraphics{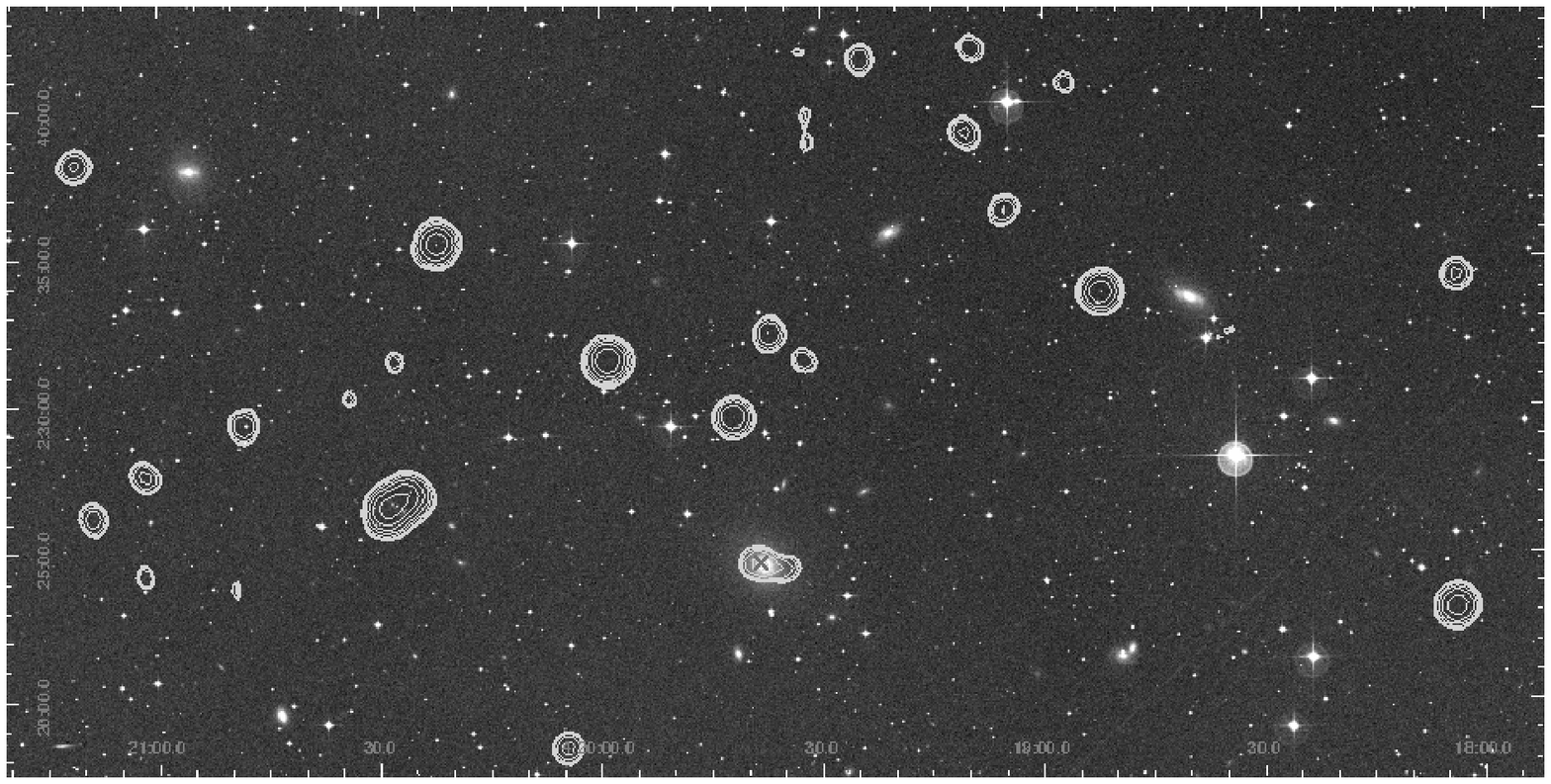}
\includegraphics{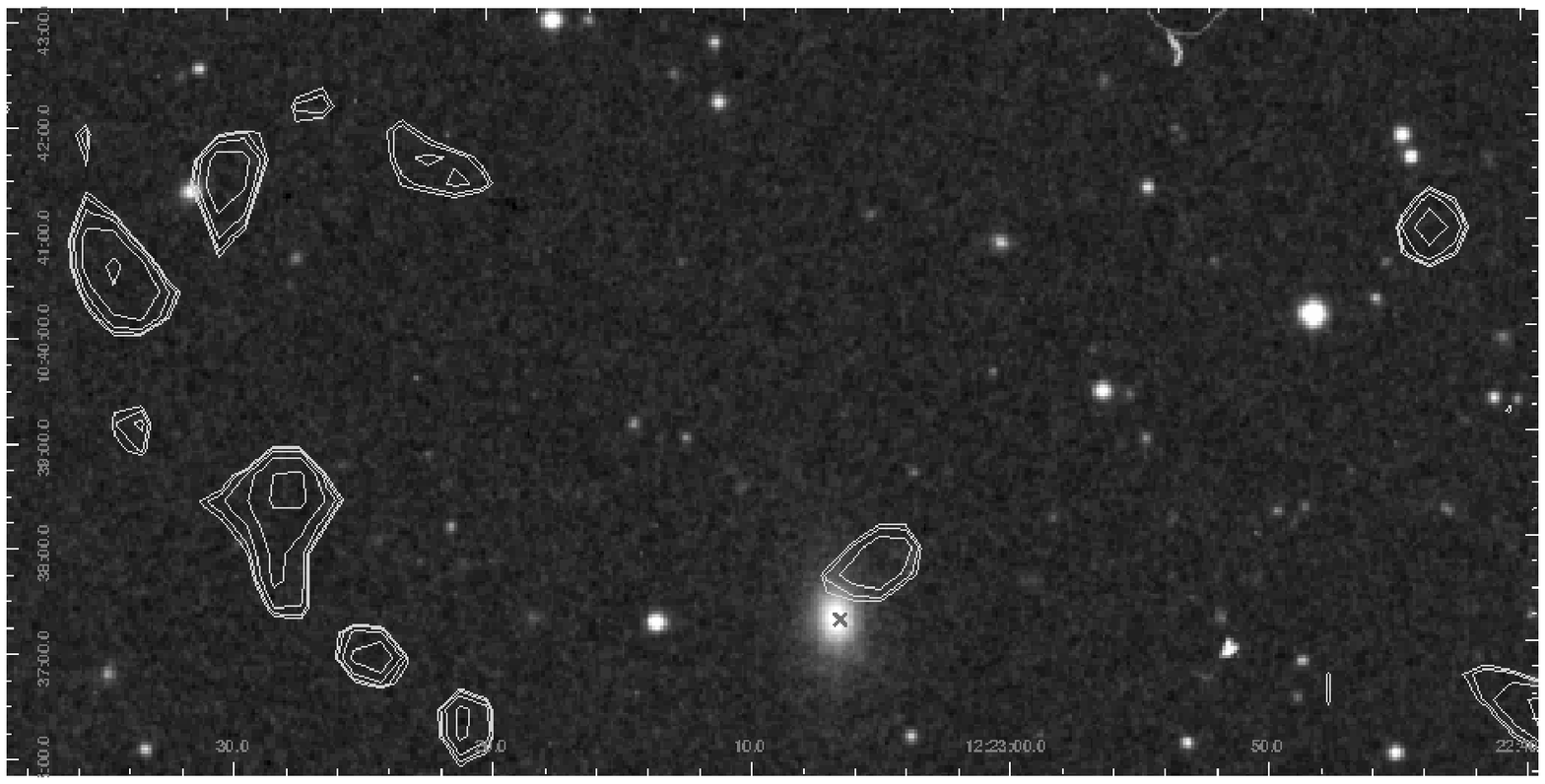}
\caption{NGC1550 and NGC4325.}
\end{figure*}

\begin{figure*}[h!]
\includegraphics{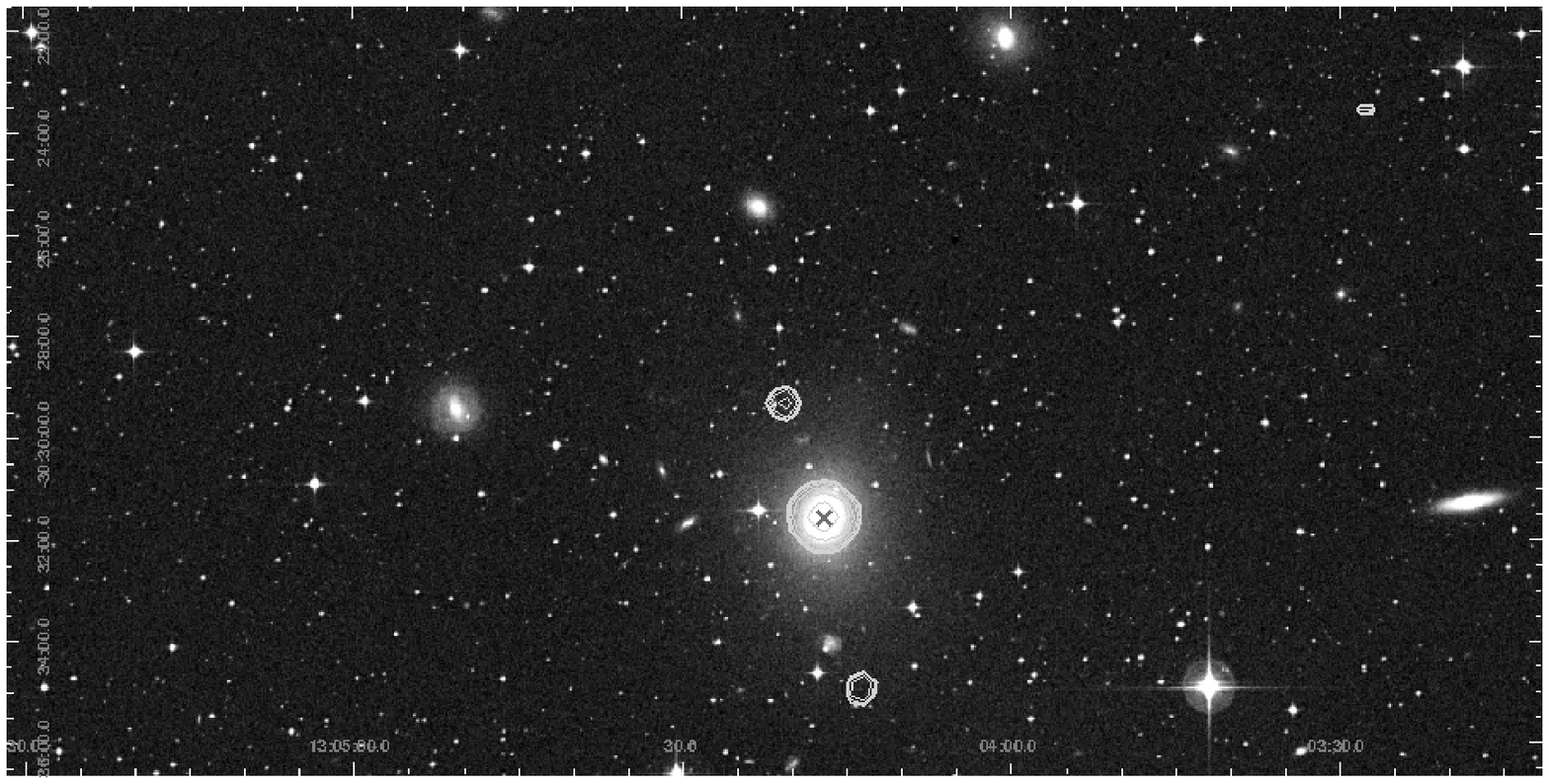}
\includegraphics{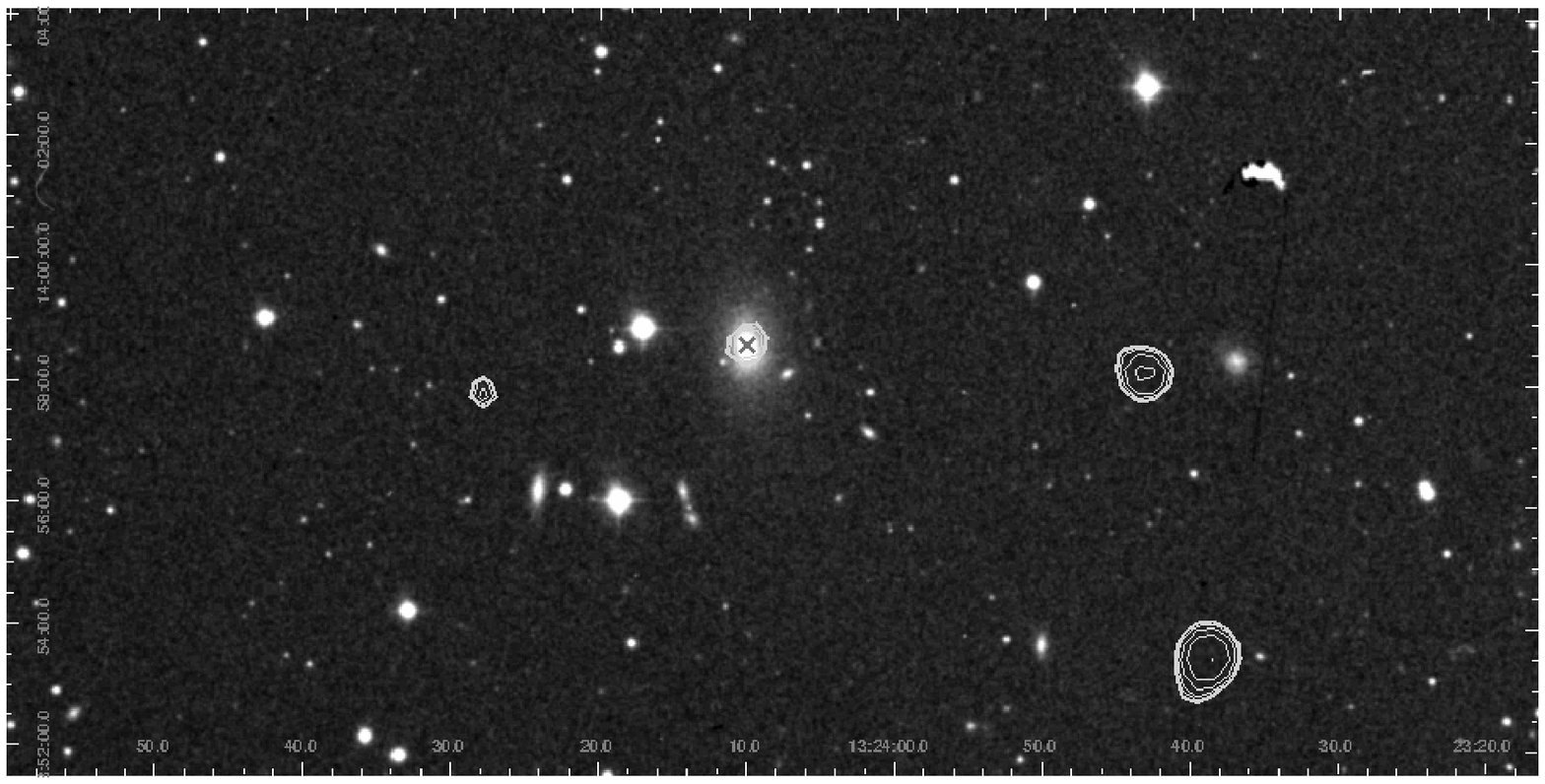}
\caption{NGC4936 and NGC5129.}
\end{figure*}

\begin{figure*}[h!]
\includegraphics{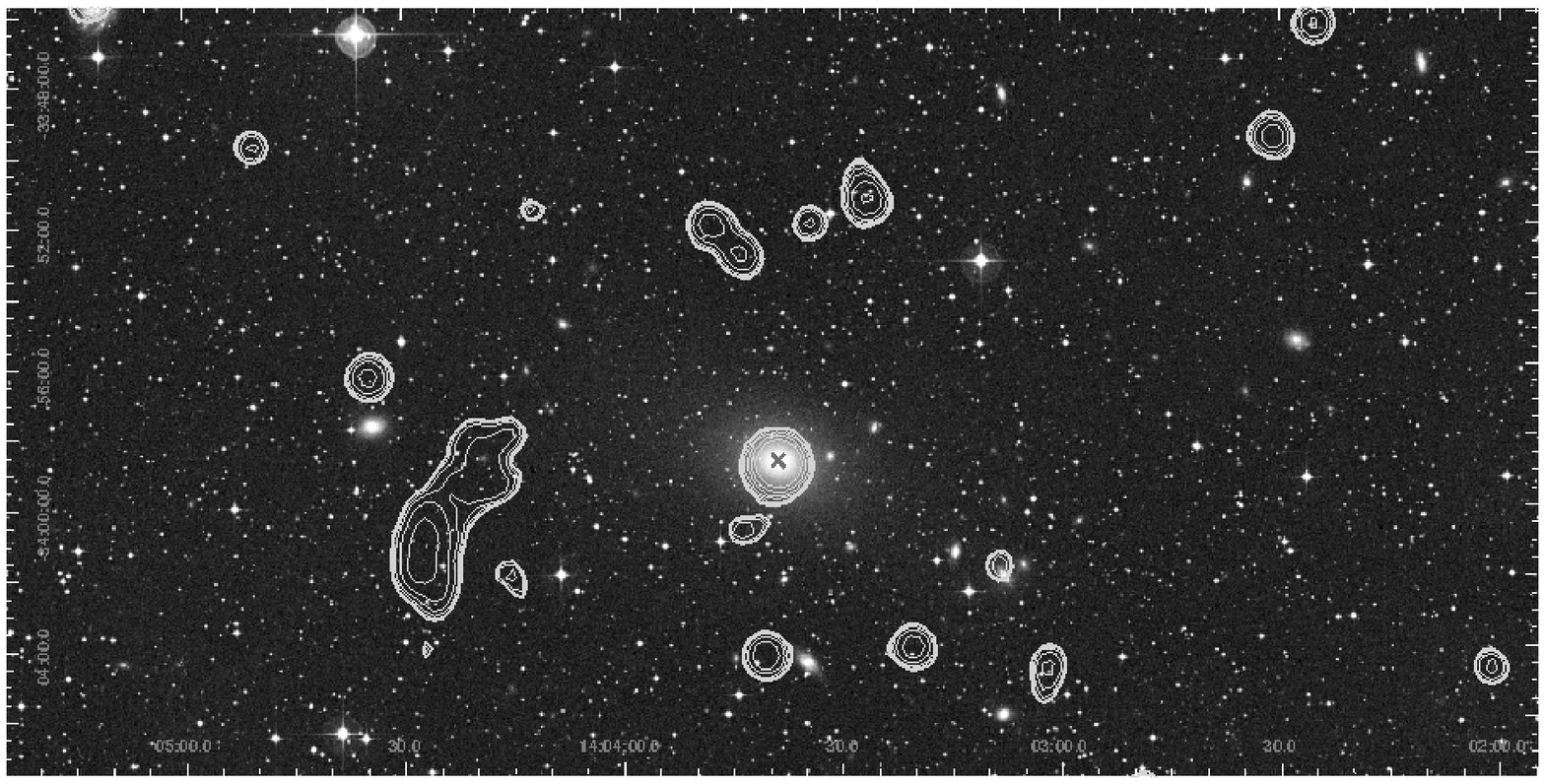}
\includegraphics{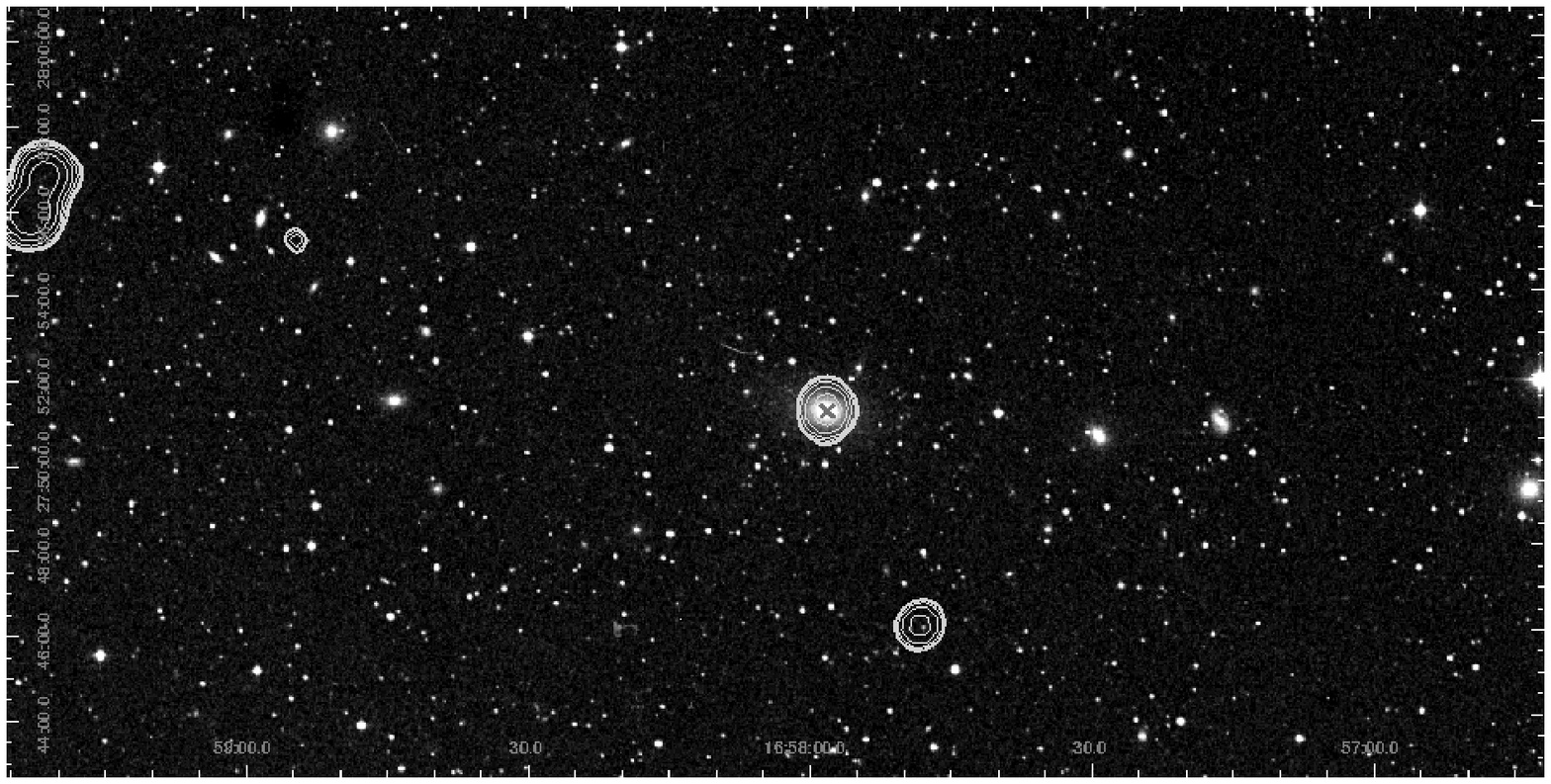}
\caption{NGC5419 and NGC6269.} 
\end{figure*}

\begin{figure*}[h!]
\includegraphics{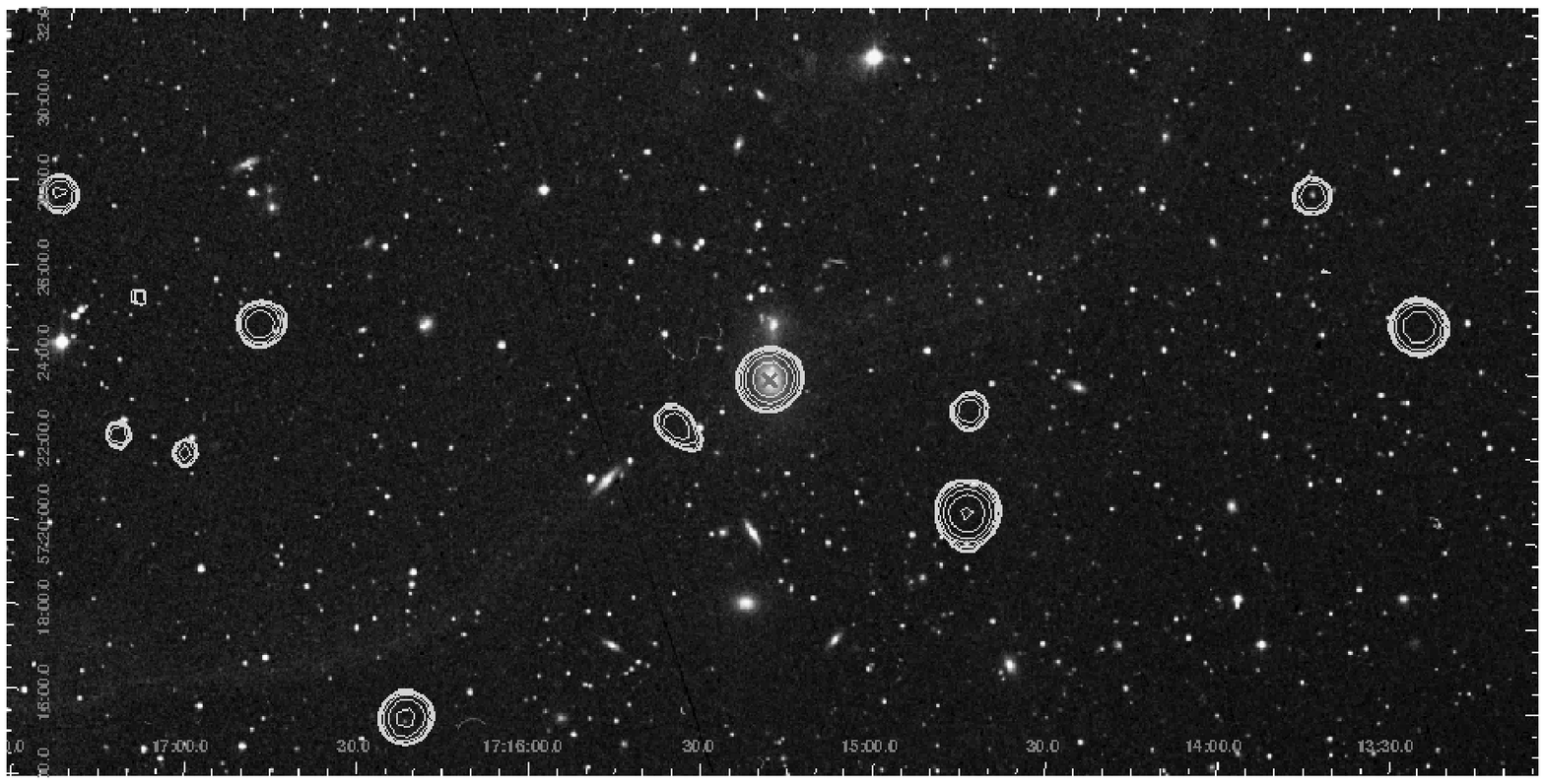}
\includegraphics{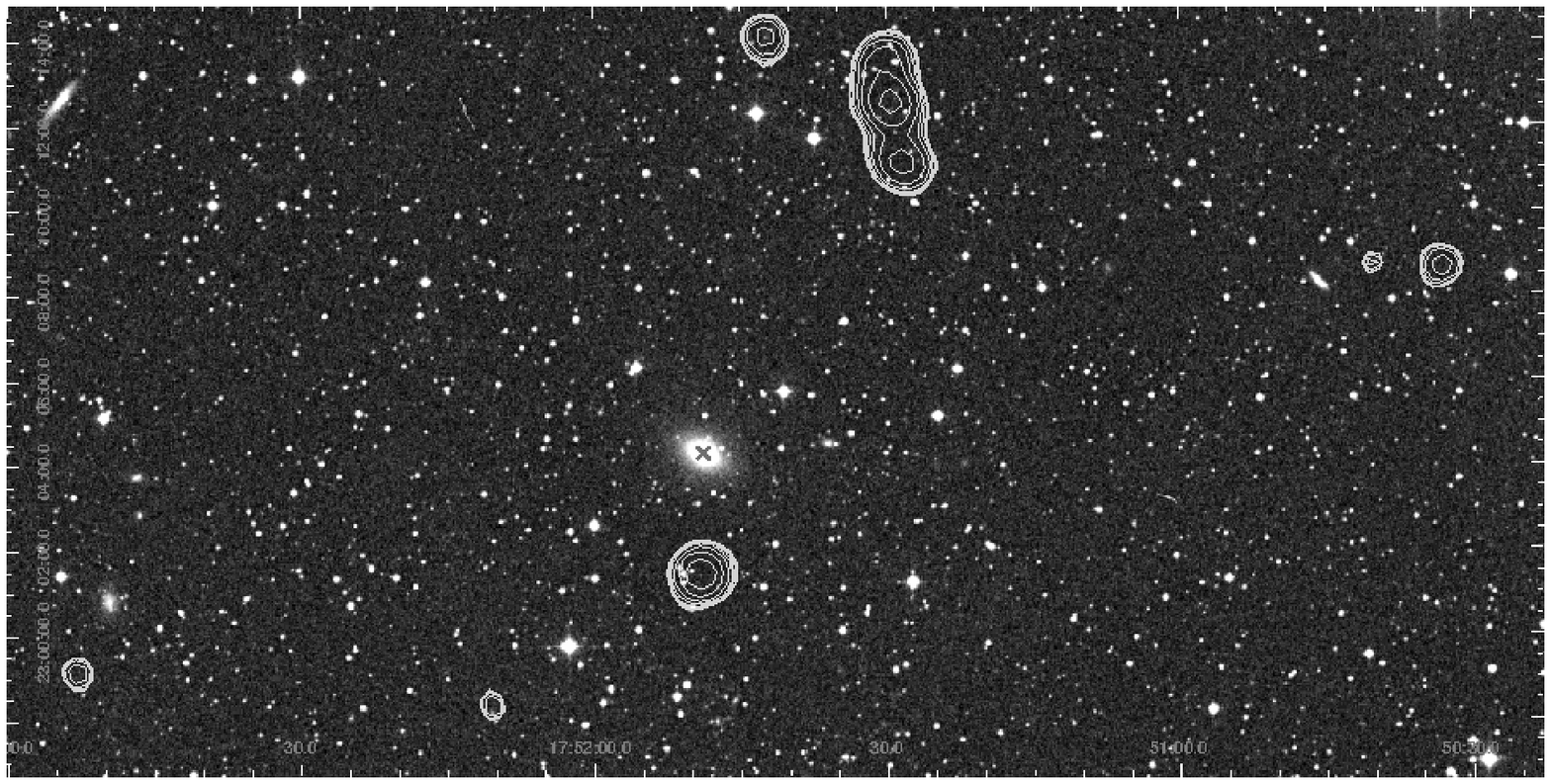}
\caption{NGC6338 and NGC6482.} 
\end{figure*}

\begin{figure*}[h!]
\includegraphics{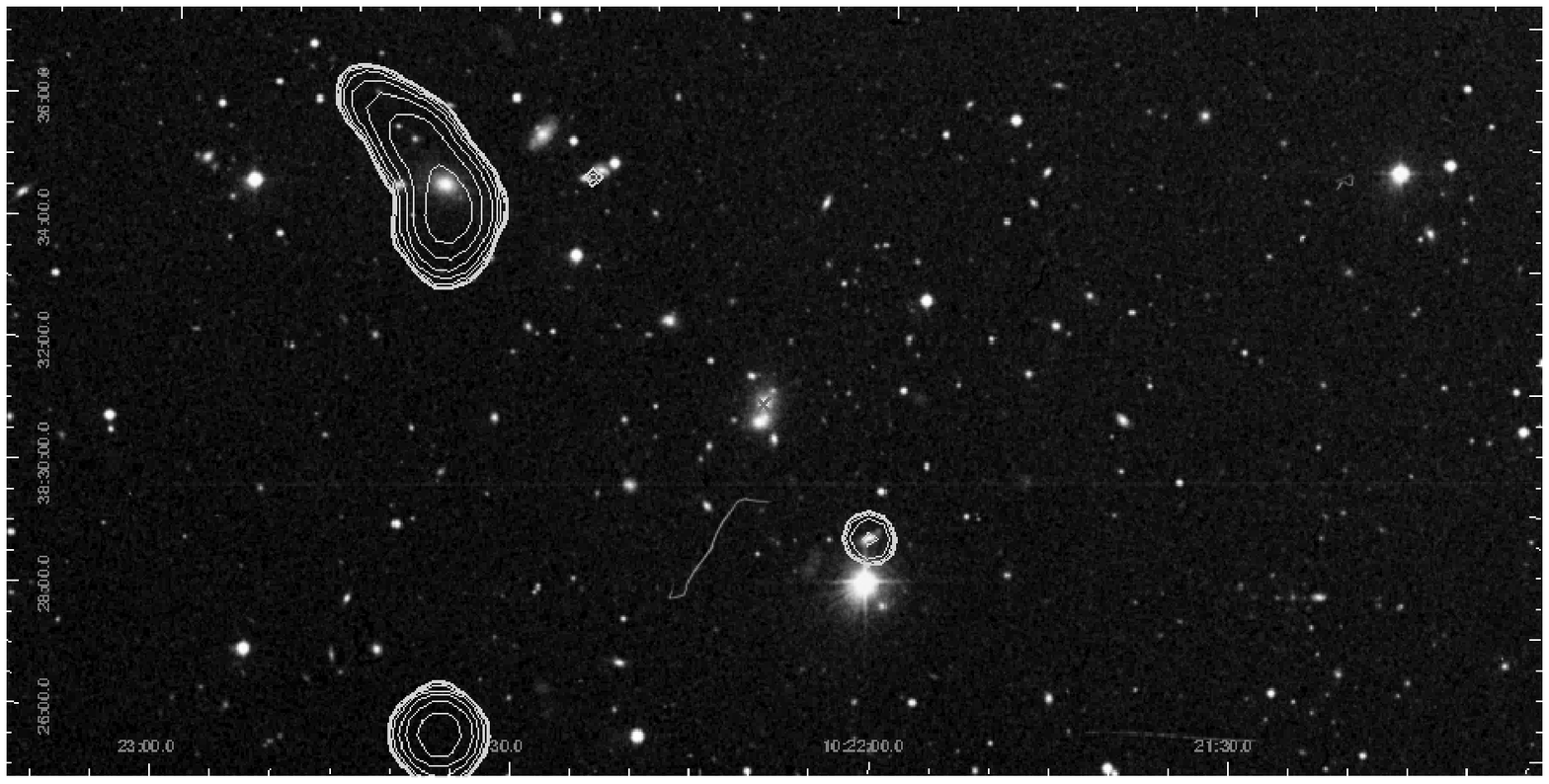}
\includegraphics{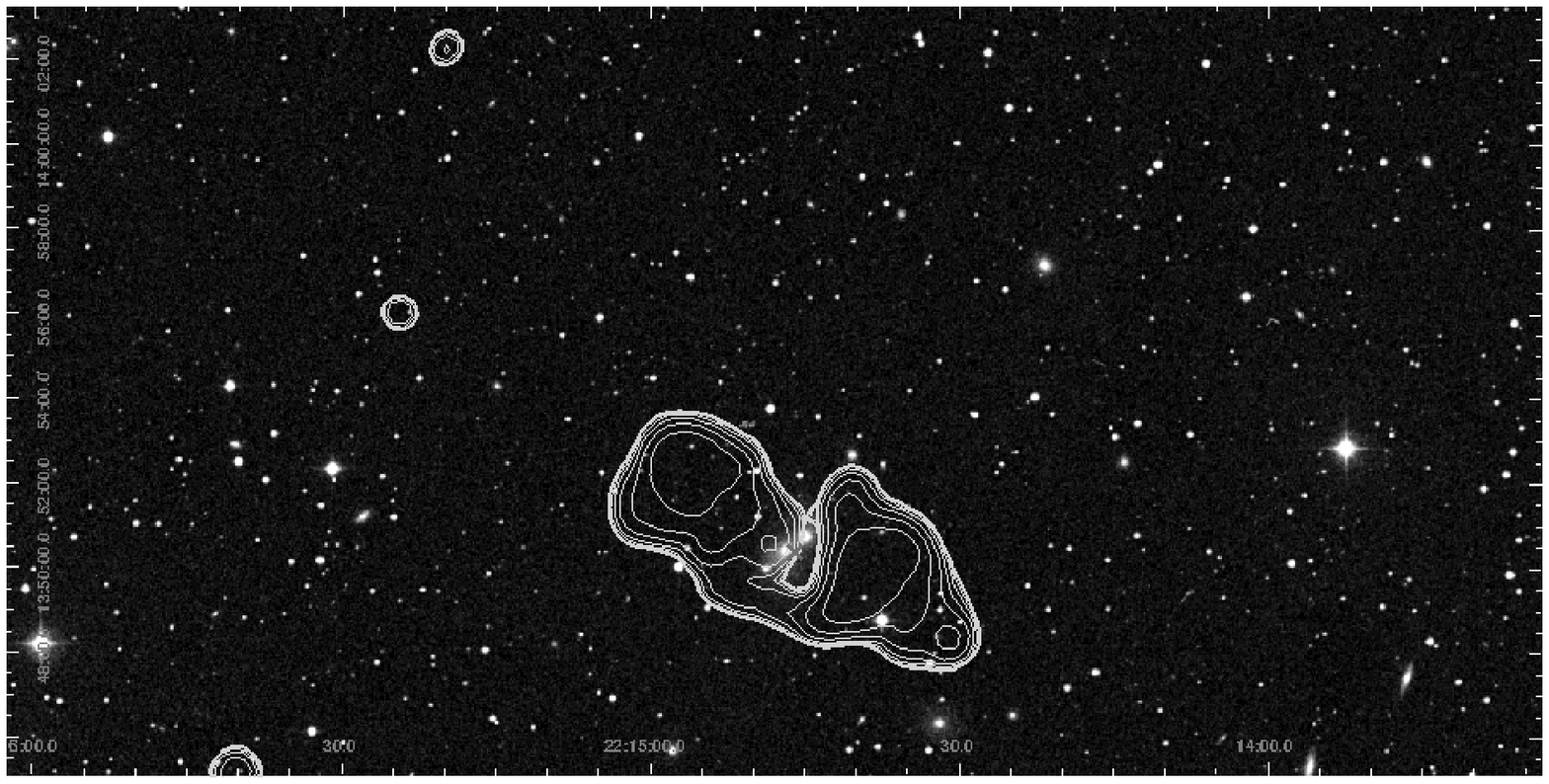}
\caption{RXCJ1022 and RXCJ2214.} 
\end{figure*}
\begin{figure*}[h!]
\includegraphics{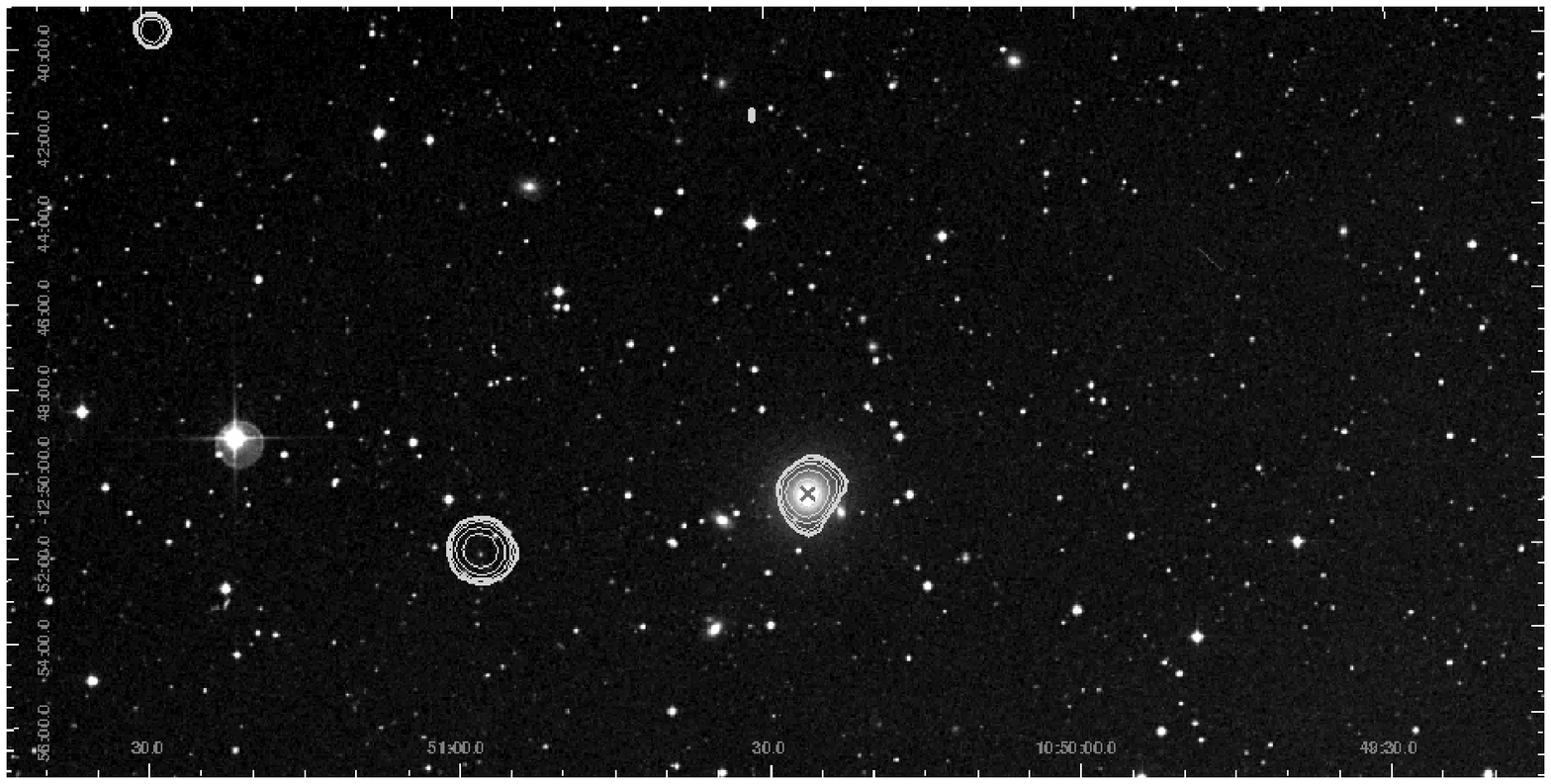}
\caption{SS2B.}
\end{figure*}
\end{document}